\documentclass[11pt]{article}

\usepackage{amssymb}
\usepackage{cite}  

\usepackage{latexsym,amsmath,amsfonts,amscd}
\usepackage{subcaption}
\usepackage{graphics}
\usepackage{graphicx}
\usepackage{caption}
\usepackage{epstopdf}
\usepackage{color}
\usepackage{changebar}
\usepackage{indentfirst}
\usepackage{verbatim}
\usepackage{xcolor}
\usepackage{hyperref}
\usepackage{amssymb}
\usepackage{lscape}
\usepackage{array,multirow}
\usepackage{amsthm}

\usepackage[margin=1.1in]{geometry}

\newcommand{\CFL}{\mathsf{CFL}}		

\theoremstyle{definition}
\newtheorem{definition}{Definition}
\theoremstyle{remark}
\newtheorem{remark}{Remark}
\theoremstyle{proposition}
\newtheorem{proposition}{Proposition}

\newcommand{\U}{{\bf{U}}}
\newcommand{\V}{{\bf{V}}}
\newcommand{\F}{{\bf{F}}}
\newcommand{\G}{{\bf{G}}}
\renewcommand{\H}{{\bf{H}}}
\newcommand{\e}{{\bf{e}}}
\renewcommand{\u}{{\bf{u}}}
\renewcommand{\v}{{\bf{v}}}
\newcommand{\f}{{\bf{f}}}
\newcommand{\g}{{\bf{g}}}

\begin{document}




\title{Hyperbolic models for the spread of epidemics on networks: kinetic description and numerical methods}


\author{Giulia Bertaglia\footnote{Department of Mathematics and Computer Science, University of Ferrara, Via Machiavelli 30, 44121 Ferrara, Italy (giulia.bertaglia@unife.it)}\and
Lorenzo Pareschi\footnote{Department of Mathematics and Computer Science, University of Ferrara, Via Machiavelli 30, 44121 Ferrara, Italy (lorenzo.pareschi@unife.it)}}

\maketitle

\begin{abstract}
We consider the development of hyperbolic transport models for the propagation in space of an epidemic phenomenon described by a classical compartmental dynamics. The model is based on a kinetic description at discrete velocities of the spatial movement and interactions of a population of susceptible, infected and recovered individuals. Thanks to this, the unphysical feature of instantaneous diffusive effects, which is typical of parabolic models, is removed. In particular, we formally show how such reaction-diffusion models are recovered in an appropriate diffusive limit. The kinetic transport model is therefore considered within a spatial network, characterizing different places such as villages, cities, countries, etc. The transmission conditions in the nodes are analyzed and defined. Finally, the model is solved numerically on the network through a finite-volume IMEX method able to maintain the consistency with the diffusive limit without restrictions due to the scaling parameters. Several numerical tests for simple epidemic network structures are reported and confirm the ability of the model to correctly describe the spread of an epidemic. 
\end{abstract}



{\bf Keywords:}
Kinetic equations, hyperbolic systems, spatial epidemic models, SIR model, network models, IMEX Runge-Kutta schemes, diffusive limit




\tableofcontents

\section{Introduction}
The ongoing COVID-19 pandemic has led to a strong interest from researchers around the world in building and studying new epidemiological models capable of describing the progress of the epidemic (see for example \cite{gatto2020,kraemer2020,colombo2020,Giordano2020,albi2020b,Franco2020,bellomo2020multiscale} and the references therein). 
Mathematical models can help make scenarios of the evolution of a pandemic and are an indispensable tool to support government decision-making on the control of infectious diseases. Together with computer simulations, they permit to build and test hypothesis, quantify the uncertainty related to random input parameter values, and estimate key parameters from collected data. 
Usually, SIR-type compartmental models, inspired by the seminal work of Kermack and McKendrick, are adopted \cite{hethcote2000,capasso1978}. In the classical setting the population is assumed to be divided in three classes: susceptible (S), who can contract the disease, infectious (I), who have contracted the disease and can transmit it, and removed or recovered (R), consisting of those with permanent infection-acquired immunity (eventually, even deceased). More complex models include additional classes of individuals, like the SEIR model, which considers also a subgroup of exposed subjects (E), individuating people in the latent period, who are infected but not yet infectious; or like the MSEIR model, which includes also the class M of infants with passive immunity \cite{hethcote2000}. The choice of which compartments to include in a model depends on the characteristics of the infectious disease analyzed and the aim of the model. Models including several compartments have been especially designed to deal with the COVID-19 pandemic (see for example \cite{gatto2020, Giordano2020}). 

All these classical compartmental models represent the spread of the epidemic only concerning the temporal evolution of the disease among the population, but not taking into account spatial effects. In many cases, the concept of the average behavior of a large population is sufficient to provide useful guidance on the development of the epidemic. However, the importance of the spatial component of many transmission systems is being increasingly recognized, especially when there is a need to consider spatially heterogeneous  interventions, as happened for COVID-19 \cite{riley2015}. Thus, to describe the spread of epidemics in a more detailed way, and design effective confinement strategies, several spatially extended models, based on systems of parabolic reaction-diffusion equations, have been proposed \cite{sun2012,wang2010,Murray}. However, the parabolic character of these models leads the disease to propagate instantaneously in space, with infinite speeds. This unphysical feature has been avoided through hyperbolic systems based on relaxation approximations\cite{barbera2013,bargmann2011} or nonlocal interactions\cite{colombo2020}.

Furthermore, networks (or graphs) are extremely flexible tools for representing complex systems of interacting components. Each component is represented by a node and each arc (or edge) between nodes describes some sort of interaction between them. Because of their flexibility, networks have been used to model infection spread in different forms. Nodes can describe single individuals, groups of individuals (e.g. households, farms, cities) or locations to which individuals are connected \cite{koch2013, pellis2015}. Links can represent infectious attempts or transmission events or simply social relationships through which the infection can spread, movements of animals between farm, flight routes, streets \cite{pellis2015,balcan2009,merler2010}. We refer to \cite{gatto2020,kraemer2020} for recent applications of these ideas to the specific case of COVID-19.

In this work, without claiming to provide an answer to the complex problem of spatial propagation of an epidemic phenomenon, we try to provide a contribution in this direction by building models based on hyperbolic partial derivative equations that allow a more realistic description of the dynamics of the individuals involved in the epidemic phenomenon. The model, inspired by two-velocity kinetic equations\cite{Toscani1997}, is based on a kinetic description of the diffusive spread of the epidemic governed by a SIR-type dynamics. In particular, it is shown that under a suitable scaling limit the model permits to recover the classical reaction-diffusion SIR-type system \cite{sun2012,wang2010}. Although, to simplify our presentation, we have focused on a simple SIR-type dynamic, the approach can be naturally extended to more realistic compartmental descriptions including more populations.

Subsequently, the hyperbolic system is considered on a spatial network where the nodes represent groups of individuals, typically villages, cities, regions or even countries that evolve through a standard SIR model. In addition, suitable transmission conditions are derived at each arc-node interface. We emphasize that the notion of network used here and in transport based models (so typically a network composed by few nodes where we are interested also on the solution over the arcs) and the notion of network in social sciences (so typically composed by a random graph with a large number of nodes) are very different. Note also, that here, unlike other network models based on hyperbolic balance laws and kinetic equations, such as in chemotaxis and traffic flows \cite{bretti2014,bretti2006,piccoli2006,Tosin2015,MR3200227}, the nodes themselves are evolving. The resulting system of equations is then solved on each arc using a suitable IMEX finite-volume approach that permits to achieve uniform second order accuracy even in stiff regimes where the diffusive behavior dominates \cite{boscarino2017,Naldi2000,Gosse2002}.

Clearly, a model on spatial network schematizes a simplified reality of a two-dimensional model. Such models find particular application in the case in which the directions of displacement are characterized essentially by monodimensional dynamics, as in the case of road traffic and related problems \cite{bretti2014,bretti2006,piccoli2006}. The model here presented aims precisely at describing the spread of an epidemic on the basis of transport dynamics between cities, towns and other inhabited centers, due to the mobility that typically takes place on spatial networks of connection. 

The rest of the manuscript is organized as follows. In Section \ref{section_mathematicalmodel} we introduce the mathematical model. First, we discuss the one-dimensional model in a bounded domain, its main properties, and formally derive  the diffusion limit. Then, in Section \ref{section_network}, the model is considered in a spatial network with appropriate transmission conditions. Section \ref{section_numericalmodel} is dedicated to the description of the adopted IMEX finite-volume scheme, which allows to preserve the limiting parabolic behavior of the system. Several numerical examples, including networks with different arcs and nodes are presented in Section \ref{section_numericalresults}, to illustrate the ability of the model to correctly describe  the spread of the epidemic. Some conclusions are reported in the last Section.

\section{Mathematical model}
\label{section_mathematicalmodel}
\subsection{A SIR-type discrete kinetic transport model}
\label{section_kinetic_SIR}
One of the standard compartmental models widely used in epidemiology is the so-called SIR model, in which the population is assumed to be divided in three classes: the susceptible (S), the infectious (I), and the removed or recovered (R). Let us assume to have a population with similar individuals, without prior immunity, which are uniformly mixed and that do not present behavioral changes during the epidemic, and neglect the vital dynamics represented by births and deaths because of the time scale considered. In the simplest one dimensional (1D) case, if we consider individuals moving in two opposite directions (indicated by signs ``+'' and ``-''), with velocities $\pm \lambda_S$ for susceptible, $\pm \lambda_I$ for infectious and $\pm \lambda_R$ for recovered, we can describe the spatio-temporal dynamics of the population through the following SIR-type discrete velocity model
\begin{equation}
\begin{split}
	\frac{\partial S^+}{\partial t} + \lambda_S \frac{\partial S^+}{\partial x} &= - f(S^+,I) - \frac{1}{2\tau_S}\left(S^+ - S^-\right)										\\ 
	\frac{\partial S^-}{\partial t} - \lambda_S \frac{\partial S^-}{\partial x} &= - f(S^-,I) + \frac{1}{2\tau_S}\left(S^+ - S^-\right)										\\ 
	\frac{\partial I^+}{\partial t} + \lambda_I \frac{\partial I^+}{\partial x} &=  f(S^+,I) -\gamma I^+ - \frac{1}{2\tau_I}\left(I^+ - I^-\right)								\\ 
	\frac{\partial I^-}{\partial t} - \lambda_I \frac{\partial I^-}{\partial x} &=  f(S^-,I) -\gamma I^- + \frac{1}{2\tau_I}\left(I^+ - I^-\right)								\\  
	\frac{\partial R^+}{\partial t} + \lambda_R \frac{\partial R^+}{\partial x} &= \gamma I^+ - \frac{1}{2\tau_R}\left(R^+ - R^-\right)									\\ 
	\frac{\partial R^-}{\partial t} - \lambda_R \frac{\partial R^-}{\partial x} &= \gamma I^- + \frac{1}{2\tau_R}\left(R^+ - R^-\right) .									
\end{split}
\label{eq.SIR_kinetic_diag}	
\end{equation}
In the above system, individuals $S(x,t)$, $I(x,t)$ and $R(x,t)$, representing, respectively, the number of susceptible, infectious and recovered, at location $x \in \Omega \subseteq \mathbb{R}$ and time $t>0$ (expressed in relative value with respect to the total reference population size $N = S + I + R$), are defined as
\begin{equation*}
 	S = S^+ + S^- , \quad 
 	I = I^+ + I^- , \quad 
 	R = R^+ + R^- .	
\end{equation*}
The parameter $\gamma$ represents the recovery rate, while the transmission of the infection is governed by an incidence function $f(S^\pm,I)$ modeling the transmission of the disease. We assume local interactions characterize the general incidence function 
\begin{equation}
f(S^\pm,I)=\beta \frac{S^\pm I^p}{1+kI},
\label{eq:incf}
\end{equation}
where the classic bilinear case corresponds to $p = 1$, $k=0$, even though it has been observed that an incidence rate that increases more than linearly with respect to the number of infectious $I$ can occur under certain circumstances \cite{capasso1978,barbera2013,korobeinikov2005,liu1986}. The parameter $\beta$ characterizes the contact rate \cite{hethcote2000}, whereas the case $k > 0$ takes into account social distancing and other control effects which occur during the progress of the disease \cite{wang2020,Franco2020}. Notice that, when $p=1$, for an infection prevalence that tends to 1 (in relative values), the corresponding incidence rate is approximately $f\approx\frac{\beta}{k} S$, giving a purely frequency-dependent transmission, with $k$ acting like a damping parameter. On the other hand, for a low infection prevalence, the transmission is purely density-dependent, resulting approximately $f\approx\beta SI$. In general, although for simplicity of notations we have omitted the spatial dependence, the parameters $\gamma$ and $\beta$ can depend on the position $x$, as well as the positive relaxation times $\tau_S$, $\tau_I$, and $\tau_R$. 
It is important to notice that in this model the class of $I$ includes also the asymptomatic population, which still acts like infected and can transmit the disease, even if never developing symptoms. Moreover, it is assumed that an individual, after recovering, always becomes immune. 

The model must then be supplemented with appropriate initial and boundary data. A more detailed discussion about boundary conditions is postponed to Section \ref{section_network} where system \eqref{eq.SIR_kinetic_diag} is considered within a spatial network. A structural property of \eqref{eq.SIR_kinetic_diag} is that the first four equations are independent of the last two. Once $I^\pm$ are known, the explicit forms of $R^\pm$ can be determined directly by solving the last two equations.

The standard threshold of epidemic models is the well-known reproduction number $R_0$, which defines the average number of secondary infections produced when one infected individual is introduced into a host population in which everyone is susceptible \cite{hethcote2000}. This number determines when an infection can invade and persist in a new host population. For many deterministic infectious disease models, an infection begin in a fully susceptible population if and only if $R_0 > 1$. 
More precisely, assuming no inflow/outflow boundary conditions in $\Omega$, summing up the third and fourth equations in \eqref{eq.SIR_kinetic_diag} and integrating over space  we have
\[
\frac{\partial}{\partial t} \int_{\Omega} I(x,t)\,dx =  \int_{\Omega} f(S,I)\,dx-\int_{\Omega} \gamma(x) I(x,t)\,dx \geq 0
\]
when
\begin{equation}
R_0(t)=\frac{\int_{\Omega} f(S,I)\,dx}{\int_{\Omega} \gamma(x) I(x,t)\,dx} \geq 1.
\end{equation}
The above quantity therefore defines the basic reproduction number for system \eqref{eq.SIR_kinetic_diag} describing the space averaged instantaneous variation of the number of infective individuals at time $t>0$.

Let us also observe that, under the same no inflow/outflow boundary conditions, summing up the equations in \eqref{eq.SIR_kinetic_diag} and integrating in $\Omega$ yields the conservation of the total population number
\[
\frac{\partial}{\partial t} \int_{\Omega} (S(x,t)+I(x,t)+R(x,t))\,dx =0. 
\] 
If we now introduce the fluxes, defined by
\begin{equation}
 	J_S = \lambda_S \left(S^+ - S^-\right) , \quad 
 	J_I = \lambda_I \left(I^+ - I^-\right) , \quad 
 	J_R = \lambda_R \left(R^+ - R^-\right) ,	  	\label{eq.fluxes_kinetic_diag}
\end{equation}
we obtain a hyperbolic model equivalent to \eqref{eq.SIR_kinetic_diag}, but presenting in the following formulation a macroscopic description of the propagation of the epidemic at finite speeds
\begin{equation}
\begin{split}
	\frac{\partial S}{\partial t} + \frac{\partial J_S}{\partial x} &= -f(S,I) 			\\ 
	\frac{\partial I}{\partial t} + \frac{\partial J_I}{\partial x} &= f(S,I)  -\gamma I	\\ 
	\frac{\partial R}{\partial t} + \frac{\partial J_R}{\partial x} &= \gamma I 			\\ 
	\frac{\partial J_S}{\partial t} + \lambda_S^2 \frac{\partial S}{\partial x} &= -f(J_S,I)  -\frac{J_S}{\tau_S}										\\ 
	\frac{\partial J_I}{\partial t} + \lambda_I^2 \frac{\partial I}{\partial x} &= \frac{\lambda_I}{\lambda_S}f(J_S,I)  -\gamma J_I -\frac{J_I}{\tau_I}	\\ 
	\frac{\partial J_R}{\partial t} + \lambda_R^2 \frac{\partial R}{\partial x} &= \frac{\lambda_R}{\lambda_I}\gamma J_I -\frac{J_R}{\tau_R} .				
\end{split}
\label{eq.SIR_kinetic}	
\end{equation}
%
Formally, the above system \eqref{eq.SIR_kinetic} is a system of balance laws which can be rewritten in a compact form as
\begin{equation}
\begin{split}
	\partial_t \U + \partial_x \V &= \F(\U) \\
	\partial_t \V + \boldsymbol{\Lambda}^2 \partial_x \U &= \G(\U,\V) + \H(\V) ,
\end{split}
\label{systcompactform}
\end{equation}
in which
	\[\U =
\begin{pmatrix} 
  	S \\ I \\ R 
\end{pmatrix}, \quad
	\V = 
\begin{pmatrix} 
 	J_S \\ J_I \\ J_R
\end{pmatrix}, \quad
	\boldsymbol{\Lambda} = 
\begin{pmatrix} 
	\lambda_S &0 &0 \\ 0 &\lambda_I &0 \\ 0 &0 &\lambda_R
\end{pmatrix}, \]\\
	\[\F(\U) = 
\begin{pmatrix} 
  	-f(S,I) \\ f(S,I) - \gamma I \\ \gamma I
\end{pmatrix}, \quad
	\G(\U,\V) = 
\begin{pmatrix}
  	-f(J_S, I) \\ \frac{\lambda_I}{\lambda_S}f(J_S, I) -\gamma J_I  \\ \frac{\lambda_R}{\lambda_I}\gamma J_I 
\end{pmatrix}, \quad
	\H(\V) = -
\begin{pmatrix}
  	{J_S}/{\tau_S} \\ {J_I}/{\tau_I} \\ {J_R}/{\tau_R}
\end{pmatrix}.\]
It is immediate to see that system \eqref{systcompactform} is symmetric hyperbolic in the sense of Friedrichs-Lax \cite{friedrichs1971}, with real finite characteristic velocities (eigenvalues) $\lambda_S$, $\lambda_I$, $\lambda_R$, and a complete set of linearly independent eigenvectors. All the eigenvectors are associated with genuinely non-linear fields, defining shocks and rarefactions, and the Riemann invariants of the system correspond to the kinetic variables
\begin{equation}
S^{\pm} = \frac{1}{2} \left( S \pm \frac{J_S}{\lambda_S} \right) , \quad 
I^{\pm} = \frac{1}{2} \left( I \pm \frac{J_I}{\lambda_I} \right) ,\quad 
R^{\pm} = \frac{1}{2} \left( R \pm \frac{J_R}{\lambda_R} \right)\hspace{0.5mm} .
\label{eq.RI}
\end{equation}
Furthermore, the symmetric hyperbolicity of the system guarantees the well-posedness of the Cauchy problem (i.e. existence, uniqueness and continuous dependence of the solutions on the initial data) for suitable smooth initial data (see \cite{Amadori1999,muller1998}).

\subsection{Diffusion limit}
Let us now consider the behavior of the model in diffusive regimes. To this aim, let us introduce the diffusion coefficients
\begin{equation}
D_S=\lambda_S^2 \tau_S,\quad D_I=\lambda_I^2 \tau_I,\quad D_R=\lambda_R^2 \tau_R,
\label{eq:diff}
\end{equation}
which characterize the diffusive transport mechanism of susceptible, infectious and recovered, respectively. The diffusion limit of the system is formally recovered letting the relaxation times $\tau_S,\tau_I,\tau_R\to 0$, and simultaneously the characteristic speeds $\lambda_S,\lambda_I,\lambda_R\to\infty$,  while keeping the diffusion coefficients \eqref{eq:diff} finite. Under this scaling, from the last three equations in \eqref{eq.SIR_kinetic} we get a proportionality relation between the fluxes and the spatial derivatives  
\begin{equation}
 	J_S = -D_S \frac{\partial S}{\partial x} , \quad 
 	J_I = -D_I \frac{\partial I}{\partial x}, \quad 
 	J_R = -D_R \frac{\partial R}{\partial x}	,  	\label{eq.fluxes}
\end{equation}
which inserted into the first three equations, lead to the parabolic reaction-diffusion system \cite{Murray}
\begin{equation}
\begin{split}
	&\frac{\partial S}{\partial t} = -f(S,I) + \frac{\partial}{\partial x}\left(D_S \frac{\partial S}{\partial x}\right)  					\\ 
	&\frac{\partial I}{\partial t} = f(S,I)  -\gamma I	+ \frac{\partial}{\partial x}\left(D_I \frac{\partial I}{\partial x}\right)	\\ 
	&\frac{\partial R}{\partial t}  = \gamma I + \frac{\partial}{\partial x}\left(D_R \frac{\partial R}{\partial x}\right).					
\end{split}
\label{eq.SIR}	
\end{equation}
We refer to \cite{Toscani1997} concerning rigorous results on the diffusion limit of discrete velocity kinetic models of type \eqref{eq.SIR_kinetic_diag}. Here we limit ourselves to note that these results cannot be straightforwardly extended to our case, since the epidemic reaction terms destroy the typical monotone behavior of solutions for such discrete kinetic systems. Although interesting this aspect goes beyond the scopes of the present manuscript and will be the subject of future investigations.

Before considering the extension of the model to the case of a spatial network, some considerations should be made.
\smallskip

\begin{remark}~
\begin{itemize}
\item The hyperbolic model \eqref{eq.SIR_kinetic} differs from the models proposed in \cite{barbera2013,bargmann2011} where a simple relaxation approximation was used to avoid the paradox of infinite propagation speed. In particular, in \cite{bargmann2011} a different asymptotic behavior was considered, corresponding to the case $\tau_S$, $\tau_I$, $\tau_R \to \infty$, in which the dynamics can be reduced to a pair of coupled second order wave equations for $S$ and $I$ eliminating the fluxes $J_S$ and $J_I$. It is easy to verify that this simplification is not possible in \eqref{eq.SIR_kinetic} due to the presence of the incidence function and recovery rate on the r.h.s. of the flux equations. 
\item Although the model here discussed, for simplicity of presentation, is based on a simple SIR structure, the approach can be extended naturally to more realistic compartmental models like SEIR/MSEIR \cite{hethcote2000} or models especially designed to deal with the COVID-19 pandemic \cite{gatto2020, Giordano2020}. Similarly, the analogous diffusion limit permits to recover the corresponding diffusive system for the specific compartmental dynamic.
\end{itemize}
\end{remark}
\begin{figure}[t]
\centering
\vskip .5cm
\includegraphics[width=0.55\linewidth]{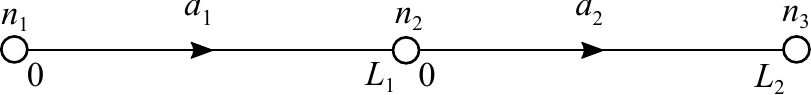}
\caption{Schematic representation of a simple network composed by 3 nodes ($n_1,n_2,n_3$) and 2 arcs ($a_1,a_2$).}
\label{fig.network_scheme}
\end{figure}
\subsection{The SIR-type kinetic transport model on networks}
\label{section_network}
A network or a connected graph $\mathcal{G}=(\mathcal{N},\mathcal{A})$ is composed of two finite sets: a set of $N$ nodes (or vertices) $\mathcal{N}$ and a set of $A$ arcs (or edges) $\mathcal{A}$, such that an arc connects a pair of nodes \cite{piccoli2006}. Arcs are bidirectional, as the graph is non-oriented, but an artificial orientation needs to be fixed in order to define a sign for the velocities and therefore the fluxes. At a node $n \in \mathcal{N}$, two different types of edges are present: incoming and outgoing \cite{bretti2006,bretti2014}. \par
For instance, in the simple network shown in Fig.~\ref{fig.network_scheme}, at node $n_2$ there is an incoming arc ($a_1$) and an outgoing one ($a_2$). The $A$ arcs of the network are parametrized as intervals $a_i =[0, L_i], i =1,\ldots,A$. For an incoming arc, $L_i$ is the abscissa of the node, whereas for an outgoing arc the same abscissa is $0$ (see Fig.~\ref{fig.network_scheme}).\par
In the model here proposed, the nodes of the network identify a particular location of interest (a municipality, a province, a region, a nation...), while each arc represents the whole set of paths connecting each location to the others. Hence, each node is considered with its own initial population and a localized social dynamics, which can be influenced by those individuals moving from and to the different locations considered in the network, traveling through the arcs. This produces a network model in which on each arc the spatial 1D system \eqref{eq.SIR_kinetic} is solved, while on each node the SIR compartmental model with speed alignment is evaluated \cite{hethcote2000}. To ensure the correct coupling between nodes and arcs, specific transmission conditions must be prescribed at nodes.\par
It is worth to highlight that, unlike other network models based on hyperbolic balance laws and kinetic equations, such as in chemotaxis and traffic flows \cite{piccoli2006,bretti2006,bretti2014}, in the here proposed model the nodes themselves are evolving. This slightly changes the transmission conditions, which result simply doubled for each node with respect to networks having non-evolutive nodes. The reason behind this duplication lays on the difference between left and right boundary states, which is due to the evolution of the epidemic at the node. Without the epidemic evolution in time at the nodes, the two boundary states would coincide.\par
\begin{figure}[t]
\centering
\includegraphics[width=0.75\linewidth]{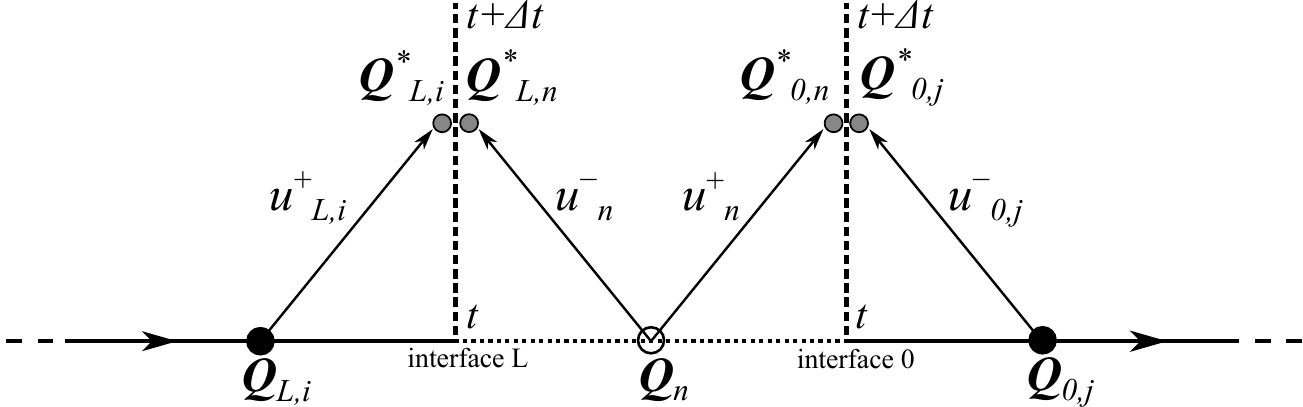}
\caption{Layout of the Riemann problems solved to calculate the states at time $t+\Delta t$, imposing the transmission conditions at the left ($L$) and right ($0$) boundary of a generic node $n \in \mathcal{N}$ at which, respectively, $a_i \in \mathcal{A}, i = 1,\ldots,N_{a,L}$ and $a_j \in \mathcal{A}, j = 1,\ldots,N_{a,0}$ arcs converge. }
\label{fig.node_conditions_scheme}
\end{figure}
\subsubsection{Transmission conditions at nodes}
\label{section_node_conditions}
The imposition of transmission conditions at arc-node interfaces is a delicate point, since the behavior
of the solution will consistently vary according to the chosen conditions \cite{bretti2014}.
To define transmission conditions at a generic node $n \in \mathcal{N}$ having $a_i \in \mathcal{A}, i = 1,\ldots,N_{a,L}$ incoming arcs and $a_j \in \mathcal{A}, j = 1,\ldots,N_{a,0}$ outgoing arcs, we need to consider $\left(1+N_{a,L}+N_{a,0}\right)$ states at time $t$, $\boldsymbol{Q}_{L,i}$,  $\boldsymbol{Q}_n$ and $\boldsymbol{Q}_{0,j}$, separated by the interface of incoming arcs (L) and the interface of outgoing arcs (0), as shown in Fig.~\ref{fig.node_conditions_scheme}. If variables are discontinuous across these interfaces, $\left(1 + N_{a,L}\right)$ new states originate at interface L, $\boldsymbol{Q}^*_{L,i}$ and $\boldsymbol{Q}^*_{L,n}$, and $\left(1 + N_{a,0}\right)$ new states originate at interface 0, $\boldsymbol{Q}^*_{0,n}$ and $\boldsymbol{Q}^*_{0,j}$, at time $t+\Delta t$ \cite{piccoli2006}. To compute them, we sought for the solution of $\left(2 + N_{a_L} + N_{a,0}\right)$ Riemann problems, recurring to the Riemann Invariants (or kinetic variables) of the system, defined in Eqs.~\eqref{eq.RI}, and to the principle of conservation of fluxes at interfaces \cite{bretti2006,bretti2014}.\par
For each one of the three compartments of individuals of the SIR-type model discussed in \S~\ref{section_kinetic_SIR}, for ease of notation, let us indicate with $u$ the number of individuals of the compartment, with $v$ the corresponding analytical flux, with $\lambda$ its characteristic velocity, and with $u^{\pm}$ the Riemann Invariants. To impose the transmission conditions at interface L, we need to solve the following system (for each compartment of the model):
\begin{equation}
\begin{split}
	&\frac{1}{2}\left(u^*_{L,i} + \frac{v^*_{L,i}}{\lambda_i}\right) = u_{L,i}^+ 	\\
	&\frac{1}{2}\left(u_{L,i}^* - \frac{v_{L,i}^*}{\lambda_i}\right)	=  \sum_{k=1}^{N_{a,L}} \alpha_{i,k} u_{L,k}^+ + \alpha_{i,n} u_{n}^-		\\
	&\frac{1}{2}\left(u^*_{L,n} - \frac{v^*_{L,n}}{\lambda_n}\right) = u_{n}^- 		\\
	&\frac{1}{2}\left(u_{L,n}^* + \frac{v_{L,n}^*}{\lambda_n}\right)	=  \sum_{k=1}^{N_{a,L}} \alpha_{n,k} u_{L,k}^+ + \alpha_{n,n} u_{n}^-	 .	
\end{split}
\label{eq.junction_L}	
\end{equation}
This linear system can be written in matrix form, resulting:
\begin{equation}
\boldsymbol{A}_L \boldsymbol{Q}^*_L = 2\,\boldsymbol{S}_L ,
\end{equation}
with
\[ \boldsymbol{Q}^*_L =
\begin{pmatrix} 
  	u^*_{L,i} \\ v^*_{L,i} \\ u^*_{L,n} \\ v^*_{L,n}
\end{pmatrix},
\quad
\boldsymbol{A}_L = 
\begin{pmatrix} 
	1 &\frac{1}{\lambda_i} &0 &0 \\
	1 &-\frac{1}{\lambda_i} &0 &0 \\
	0 &0 &1 &-\frac{1}{\lambda_n} \\ 
	0 &0 &1 &\frac{1}{\lambda_n}
\end{pmatrix},
\quad
\boldsymbol{S}_L =
\begin{pmatrix} 
  	u_{L,i}^+ \\ \sum_{k=1}^{N_{a,L}} \alpha_{i,k} u_{L,k}^+ + \alpha_{i,n} u_{n}^-	 \\ u_{n}^- \\ \sum_{k=1}^{N_{a,L}} \alpha_{n,k} u_{L,k}^+ + \alpha_{n,n} u_{n}^-
\end{pmatrix}. \]
On the other hand, to impose the transmission conditions at interface 0, we need to solve the following system:
\begin{equation}
\begin{split}
	&\frac{1}{2}\left(u^*_{0,n} + \frac{v^*_{0,n}}{\lambda_n}\right) = u_{n}^+ 	\\
	&\frac{1}{2}\left(u_{0,n}^* - \frac{v_{0,n}^*}{\lambda_n}\right)	=  \sum_{k=1}^{N_{a,0}} \alpha_{n,k} u_{0,k}^- + \alpha_{n,n} u_{n}^+ 		\\
	&\frac{1}{2}\left(u^*_{0,j} - \frac{v^*_{0,j}}{\lambda_j}\right) = u_{0,j}^- 		\\
	&\frac{1}{2}\left(u_{0,j}^* + \frac{v_{0,j}^*}{\lambda_j}\right)	=  \sum_{k=1}^{N_{a,0}} \alpha_{j,k} u_{0,k}^- + \alpha_{j,n} u_{n}^+	 ,	
\end{split}
\label{eq.junction_0}	
\end{equation}
which, written in compact form, reads:
\begin{equation}
\boldsymbol{A}_0 \boldsymbol{Q}^*_0 = 2\,\boldsymbol{S}_0 ,
\end{equation}
with
\[ \boldsymbol{Q}^*_0 =
\begin{pmatrix} 
  	u^*_{0,j} \\ v^*_{0,j} \\ u^*_{0,n} \\ v^*_{0,n}
\end{pmatrix},
\quad
\boldsymbol{A}_0 = 
\begin{pmatrix} 
	1 &-\frac{1}{\lambda_j} &0 &0 \\
	1 &\frac{1}{\lambda_j} &0 &0 \\
	0 &0 &1 &\frac{1}{\lambda_n} \\ 
	0 &0 &1 &-\frac{1}{\lambda_n}
\end{pmatrix},
\quad
\boldsymbol{S}_0 =
\begin{pmatrix} 
  	u_{0,j}^- \\  \sum_{k=1}^{N_{a,0}} \alpha_{j,k} u_{0,k}^- + \alpha_{j,n} u_{n}^+ \\ u_{n}^+ \\ \sum_{k=1}^{N_{a,0}} \alpha_{n,k} u_{0,k}^- + \alpha_{n,n} u_{n}^+
\end{pmatrix}. \]
Constants $\alpha_{i,j} \in [0,1]$ are the transmission coefficients and represent the probability that an individual at a generic interface decides to move across that interface from the $j$-$th$ location to the $i$-$th$ location, also including the turnabout on the same location. The choice of the transmission coefficients is very relevant since these coefficients deeply affect the mobility flows of the network. It is worth to notice that the condition differs when considering an incoming or an outgoing flux, due to the artificial orientation that has been fixed. Indeed, for each incoming arc, we need to use $u_{L,i}^+$ from the arc and $u_n^-$ from the node; while for each outgoing arc we consider $u_{0,j}^-$ from the arc and $u_n^+$ from the node \cite{bretti2014}.\par
Furthermore, to guarantee the conservation of fluxes at the interface, therefore ensuring that the global mass of the system is conserved over time, the following must hold \cite{bretti2006,bretti2014}:
\begin{equation}
v_{L,n}^*=\sum_{i=1}^{N_{a,L}} v_{L,i}^*, \qquad \qquad v_{0,n}^*=\sum_{j=1}^{N_{a,0}} v_{0,j}^* .
\end{equation}
To respect these conditions, it is enough to impose at interface L
\begin{equation}
	\lambda_i = \sum_{k=1}^{N_{a,L}} \alpha_{k,i} \lambda_k + \alpha_{n,i} \lambda_n , \qquad \qquad
	\lambda_n = \sum_{k=1}^{N_{a,L}} \alpha_{k,n} \lambda_k + \alpha_{n,n} \lambda_n ,
\label{eq.conservationFluxes_L}
\end{equation}
and at interface 0:
\begin{equation}
	\lambda_n = \sum_{k=1}^{N_{a,0}} \alpha_{k,n} \lambda_k + \alpha_{n,n} \lambda_n , \qquad \qquad
	\lambda_j = \sum_{k=1}^{N_{a,0}} \alpha_{k,j} \lambda_k + \alpha_{n,j} \lambda_n .
\label{eq.conservationFluxes_0}
\end{equation}\par
It is straightforward to analytically solve system \eqref{eq.junction_L}, obtaining
\begin{equation}
\begin{split}
	&u^*_{L,i} = u^+_{L,i} + \sum_{k=1}^{N_{a,L}} \alpha_{i,k} u_{L,k}^+ + \alpha_{i,n} u_{n}^- \\
	&v^*_{L,i} = \lambda_i \left(u^+_{L,i} - \sum_{k=1}^{N_{a,L}} \alpha_{i,k} u_{L,k}^+ - \alpha_{i,n} u_{n}^- \right)  \\
	&u^*_{L,n} =  u_{n}^- + \sum_{k=1}^{N_{a,L}} \alpha_{n,k} u_{L,k}^+ + \alpha_{n,n} u_{n}^-  \\
	&v^*_{L,n} = - \lambda_n \left( u_{n}^- - \sum_{k=1}^{N_{a,L}} \alpha_{n,k} u_{L,k}^+ - \alpha_{n,n} u_{n}^- \right) .
\end{split}
\label{sol.junction_L}	
\end{equation}
The same applies for system \eqref{eq.junction_0}, which gives
\begin{equation}
\begin{split}
	&u^*_{0,n} =  u_{n}^+ + \sum_{k=1}^{N_{a,0}} \alpha_{n,k} u_{0,k}^- + \alpha_{n,n} u_{n}^+  \\
	&v^*_{0,n} = \lambda_n \left(u_{n}^+ - \sum_{k=1}^{N_{a,0}} \alpha_{n,k} u_{0,k}^- - \alpha_{n,n} u_{n}^+ \right)  \\
	&u^*_{0,j} =  u_{0,j}^-  + \sum_{k=1}^{N_{a,0}} \alpha_{j,k} u_{0,k}^- + \alpha_{j,n} u_{n}^+  \\
	&v^*_{0,j} = - \lambda_j\left( u_{0,j}^-  - \sum_{k=1}^{N_{a,0}} \alpha_{j,k} u_{0,k}^- - \alpha_{j,n} u_{n}^+\right) .
\end{split}
\label{sol.junction_0}	
\end{equation}
\begin{remark}
We underline that the hypothesis of constant transmission coefficients $\alpha_{i,j}$ (which corresponds to imposing fixed transport velocities) may be realistic only over relatively short time spans, after which the prevailing direction of movement might change or even reverse. In fact, it is possible to consider time-varying mobility fluxes acting on the transmission coefficients at nodes of the proposed model, simply imposing a time dependence on them. 
However, since this aspect is out of the scope of this work, in the following only constant transmission coefficients will be taken into account.
\end{remark}
\subsubsection{Boundary conditions}
\label{section_BC}
Nodes located at the extremes of the domain are without any incoming arc, when concerning inlet boundaries, or without any outgoing arc, when concerning outlet boundaries. At these nodes, in order to ensure that there are no individuals entering to or exiting from the network (hence considering a population which remains constant in time), we just recover the standard zero-flux boundary conditions \cite{piccoli2006}, which consists in imposing at inlet nodes
\begin{equation}
v^*_{L,n} = 0, 	\qquad \qquad u^*_{L,n} = u_n - \frac{v_{n}}{\lambda_n} ;
\label{eq.zeroflux_inlet}	
\end{equation}
while at outlet nodes:
\begin{equation}
v^*_{0,n} = 0, 	\qquad \qquad	u^*_{0,n} = u_n + \frac{v_{n}}{\lambda_n} .
\label{eq.zeroflux_outlet}	
\end{equation}
\section{Numerical method}
\label{section_numericalmodel}
Numerical methods for hyperbolic balance laws and kinetic equations in the diffusive limit have a long history (see \cite{Naldi2000, Jin1998, Gosse2002, boscarino2017} and the references therein). Here, to design a numerical scheme for system \eqref{eq.SIR_kinetic},  we follow the Implicit-Explicit (IMEX) Runge-Kutta approach recently proposed in \cite{boscarino2017} for hyperbolic systems with multiscale relaxation.

\subsection{IMEX finite volume schemes}
IMEX Runge-Kutta schemes can be easily represented by a double tableau (explicit on the left, implicit on the right) in the usual Butcher notation \cite{boscarino2017,pareschi2005}
\begin{center}
\begin{tabular}{c | c}
$\tilde c$ & $\tilde A$ \\  
\hline \\[-1.0em] 
 &  $\tilde b^T$ \\ 
\end{tabular}
\hspace{2.0cm}
\begin{tabular}{c | c}
$c$ & $A$ \\  
\hline \\[-1.0em] 
 &  $b^T$ \\ 
\end{tabular}\,.
\end{center}
Matrices $\tilde A = (\tilde a_{kj})$, with $\tilde a_{kj} = 0 $ for $ j\geq k$, and $A = (a_{kj})$ are $s \times s$ matrices, with $s$ number of Runge-Kutta stages. It is always preferable in terms of computational efficiency to deal with diagonally implicit Runge-Kutta (DIRK) schemes, which ensure that the explicit part of the scheme term is always evaluated explicitly \cite{ascher1997}, hence $a_{kj} = 0 $ for $ j > k$. The coefficient vectors $\tilde c$ and $c$ are given by
\begin{equation*}
\tilde c_k = \sum^{k-1}_{j=1} \tilde a_{kj} , \qquad \qquad c_k = \sum^{k}_{j=1} a_{kj},
\end{equation*}
and vectors $\tilde b = (\tilde b_1, ...,\tilde b_s)^T$ and $b = (b_1, ...,b_s)^T$ are the quadrature weights that permit to combine the internal Runge-Kutta stages. Furthermore, referring to \cite{boscarino2017}, if the following relations hold, the method is said to be globally stiffly accurate (GSA)
\begin{definition}
An IMEX-RK method is said to be globally stiffly accurate (GSA) if not only the corresponding diagonally implicit Runge-Kutta (DIRK) method is stiffly accurate, namely
\begin{equation*}
a_{kj} = b_j, \qquad j = 1,\ldots,s ,
\end{equation*}
but also the explicit method satisfies
\begin{equation*}
\tilde a_{kj} = \tilde b_j, \qquad j = 1,\ldots,s-1 .
\end{equation*}
\end{definition}
It is worth to notice that this definition states also that the numerical solution of a GSA IMEX-RK scheme coincides exactly with the last internal stage of the scheme. 

In system \eqref{eq.SIR_kinetic}, once the diffusion coefficients in \eqref{eq:diff} have been fixed, the scaling depends on the relaxation times $\tau_S$, $\tau_I$, $\tau_R$. Indeed, these relaxation terms modify the nature of the behavior of the solution, which can result either hyperbolic or parabolic. Standard IMEX Runge-Kutta methods for hyperbolic systems with relaxation terms loose their efficiency \cite{Naldi2000, Jin1998, Gosse2002, boscarino2017} and a different approach must be adopted to guarantee asymptotic preservation (AP) in stiff regimes (i.e. the consistency of the scheme with the equilibrium system is guaranteed and the order of accuracy is maintained in the stiff limit).

Following \cite{boscarino2017}, the IMEX Runge-Kutta approach that we consider for system \eqref{systcompactform} consists in computing the internal stages
\begin{equation}
\begin{split}
	&\U^{(k)} = \U^n -  \Delta t \sum_{j=1}^{k} a_{kj} \partial_x \V^{(j)}  + \Delta t \sum_{j=1}^{k-1} \tilde{a}_{kj} \F\left(\U^{(j)}\right)
	\\
	&\V^{(k)} = \V^n -  \Delta t \sum_{j=1}^{k-1} \tilde{a}_{kj} \left( \boldsymbol{\Lambda}^2 \partial_x \U^{(j)} - \G\left(\U^{(j)},\V^{(j)}\right)\right)  + \Delta t \sum_{j=1}^{k} a_{kj} \H\left(\V^{(j)}\right),
	\end{split}
\label{eq.iterIMEX}
\end{equation}
followed by the numerical solution
\begin{equation}
\begin{split}		
	& \U^{n+1} = \U^n - \Delta t \sum_{k=1}^{s} b_{k} \partial_x \V^{(k)} + \Delta t \sum_{k=1}^{s} \tilde{b}_{k} \F\left(\U^{(k)}\right)
	\\
	& \V^{n+1} = \V^n - \Delta t \sum_{k=1}^{s} \tilde{b}_{k} \left( \boldsymbol{\Lambda}^2 \partial_x \U^{(k)} - \G\left(\U^{(k)},\V^{(k)}\right) \right) + \Delta t \sum_{k=1}^{s} b_{k} \H\left(\V^{(k)}\right),
\end{split}
\label{eq.finalIMEX}
\end{equation}
with a time step size \(\Delta t = t^{n+1}-t^{n}\) that, ignoring the reaction terms, for small values of the relaxation time of each compartment $\tau_h, h = 1,2,3,$ follows the less restrictive between the standard hyperbolic Courant-Friedrichs-Levy condition $\Delta t \leq \CFL \frac{\Delta x}{\max\left\{\lambda_S,\lambda_I,\lambda_R\right\}}$, and the parabolic stability restriction $\Delta t \leq \nu \frac{\Delta x^2}{\max\left\{D_S,D_I,D_R\right\}}$ given by the diffusive components of the system, where $\Delta x$ is the size of the space grid, and $\CFL$ and $\nu$ are suitable stability constants (see \cite{boscarino2017}). Note that the non stiff reaction terms, which are evaluated explicitly, impose a weaker restriction on the time step of the type $\Delta t \leq \frac1{\max\{\beta,\gamma\}}$. 

\subsection{Numerical diffusion limit}
The scheme \eqref{eq.iterIMEX}-\eqref{eq.finalIMEX} permits to treat implicitly the stiff terms and explicitly all the rest, maintaining a consistent discretization of the limit system in the diffusive regime, represented by system~\eqref{eq.SIR}, i.e. the AP property. To verify the AP property, let us denote by $u_h^{(j)}$, $v_h^{(j)}$, $f_h^{(j)}$, $g_h^{(j)}$, $u_h^{n}$, and $v_h^{n}$, $h=1,2,3$, the SIR components of $\U^{(j)}$, $\V^{(j)}$, $\F(\U^{(j)})$, $\G(\U^{(j)},\V^{(j)})$, $\U^{n}$, and $\V^{n}$, respectively. Then, the IMEX scheme \eqref{eq.iterIMEX}-\eqref{eq.finalIMEX} can be rewritten highlighting the scale parameters as follows 
\begin{equation}
\begin{split}	
	& \u_h = u_h^n \e - \Delta t A \partial_x \v_h + \Delta t \tilde A \f_h \\
	& \v_h = v_h^n \e - \Delta t \tilde A \left( \frac{D_h}{\tau_h} \partial_x \u_h - \g_h \right) - \frac{\Delta t}{\tau_h} A \v_h ,
\end{split}	
\label{eq.iterIMEX_ex}
\end{equation}
where $\u_h=(u_h^{(1)},\ldots,u_h^{(s)})^T$, $\v_h=(v_h^{(1)},\ldots,v_h^{(s)})^T$, $\f_h=(f_h^{(1)},\ldots,f_h^{(s)})^T$, $\g_h=(g_h^{(1)},\ldots,g_h^{(s)})^T$, $\e = (1,\ldots,1)^T \in \mathbb{R}^s$, and we use notations $\tau_h$ and $D_h$ to denote the relaxation times and diffusion constant of each compartment. The final solution therefore reads
\begin{equation}
\begin{split}	
	& u_h^{n+1} = u_h^n - \Delta t b^T \partial_x \v_h + \Delta t \tilde b^T \f_h 
	 \\
	& v_h^{n+1} = v_h^n - \Delta t \tilde b^T \left( \frac{D_h}{\tau_h} \partial_x \u_h - \g_h \right) - \frac{\Delta t}{\tau_h} b^T \v_h .
\end{split}	
\label{eq.finalIMEX_ex}
\end{equation}
From the second equation in \eqref{eq.iterIMEX_ex} we obtain
\begin{equation}
\v_h = \left( \frac{\tau_h}{\Delta t} I_d + A\right)^{-1} \left( \frac{\tau_h}{\Delta t} v_h^n \e - \tilde A \left(D_h \partial_x \u_h - \tau_h \g_h\right) \right),
\label{eq:vh}
\end{equation}
which substituted into the first equation of \eqref{eq.iterIMEX_ex} gives 
\begin{equation}
\begin{split}
\u_h =& u_h^n \e - \Delta t A \left( \frac{\tau_h}{\Delta t} I_d + A\right)^{-1} \partial_x  \left( \frac{\tau_h}{\Delta t} v_h^n \e + \tilde A \tau_h \g_h\right) \\
&+ \Delta t A \left( \frac{\tau_h}{\Delta t} I_d + A\right)^{-1} \tilde A \partial_x \left(D_h \partial_x \u_h\right) + \Delta t \tilde A \f_h. 
\end{split}
\label{eq:imexcs}
\end{equation}
Now, as $\tau_h\to 0$ we get
\begin{equation}
\u_h = u^n_h \e+\Delta t \tilde A \partial_{x} (D_h \partial_{x} \u_h) + \Delta t\tilde A \f_h, 
\label{eq:apstages}
\end{equation}
and thus, the internal stages correspond to the stages of the explicit scheme applied to the reaction-diffusion system \eqref{eq.SIR}.

However, this is not enough to guarantee that the scheme satisfies the AP property, since we need to verify the same consistency property on the final numerical solution.

To this aim, let us rewrite \eqref{eq:vh} as
\begin{equation*}
\v_h = \frac{\tau_h}{\Delta t} A^{-1} v_h^n \e + A^{-1}\left( I_d - \frac{\tau_h}{\Delta t} A^{-1} \right) \tilde A D_h \partial_x \u_h + \tau_h A^{-1} \tilde A \g_h + \mathcal{O}\left( \tau_h^2\right),
\end{equation*}
which substituted into the second equation of \eqref{eq.finalIMEX_ex} leads to
\begin{equation*}
\begin{split}
v_h^{n+1} = &\left(1-b^TA^{-1}e\right) v_h^n - \left( b^TA^{-1} \tilde A - \tilde b^T \right) \frac{\Delta t}{\tau_h} D_h \partial_x \u_h\\
& - b^T A^{-2} \tilde A D_h \partial_x \u_h - \Delta t \left( b^TA^{-1} \tilde A - \tilde b^T \right) \g_h+ \mathcal{O}\left( \tau_h\right).
\end{split}
\end{equation*}
In order to pass to the limit $\tau_h \rightarrow 0$, we must require that $b^TA^{-1} \tilde A - \tilde b^T = 0$, which is satisfied if the IMEX scheme is GSA. Indeed, in GSA methods, $b^T = \e_s^T A$ and $\tilde b^T = \e_s^T \tilde A$, therefore $b^TA^{-1} \tilde A - \tilde b^T = \e_s^T \tilde A - \tilde b^T = 0$.
Thus, in the limit $\tau_h\to 0$ we finally get
\begin{subequations}
\begin{align*}
		& v_h^{n+1} = - b^T A^{-2} \tilde A D_h \partial_x \u_h,
\end{align*}	
\end{subequations}
and the resulting numerical solution for $u_h^{n+1}$ satisfies 
\begin{equation}
\begin{split}
	& u_h^{n+1} = u^n_h+ \Delta t\tilde b^T \partial_{x} ( D_h \partial_{x}\u_h) + \Delta t \tilde b^T \f_h,
\end{split}
\label{eq:apfinal}	
\end{equation}
which, combined with \eqref{eq:apstages}, proves the AP-property of the IMEX scheme with respect to the reaction-diffusion system \eqref{eq.SIR}.

Therefore, we have shown the following result.
\begin{proposition}
If the IMEX Runge-Kutta method satisfies the GSA property, then the scheme \eqref{eq.iterIMEX_ex}-\eqref{eq.finalIMEX_ex} applied to system \eqref{systcompactform} for $\tau_h\to 0$, $h=1,2,3$, provides the explicit Runge-Kutta scheme \eqref{eq:apstages}-\eqref{eq:apfinal} for the limiting reaction-diffusion system \eqref{eq.SIR}.
\end{proposition}

\subsection{Removing the parabolic stiffness}
For ${\mathcal O}(1)$ values of the diffusion coefficients $D_S, D_I, D_R$ the classical parabolic stability condition $\Delta t = {\mathcal O}(\Delta x^2)$ of the explicit scheme applied to the limiting reaction-diffusion system may be too restrictive. This drawback can be avoided by modifying the IMEX Runge-Kutta approach, taking the following partitioning for the equations for $\V$ in \eqref{eq.iterIMEX}-\eqref{eq.finalIMEX} \begin{equation}
\begin{split}
	&\V^{(k)} = \V^n +  \Delta t \sum_{j=1}^{k-1} \tilde{a}_{kj} \G\left(\U^{(j)},\V^{(j)}\right)  - \Delta t \sum_{j=1}^{k} a_{kj} \left( \boldsymbol{\Lambda}^2 \partial_x \U^{(j)}-\H\left(\V^{(j)}\right)\right),\\
	& \V^{n+1} = \V^n + \Delta t \sum_{k=1}^{s} \tilde{b}_{k} \G\left(\U^{(k)},\V^{(k)}\right) - \Delta t \sum_{k=1}^{s} b_{k} \left(\boldsymbol{\Lambda}^2 \partial_x \U^{(k)} -\H\left(\V^{(k)}\right)\right).
\end{split}
\label{eq.IMEX2}
\end{equation}
Now, using the same notations as in the previous section, the IMEX-scheme leads to 
\begin{equation}
\v_h = \left( \frac{\tau_h}{\Delta t} I_d + A\right)^{-1} \left( \frac{\tau_h}{\Delta t} v_h^n \e - D_h A \partial_x \u_h + \tau_h \tilde A \g_h \right),
\label{eq:vh2}
\end{equation}
and
\begin{equation}
\begin{split}
\u_h =& u_h^n \e - \Delta t A \partial_x \left( \frac{\tau_h}{\Delta t} I_d + A\right)^{-1} \left( \frac{\tau_h}{\Delta t} v_h^n \e + \tilde A \tau_h \g_h\right) \\
&+ \Delta t A \left( \frac{\tau_h}{\Delta t} I_d + A\right)^{-1} A \partial_x \left(D_h \partial_x \u_h\right) + \Delta t \tilde A \f_h. 
\end{split}
\label{eq:imexcs2}
\end{equation}
The above equation, as $\tau_h \to 0$, yields an IMEX-scheme for the reaction-diffusion system \eqref{eq.SIR}
\begin{equation}
\begin{split}
& \u_h = u^n_h \e+\Delta t A \partial_{x} (D_h \partial_{x} \u_h) + \Delta t \tilde A \f_h.
\end{split}	
\label{eq:apstages2}
\end{equation}
In a similar way, we can analyze the numerical solution, and show that under the GSA assumption in the limit $\tau_h\to 0$ it reduces to 
\begin{equation}
\begin{split} 
	& u_h^{n+1} = u^n_h+ \Delta t b^T \partial_{x} ( D_h \partial_{x}\u_h) + \Delta t \tilde b^T \f_h.
\end{split}	
\label{eq:apfinal2}
\end{equation}
Summarizing we have now the following result.
\begin{proposition}
If the IMEX Runge-Kutta method satisfies the GSA property, then the scheme \eqref{eq.iterIMEX_ex}-\eqref{eq.finalIMEX_ex} applied to system \eqref{systcompactform}, where the equations for $\V^{(k)}$ and $\V^{n+1}$ are replaced by \eqref{eq.IMEX2}, for $\tau_h\to 0$, $h=1,2,3$, provides the IMEX-scheme \eqref{eq:apstages2}-\eqref{eq:apfinal2} for the limiting reaction-diffusion system \eqref{eq.SIR}.
\end{proposition}
Since the GSA property is essential to preserve the correct diffusion limit, in the sequel, the GSA BPR(4,4,2) scheme presented in \cite{boscarino2017} is chosen, characterized by $s=4$ stages for the implicit part, 4 stages for the explicit part and 2nd order of accuracy, which is defined by the following tableau (explicit on the left and implicit on the right)
\begin{equation}
\begin{tabular}{c | c c c c c}
0 & 0 & 0 & 0 & 0 & 0 \\
1/4 & 1/4 & 0 & 0 & 0 & 0 \\
1/4 & 13/4 & -3 & 0 & 0 & 0 \\
3/4 & 1/4 & 0 & 1/2 & 0 & 0 \\
1 & 0 & 1/3 & 1/6 & 1/2 & 0 \\ \hline
  & 0 & 1/3 & 1/6 & 1/2 & 0  \\
\end{tabular}
\hspace{1.0cm}
\begin{tabular}{c | c c c c c}
0 & 0 & 0 & 0 & 0 & 0 \\
1/4 & 0 & 1/4 & 0 & 0 & 0 \\
1/4 & 0 & 0 & 1/4 & 0 & 0 \\
3/4 & 0 & 1/24 & 11/24 & 1/4 & 0 \\
1 & 0 & 11/24 & 1/6 & 1/8 & 1/4 \\ \hline
  & 0 & 11/24 & 1/6 & 1/8 & 1/4   \\
\end{tabular}
\label{eq:tables}
\end{equation}
\subsubsection{Choice of the space discretization}
To obtain a fully discrete scheme, we consider a finite volume method for the spatial discretization, and uniform grid with mesh spacing \(\Delta x = x_{i+{1}/{2}}-x_{i-{1}/{2}}\). For each internal step of the IMEX scheme, numerical fluxes are evaluated following the Dumbser-Osher-Toro (DOT) solver, which coincides with the Godunov flux based on the exact Riemann solver for linear hyperbolic systems with constant Jacobian matrix \cite{dumbser2011,bertaglia2018,bertaglia2020}. Boundary-extrapolated values on the two sides of the interface within cell \(i\) are computed by piecewise linear reconstruction, recurring to the minmod slope limiter to obtain a TVD scheme \cite{toro2009} and achieve second order of accuracy for smooth solutions also in space. We point out that in the case of scheme \eqref{eq.IMEX2}, the stages of the IMEX scheme are implemented in the form \eqref{eq:imexcs2}, where the second order derivative is discretized directly using a second order accurate central scheme of the form
\[
\partial_{x} ( D_h \partial_{x}\u_h) \approx \frac{D_h(x_{1+1/2})(\u_h(x_{i+1})-\u_h(x_i))-D_h(x_{1-1/2})(\u_h(x_i)-\u_h(x_{i-1}))}{(\Delta x)^2}.
\]
This permits an efficient inversion of the corresponding linear system, thus avoiding the adoption of iterative procedures (e.g. Newton-Raphson method or similar ones), and the creation of large non compact stencils in the discretization of the resulting second order terms. 
\section{Numerical results}
\label{section_numericalresults}
In this Section, we present some numerical results to support the validity of the proposed model, both in the simple 1D configuration and in the network characterization. The accuracy of the scheme is verified for different values of the relaxation times, including the stiff regime of the purely diffusive system. Furthermore, we analyze the behavior of the model concerning spatially heterogeneous environments, taking into account a spatially variable contact rate, with respect to two different scenarios: $R_0 < 1$ and $R_0 > 1$. Two test cases are then designed and simulated to observe the spread of infectious diseases with respect to the mobility of individuals on networks. In all test cases we assume $p=1$ in \eqref{eq:incf}.
\begin{table}[p!]
\caption{Results of the accuracy analysis performed using the AP-explicit form of the scheme. $L^1$ error norms and empirical order of accuracy of the variables $S$, $I$, $J_S$ and $J_I$ resulting from accuracy analysis performed choosing different values of relaxation times (and characteristic velocities). $N_x$ indicates the number of cells in the computational domain.} \label{tab:accuracy_1} 
\centering
\begin{tabular}{c | l | c c | c c | c c }
&\multirow{2}{*}{$N_x$} &\multicolumn{2}{c|}{$\tau = 1.0$} &\multicolumn{2}{c|}{$\tau = 10^{-2}$} &\multicolumn{2}{c}{$\tau = 10^{-6}$}\\
\cline{3-8}
& & $L^1$ & $\mathcal{O}\left( L^1\right)$ & $L^1$ & $\mathcal{O}\left( L^1\right)$ & $L^1$ & $\mathcal{O}\left( L^1\right)$\\
\hline
\multirow{4}{*}{$S$}
     &15 & 7.8183e-02 &        & 6.3566e-02 &         & 5.7368e-02 &                  \\ 
     &45 & 1.1983e-02 &  1.7072 & 6.8991e-03 &  2.0214 & 6.1980e-03 &  2.0255  \\ 
    &135 & 1.4804e-03 &  1.9035 & 7.5256e-04 &  2.0168 & 6.7867e-04 &  2.0133 \\ 
    &405 & 1.5235e-04 &  2.0698 & 7.4838e-05 &  2.1010 & 6.7836e-05 &  2.0963 \\ 
\hline 
\multirow{4}{*}{$I$}
     &15 & 7.9779e-02 &         & 4.6081e-02 &         & 3.9811e-02 &                  \\ 
     &45 & 1.1861e-02 &  1.7349 & 4.9662e-03 &  2.0278 & 4.0831e-03 &  2.0729 \\ 
    &135 & 1.4852e-03 &  1.8912 & 5.4088e-04 &  2.0182 & 4.4396e-04 &  2.0197 \\ 
    &405 & 1.5351e-04 &  2.0658 & 5.4262e-05 &  2.0930 & 4.4342e-05 &  2.0970  \\ 
\hline
\multirow{4}{*}{$J_S$}
	&15 & 6.7762e-02 &         & 7.8869e-02 &         & 8.5108e-02 &          \\ 
     &45 & 1.0832e-02 &  1.6689 & 9.1743e-03 &  1.9583 & 8.9856e-03 &  2.0465 \\ 
    &135 & 1.2654e-03 &  1.9544 & 1.0350e-03 &  1.9861 & 8.4277e-04 &  2.1542 \\ 
    &405 & 1.2807e-04 &  2.0849 & 1.1343e-04 &  2.0126 & 9.0422e-05 &  2.0318 \\ 
\hline
\multirow{4}{*}{$J_I$}
     &15 & 1.1164e-01 &         & 6.2077e-02 &         & 6.4906e-02 &         \\ 
     &45 & 1.7386e-02 &  1.6926 & 8.3184e-03 &  1.8295 & 6.4301e-03 &  2.1044 \\ 
    &135 & 2.1539e-03 &  1.9009 & 9.6011e-04 &  1.9654 & 5.9580e-04 &  2.1653 \\ 
    &405 & 2.2147e-04 &  2.0706 & 1.0974e-04 &  1.9742 & 6.3867e-05 &  2.0327 \\ 
\hline
\end{tabular} 
\end{table}
\begin{table}[p!]
\caption{Results of the accuracy analysis performed using the AP-implicit form of the scheme. $L^1$ error norms and empirical order of accuracy of the variables $S$, $I$, $J_S$ and $J_I$ resulting from accuracy analysis performed choosing different values of relaxation times (and characteristic velocities). $N_x$ indicates the number of cells in the computational domain.} \label{tab:accuracy_2} 
\centering
\begin{tabular}{c | l | c c | c c | c c }
&\multirow{2}{*}{$N_x$} &\multicolumn{2}{c|}{$\tau = 1.0$} &\multicolumn{2}{c|}{$\tau = 10^{-2}$} &\multicolumn{2}{c}{$\tau = 10^{-6}$}\\
\cline{3-8}
& & $L^1$ & $\mathcal{O}\left( L^1\right)$ & $L^1$ & $\mathcal{O}\left( L^1\right)$ & $L^1$ & $\mathcal{O}\left( L^1\right)$\\
\hline
\multirow{4}{*}{$S$}
     &15 & 8.2209e-02 &         & 3.2899e-02 &         & 3.5143e-02 &                  \\ 
     &45 & 1.0171e-02 &  1.9022 & 3.4339e-03 &  2.0569 & 2.2120e-03 &  2.5173  \\ 
    &135 & 1.1178e-03 &  2.0099 & 5.0689e-04 &  1.7414 & 2.3935e-04 &  2.0241 \\ 
    &405 & 1.1118e-04 &  2.1008 & 5.6847e-05 &  1.9915 & 2.5405e-05 &  2.0416 \\ 
\hline
\multirow{4}{*}{$I$}
     &15 & 7.6661e-02 &         & 4.8115e-02 &         & 4.4768e-02 &                  \\ 
     &45 & 1.0756e-02 &  1.7876 & 5.8666e-03 &  1.9154 & 4.4697e-03 &  2.0973  \\ 
    &135 & 1.2890e-03 &  1.9312 & 7.6165e-04 &  1.8583 & 5.0683e-04 &  1.9815  \\ 
    &405 & 1.3126e-04 &  2.0794 & 8.1642e-05 &  2.0327 & 5.1867e-05 &  2.0749  \\ 
\hline
\multirow{4}{*}{$J_S$}
     &15 & 4.9000e-02 &         & 8.4703e-02 &         & 3.6605e-02 &         \\ 
     &45 & 6.9525e-03 &  1.7774 & 1.7916e-02 &  1.4140 & 1.1223e-02 &  1.0761  \\ 
    &135 & 7.7558e-04 &  1.9964 & 2.2137e-03 &  1.9033 & 1.6154e-03 &  1.7644 \\ 
    &405 & 7.7205e-05 &  2.1001 & 2.4000e-04 &  2.0224 & 1.7523e-04 &  2.0219  \\ 
\hline
\multirow{4}{*}{$J_I$}
     &15 & 6.5263e-02 &         & 5.5500e-02 &         & 7.5371e-02 &         \\ 
     &45 & 1.0642e-02 &  1.6508 & 2.1501e-02 &  0.8632 & 1.1732e-02 &  1.6932 \\ 
    &135 & 1.2714e-03 &  1.9340 & 3.1556e-03 &  1.7467 & 1.6338e-03 &  1.7944  \\ 
    &405 & 1.2910e-04 &  2.0820 & 3.7132e-04 &  1.9478 & 1.7689e-04 &  2.0236  \\ 
\hline
\end{tabular} 
\end{table}
\begin{figure}[t!]
\centering
\begin{subfigure}{0.32\textwidth}
\includegraphics[width=1\linewidth]{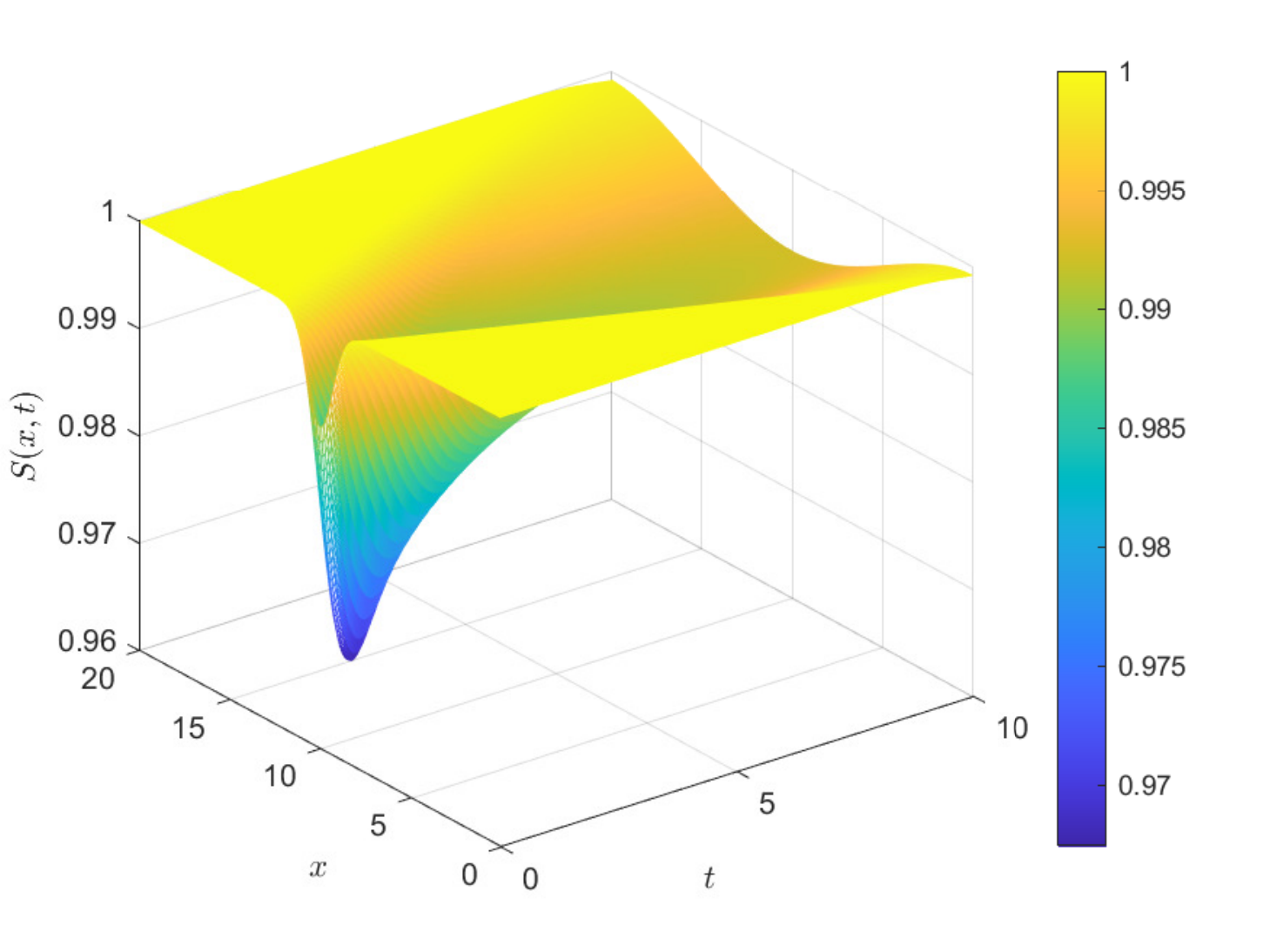}
\caption{(a)}
\label{fig.TC4_tau1_t10_xt_S}
\end{subfigure}
\begin{subfigure}{0.32\textwidth}
\includegraphics[width=1\linewidth]{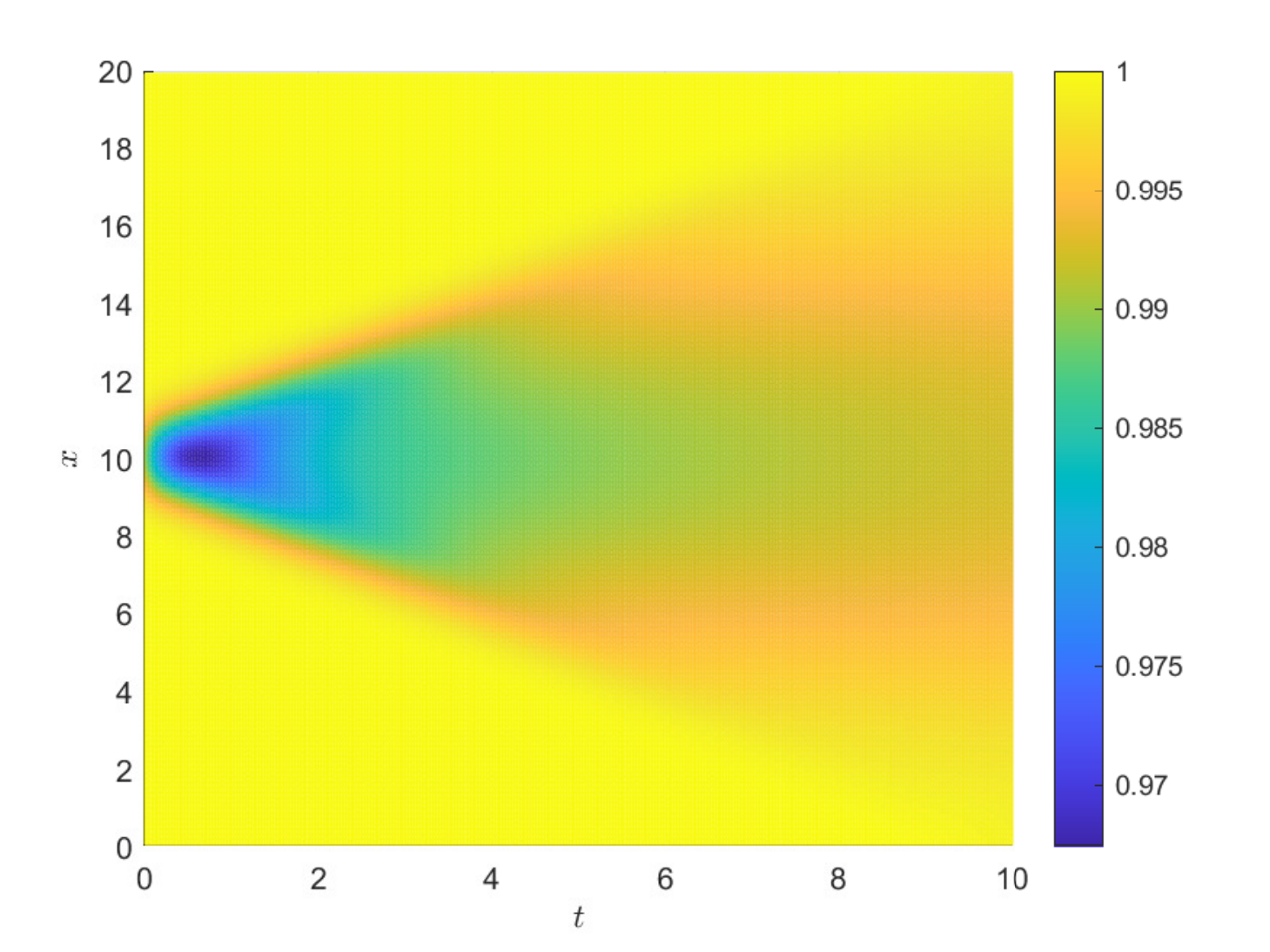}
\caption{(b)}
\label{fig.TC4_tau1_t10_xt_S_top}
\end{subfigure}
\begin{subfigure}{0.32\textwidth}
\includegraphics[width=1\linewidth]{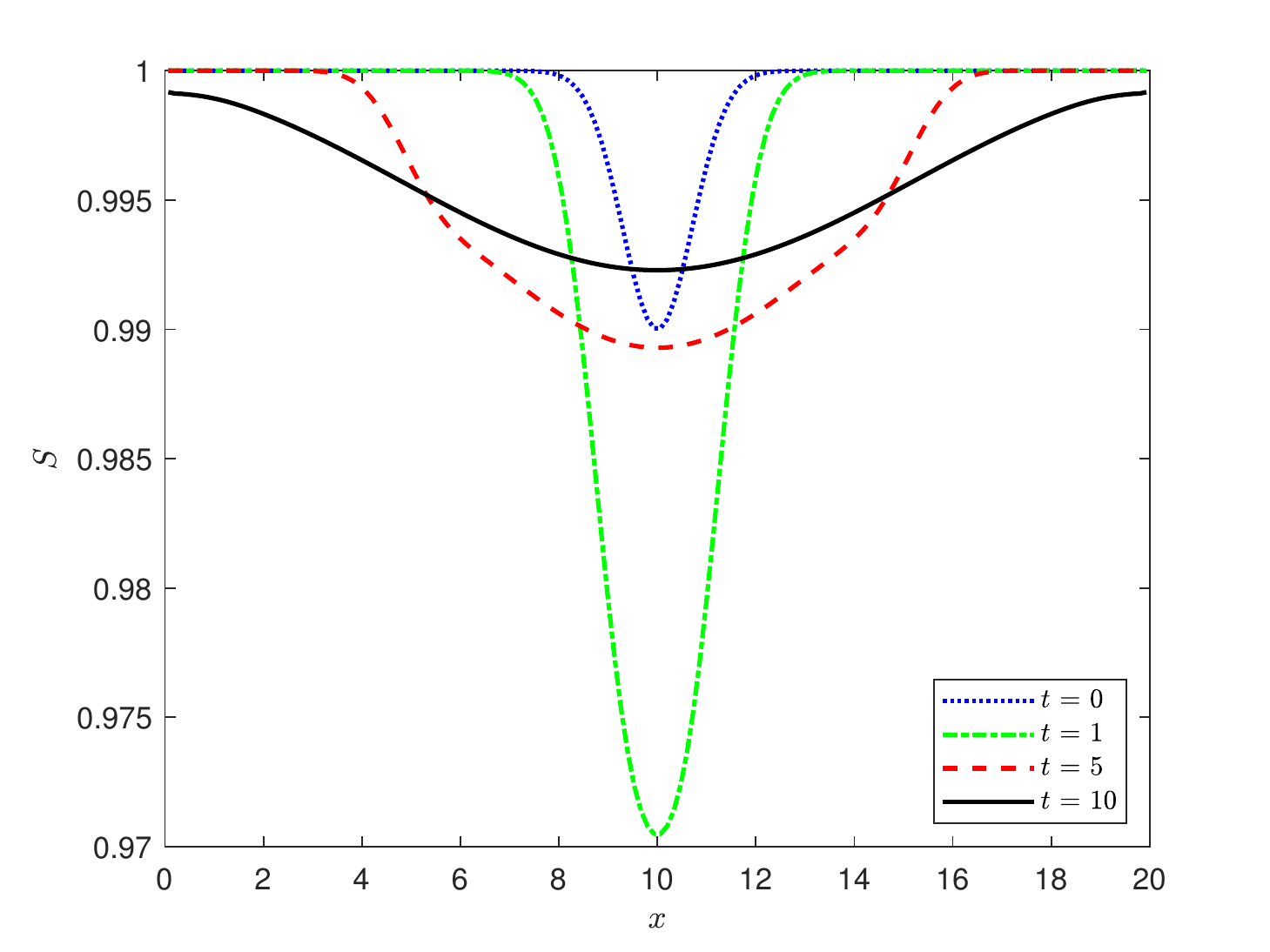}
\caption{(c)}
\label{fig.TC4_tau1_x_S}
\end{subfigure}
\begin{subfigure}{0.32\textwidth}
\includegraphics[width=1\linewidth]{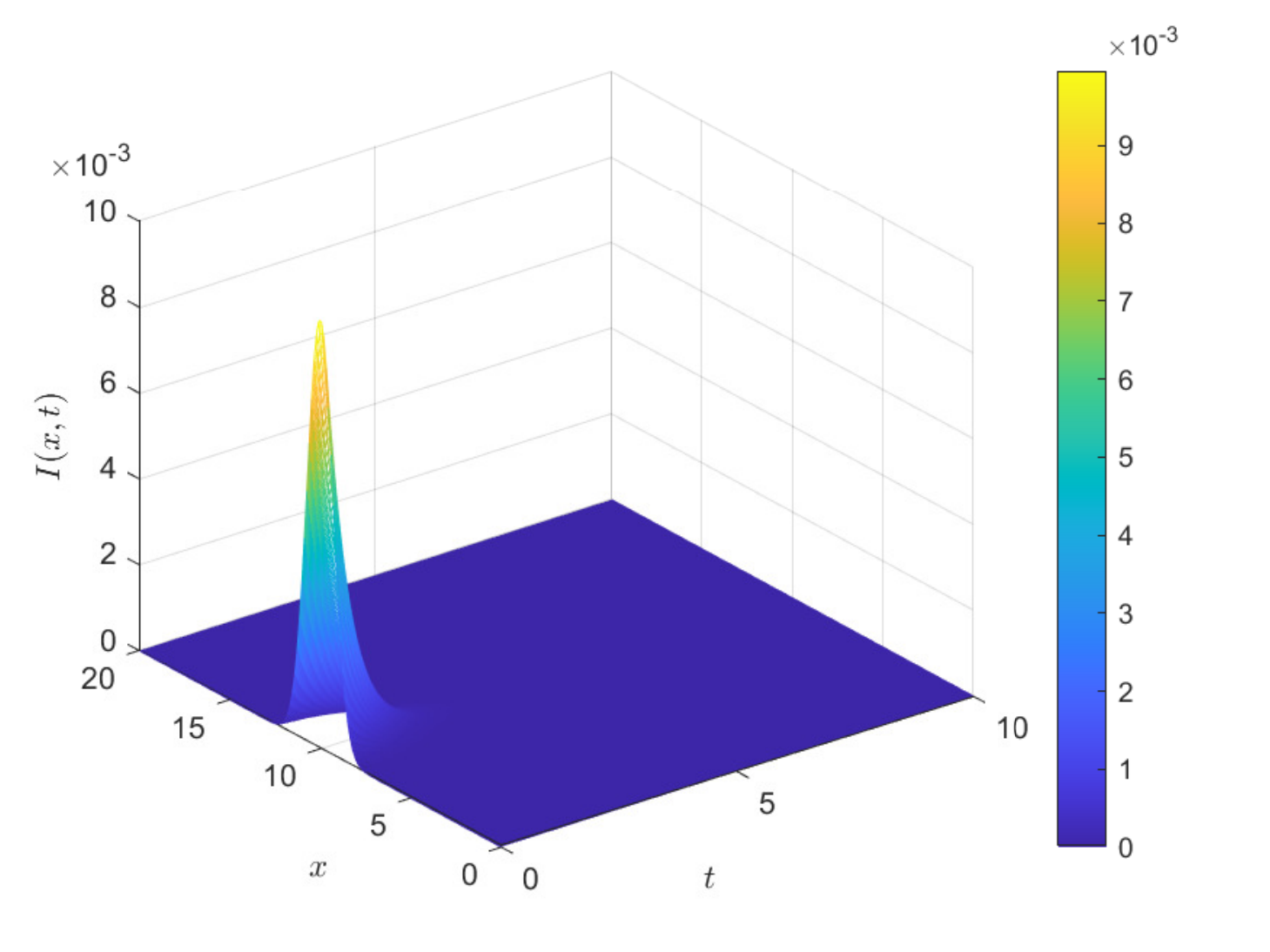}
\caption{(d)}
\label{fig.TC4_tau1_t10_xt_I}
\end{subfigure}
\begin{subfigure}{0.32\textwidth}
\includegraphics[width=1\linewidth]{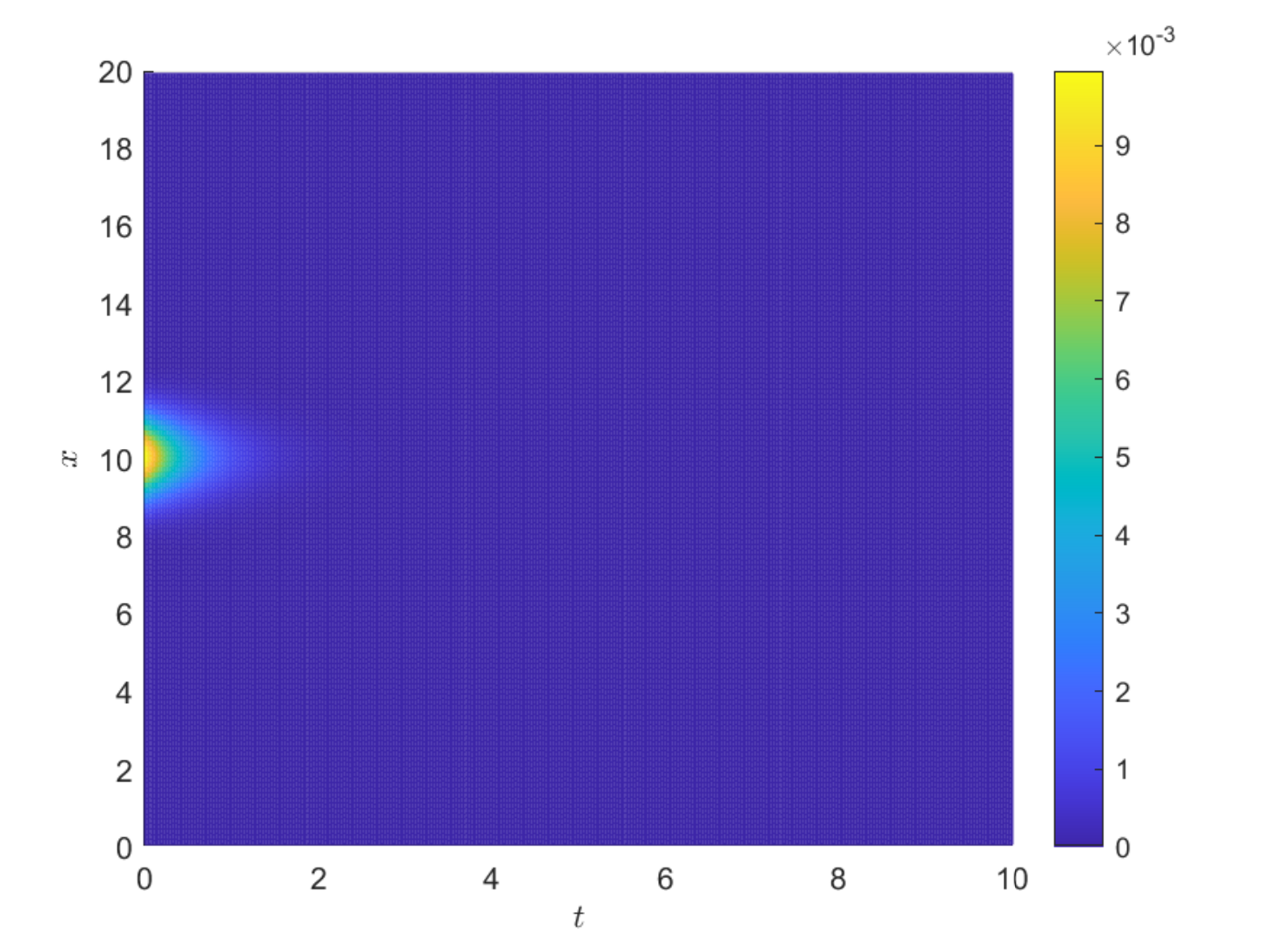}
\caption{(e)}
\label{fig.TC4_tau1_t10_xt_I_top}
\end{subfigure}
\begin{subfigure}{0.32\textwidth}
\includegraphics[width=1\linewidth]{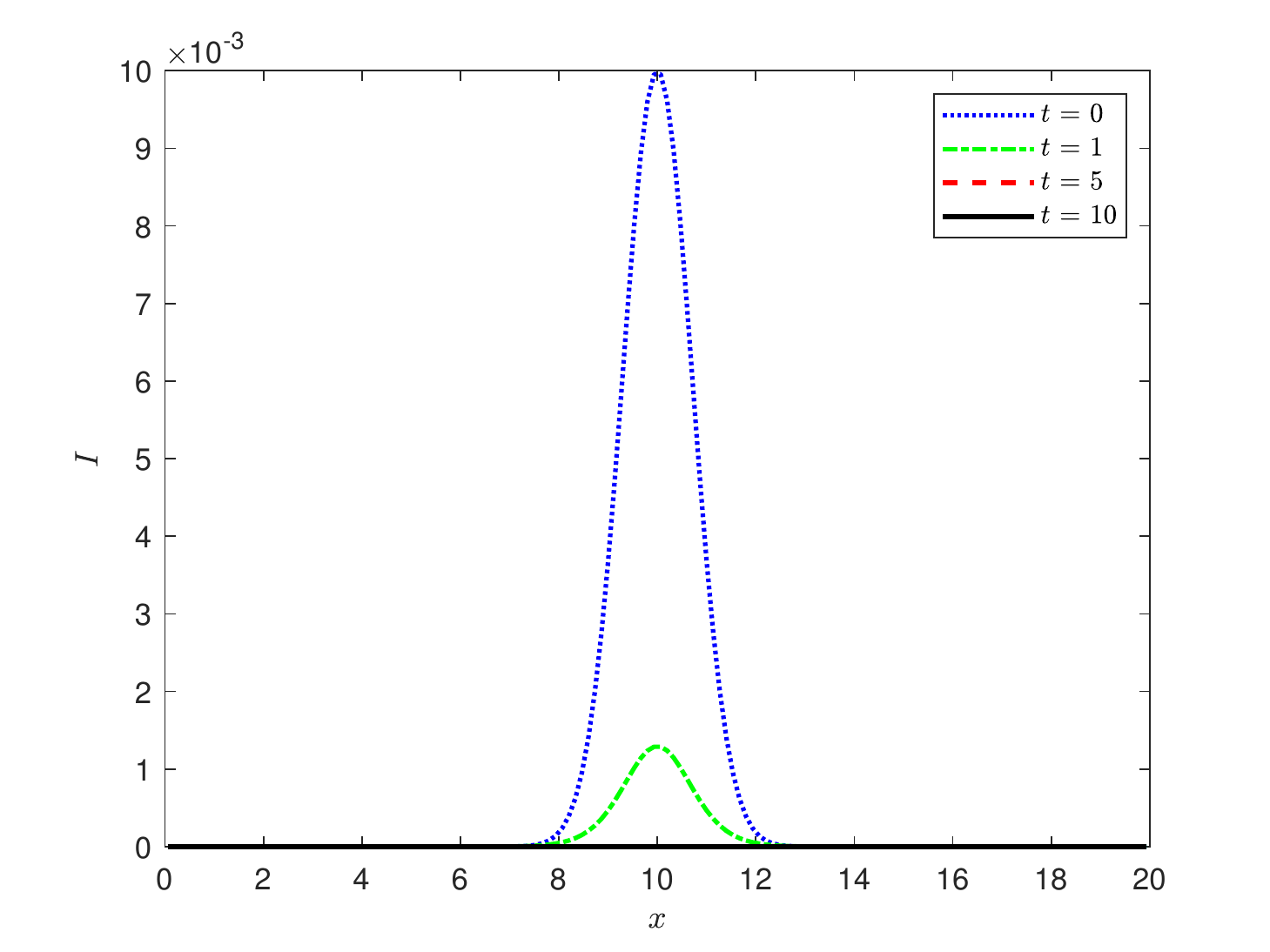}
\caption{(f)}
\label{fig.TC4_tau1_x_I}
\end{subfigure}
\caption{Numerical results of the spatially heterogeneous test case with hyperbolic configuration of relaxation times and characteristic velocities ($\tau = 1.0, \lambda^2 = 1.0$) and reproduction number $R_0 <1$. Time and spatial evolution of $S$ presented in (a), (b) and (c); evolution of $I$ shown in (d), (e) and (f).}
\label{fig.TC4_tau1}
\end{figure}
\begin{figure}[t!]
\centering
\begin{subfigure}{0.32\textwidth}
\includegraphics[width=1\linewidth]{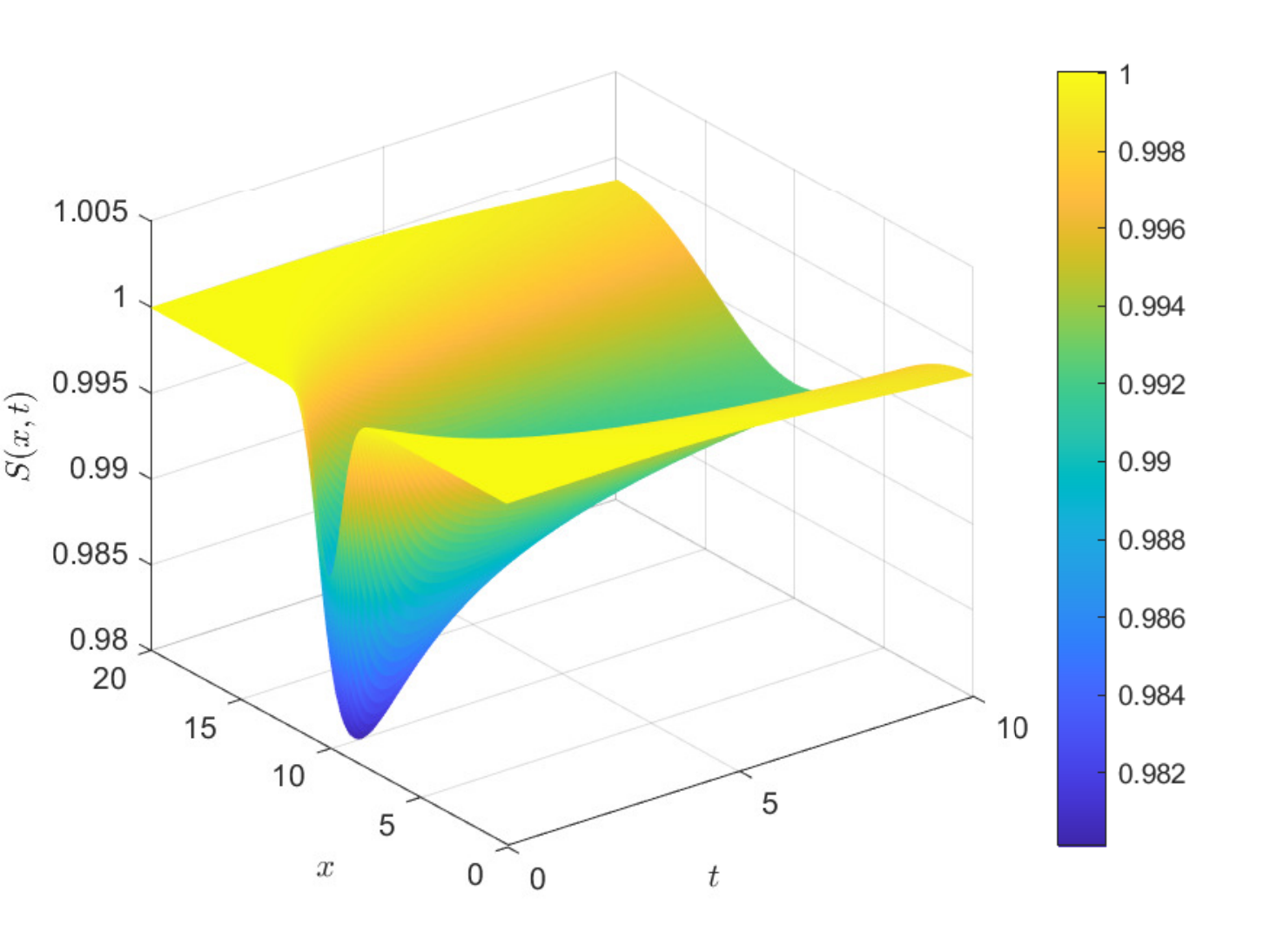}
\caption{(a)}
\label{fig.TC4_tau1e5_t10_xt_S}
\end{subfigure}
\begin{subfigure}{0.32\textwidth}
\includegraphics[width=1\linewidth]{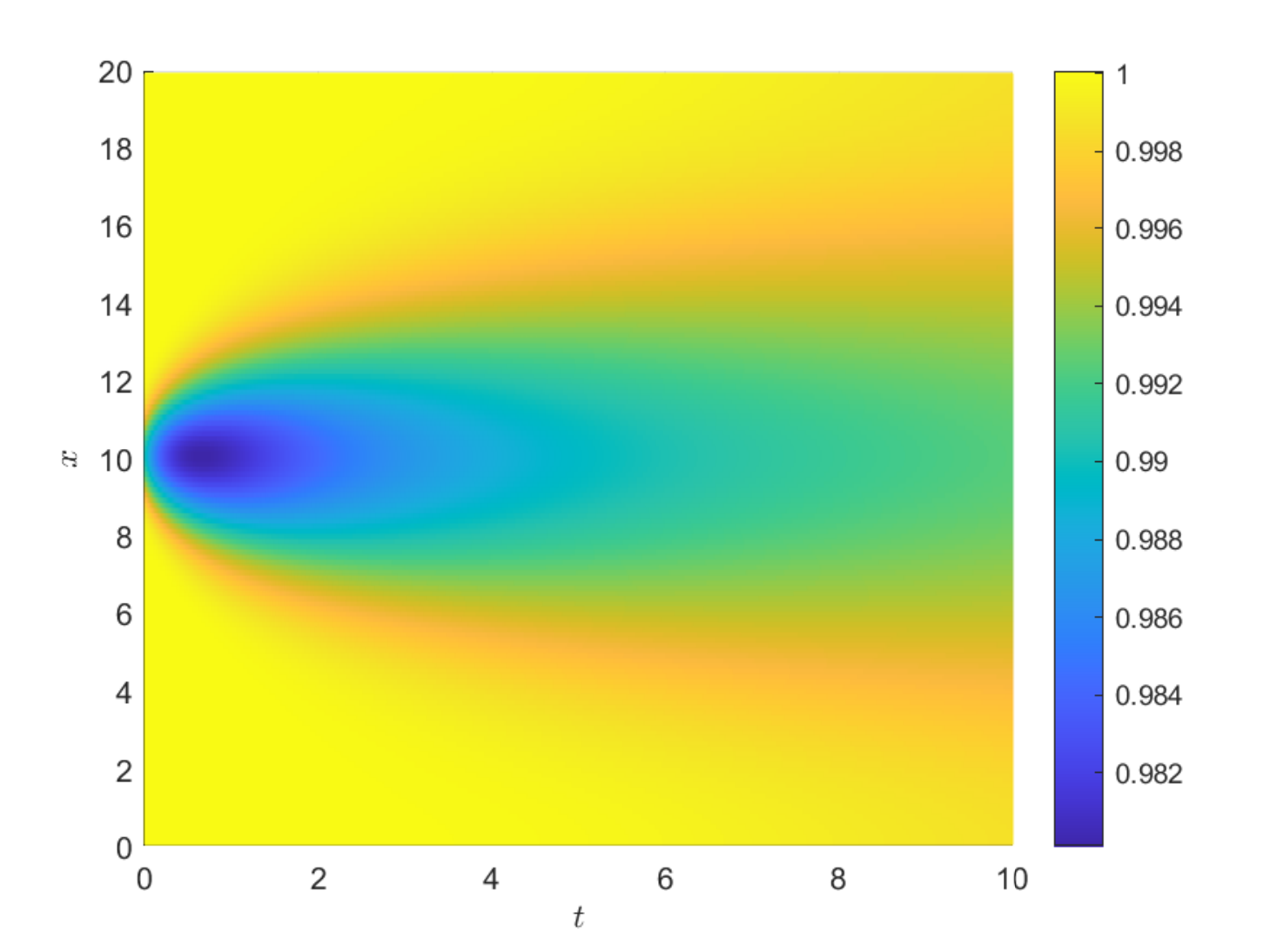}
\caption{(b)}
\label{fig.TC4_tau1e5_t10_xt_S_top}
\end{subfigure}
\begin{subfigure}{0.32\textwidth}
\includegraphics[width=1\linewidth]{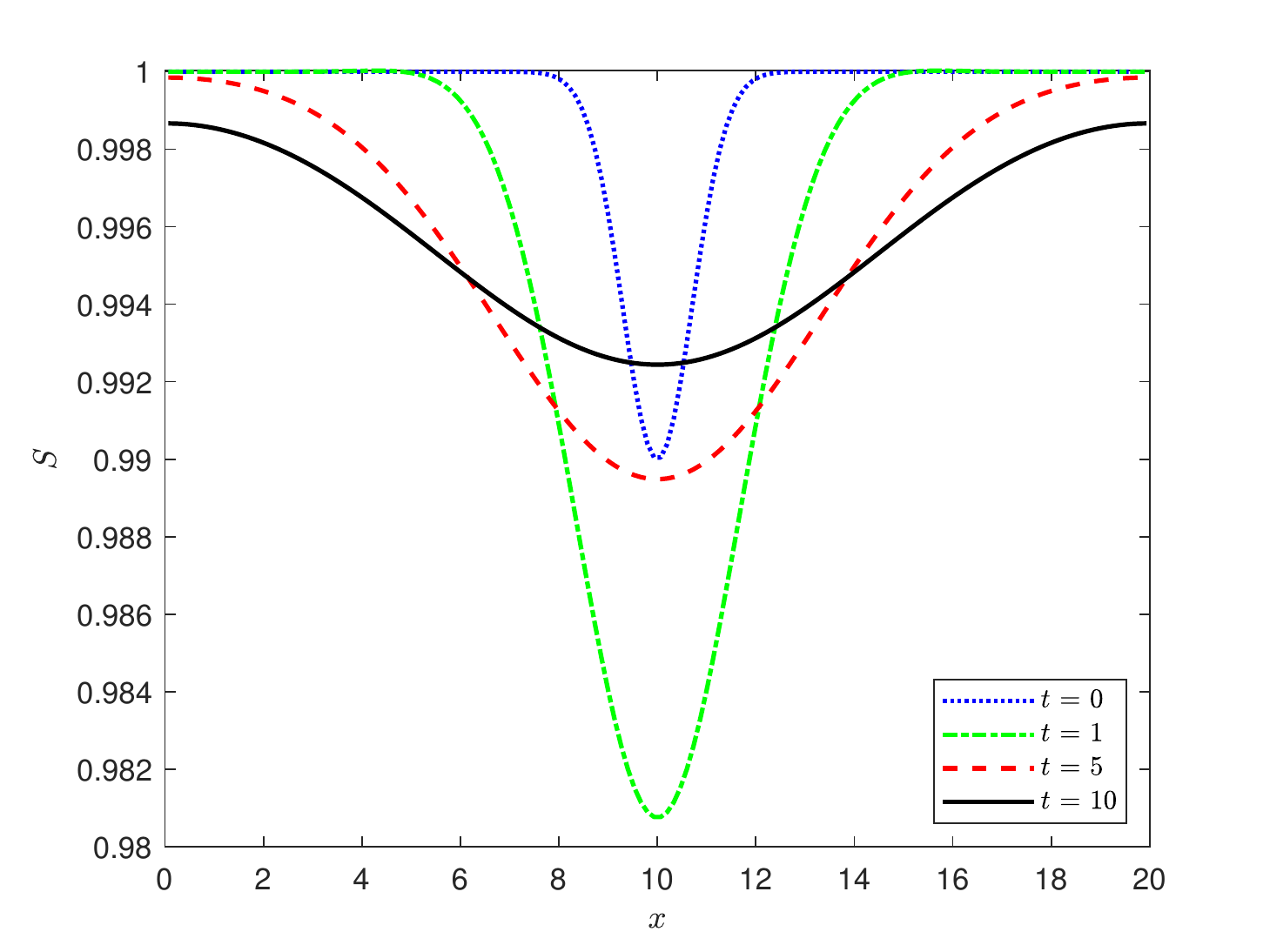}
\caption{(c)}
\label{fig.TC4_tau1e5_x_S}
\end{subfigure}
\begin{subfigure}{0.32\textwidth}
\includegraphics[width=1\linewidth]{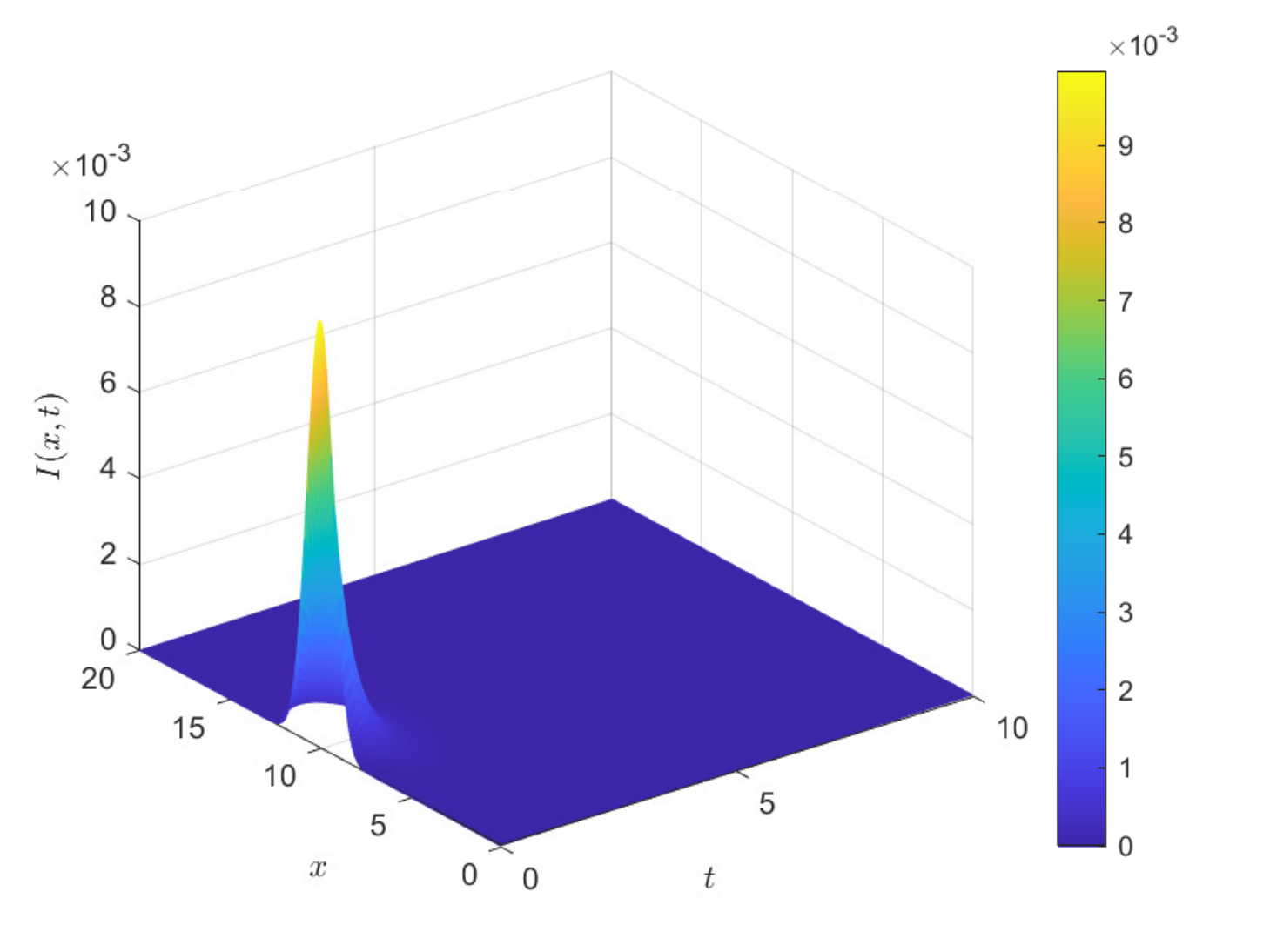}
\caption{(d)}
\label{fig.TC4_tau1e5_t10_xt_I}
\end{subfigure}
\begin{subfigure}{0.32\textwidth}
\includegraphics[width=1\linewidth]{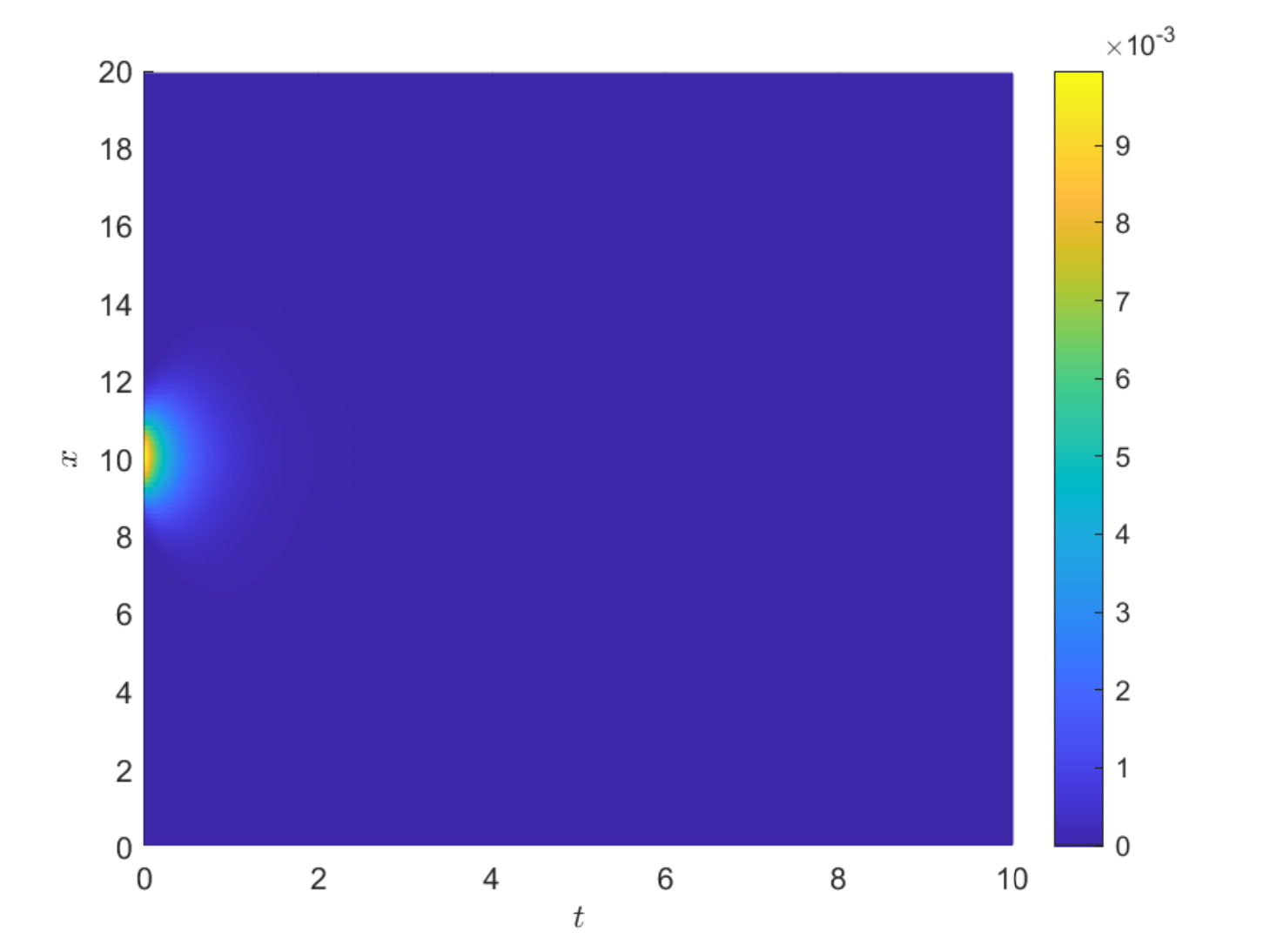}
\caption{(e)}
\label{fig.TC4_tau1e5_t10_xt_I_top}
\end{subfigure}
\begin{subfigure}{0.32\textwidth}
\includegraphics[width=1\linewidth]{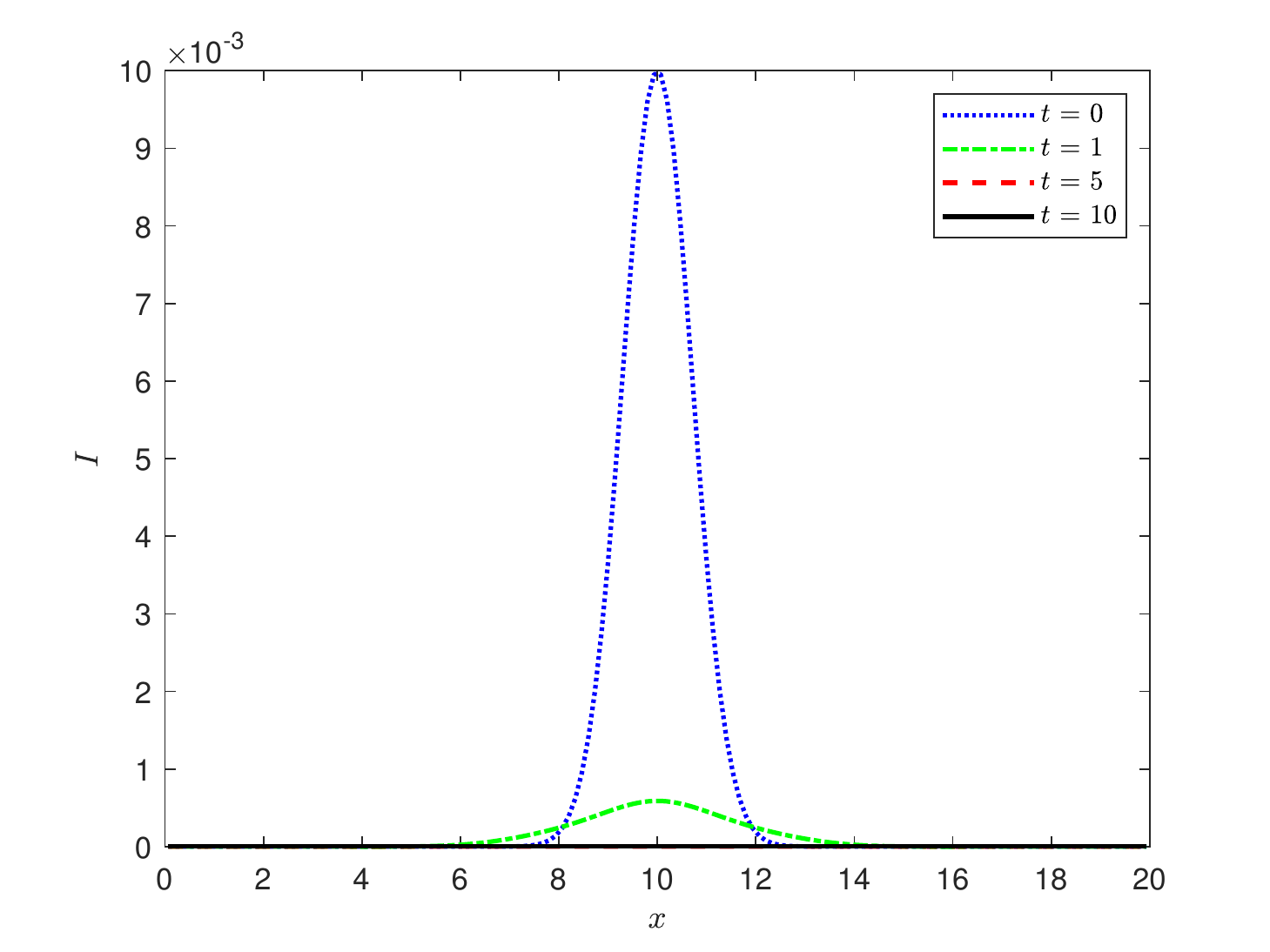}
\caption{(f)}
\label{fig.TC4_tau1e5_x_I}
\end{subfigure}
\caption{Numerical results of the spatially heterogeneous test case with parabolic configuration of relaxation times and characteristic velocities ($\tau = 10^{-5}, \lambda^2 = 10^5$) and reproduction number $R_0 <1$. Time and spatial evolution of $S$ presented in (a), (b) and (c); evolution of $I$ shown in (d), (e) and (f).}
\label{fig.TC4_tau1e5}
\end{figure}
\begin{figure}[t!]
\centering
\begin{subfigure}{0.32\textwidth}
\includegraphics[width=1\linewidth]{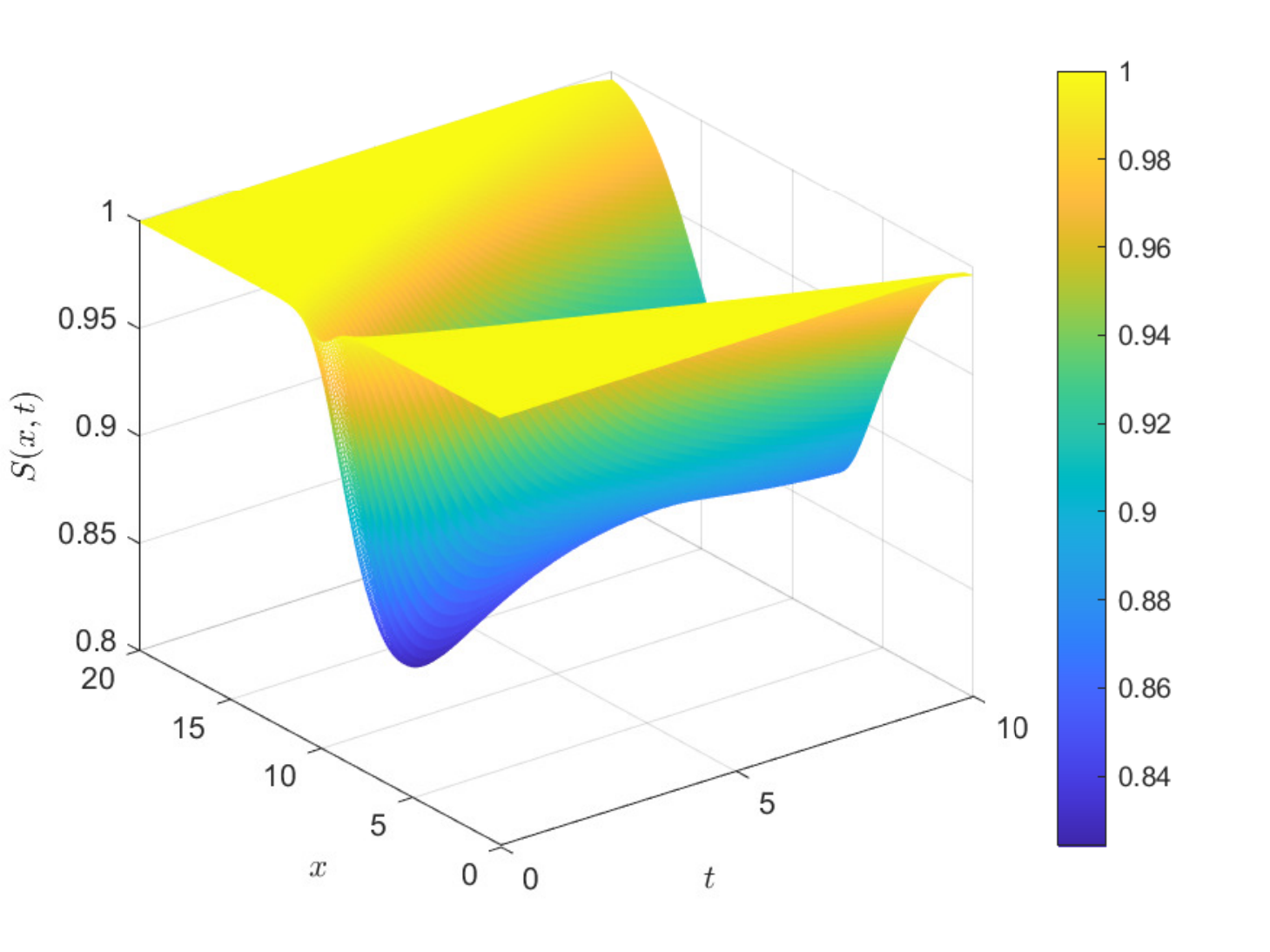}
\caption{(a)}
\label{fig.TC4.1_tau1_t10_xt_S}
\end{subfigure}
\begin{subfigure}{0.32\textwidth}
\includegraphics[width=1\linewidth]{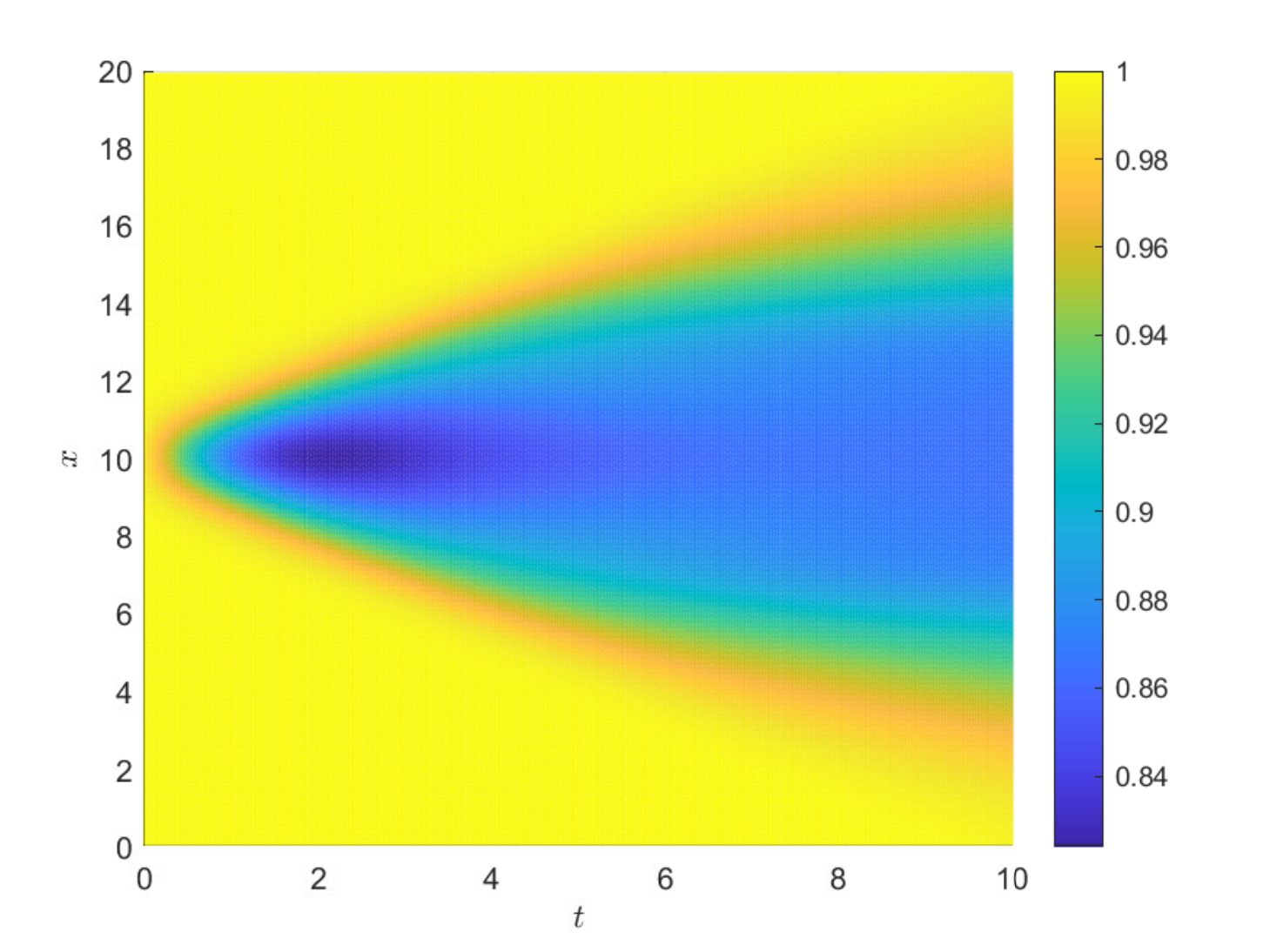}
\caption{(b)}
\label{fig.TC4.1_tau1_t10_xt_S_top}
\end{subfigure}
\begin{subfigure}{0.32\textwidth}
\includegraphics[width=1\linewidth]{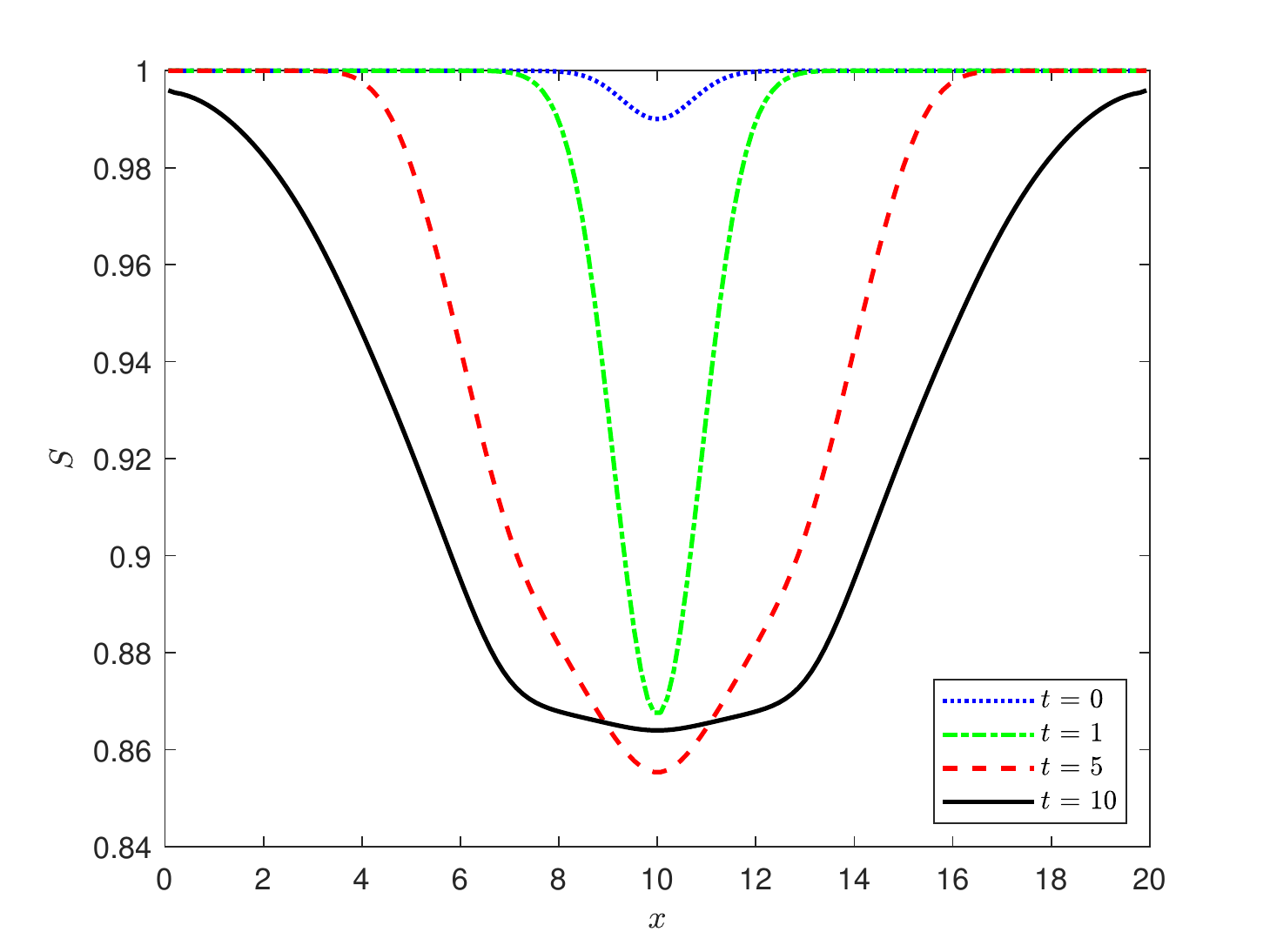}
\caption{(c)}
\label{fig.TC4.1_tau1_x_S}
\end{subfigure}
\begin{subfigure}{0.32\textwidth}
\includegraphics[width=1\linewidth]{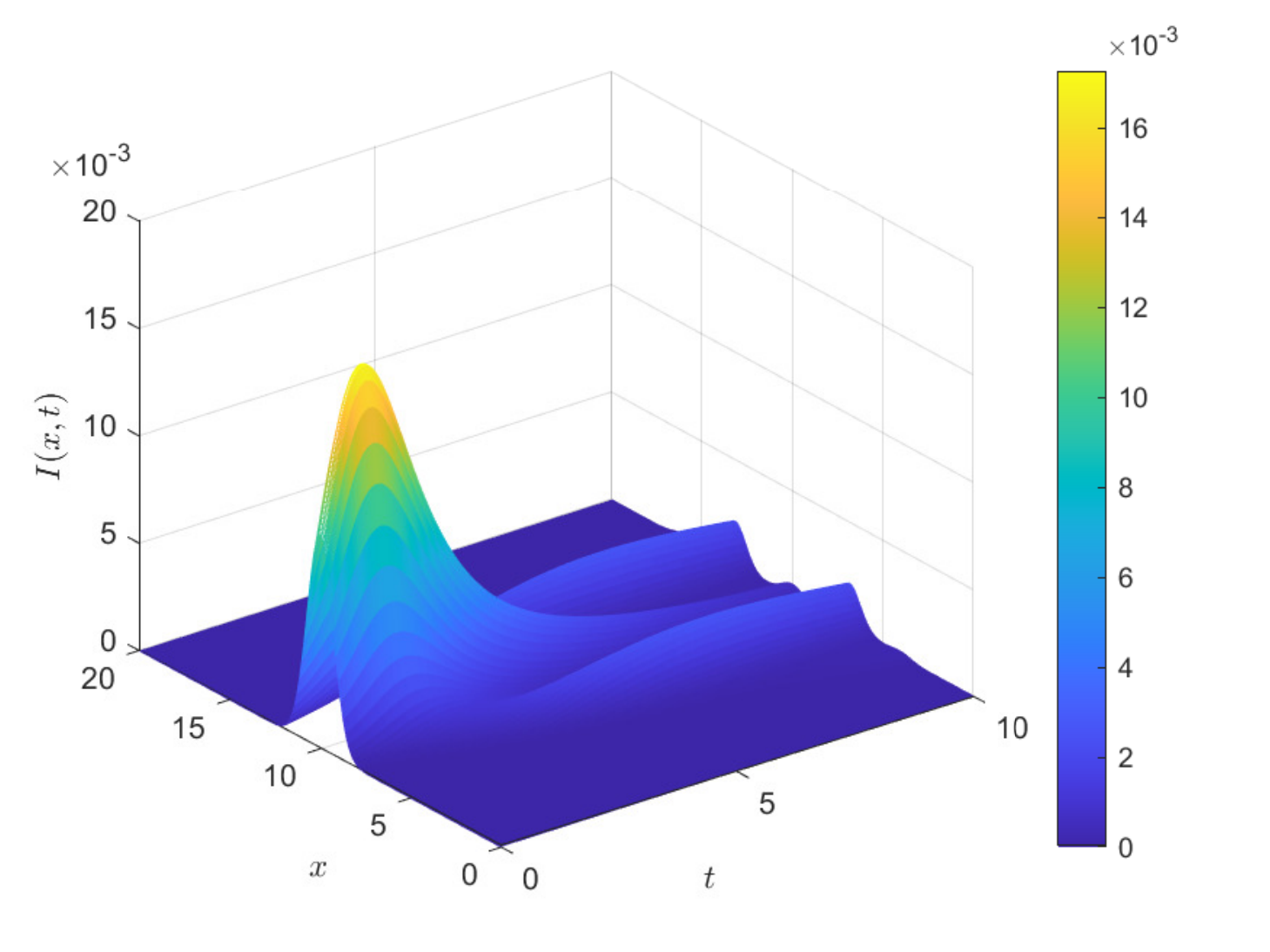}
\caption{(d)}
\label{fig.TC4.1_tau1_t10_xt_I}
\end{subfigure}
\begin{subfigure}{0.32\textwidth}
\includegraphics[width=1\linewidth]{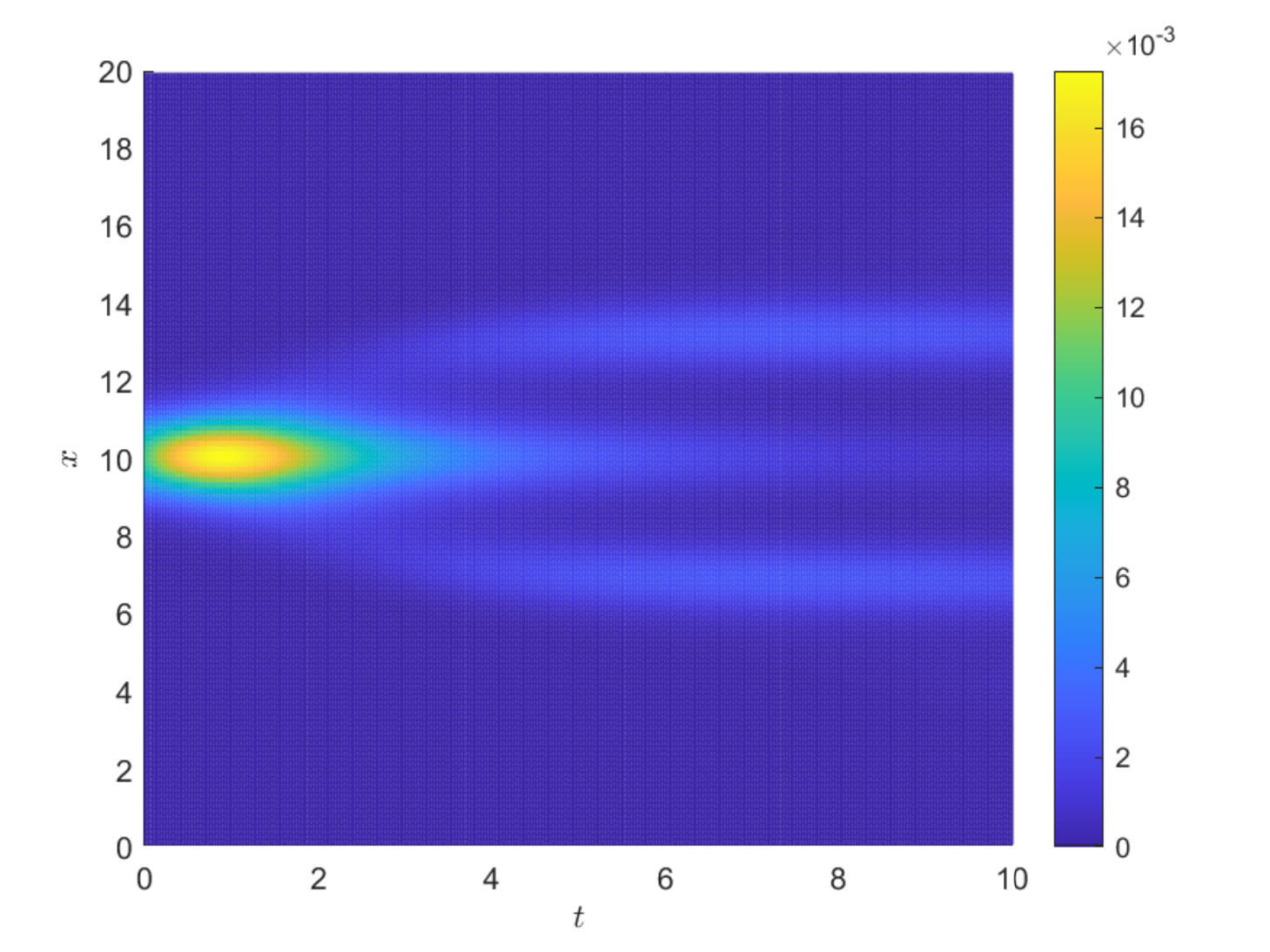}
\caption{(e)}
\label{fig.TC4.1_tau1_t10_xt_I_top}
\end{subfigure}
\begin{subfigure}{0.32\textwidth}
\includegraphics[width=1\linewidth]{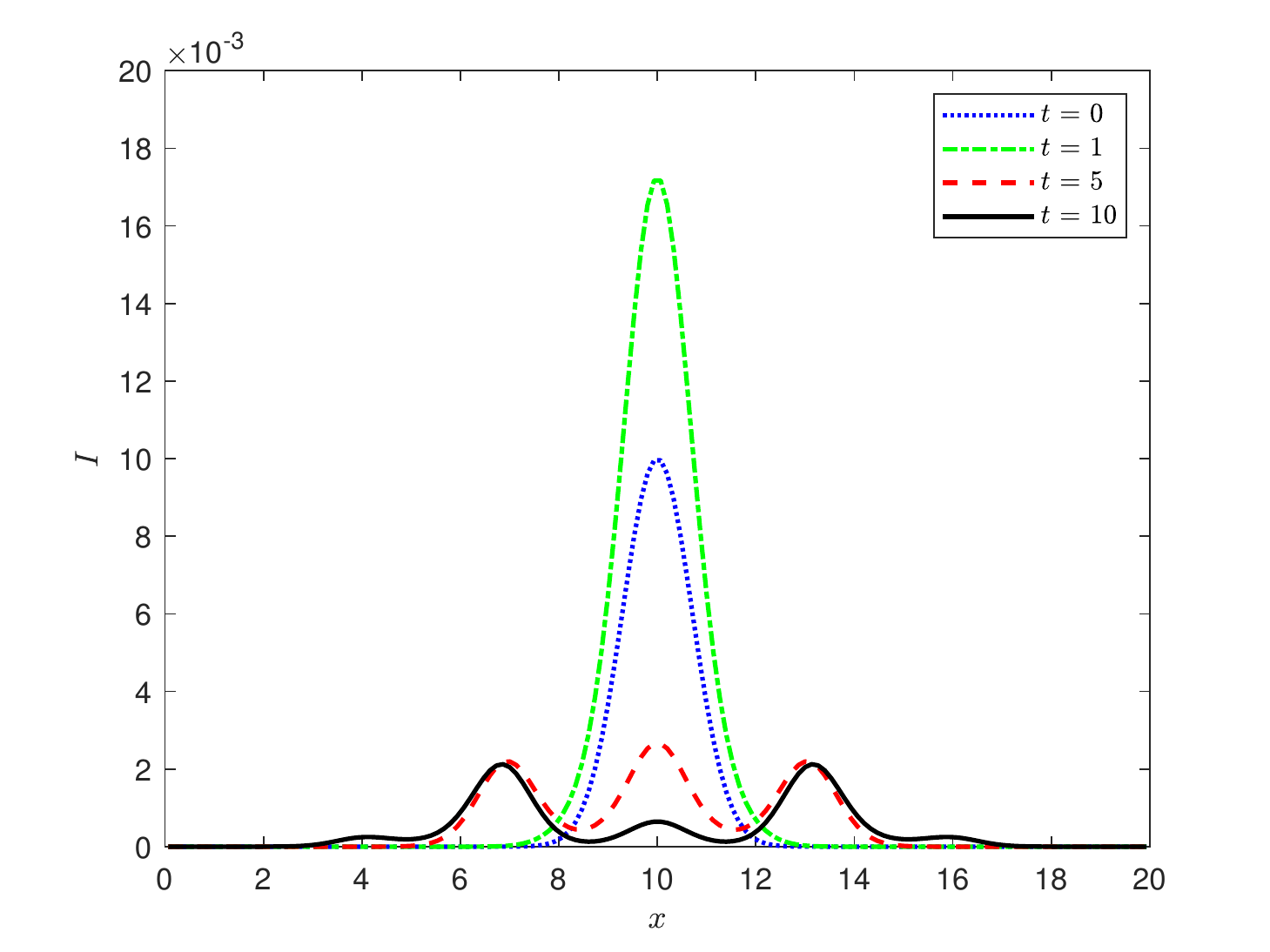}
\caption{(f)}
\label{fig.TC4.1_nc135_tau1_x_I}
\end{subfigure}
\caption{Numerical results of the spatially heterogeneous test case with hyperbolic configuration of relaxation times and characteristic velocities ($\tau = 1.0, \lambda^2 = 1.0$) and reproduction number $R_0 >1$. Time and spatial evolution of $S$ presented in (a), (b) and (c); evolution of $I$ shown in (d), (e) and (f).}
\label{fig.TC4.1_tau1}
\end{figure}
\begin{figure}[t!]
\centering
\begin{subfigure}{0.32\textwidth}
\includegraphics[width=1\linewidth]{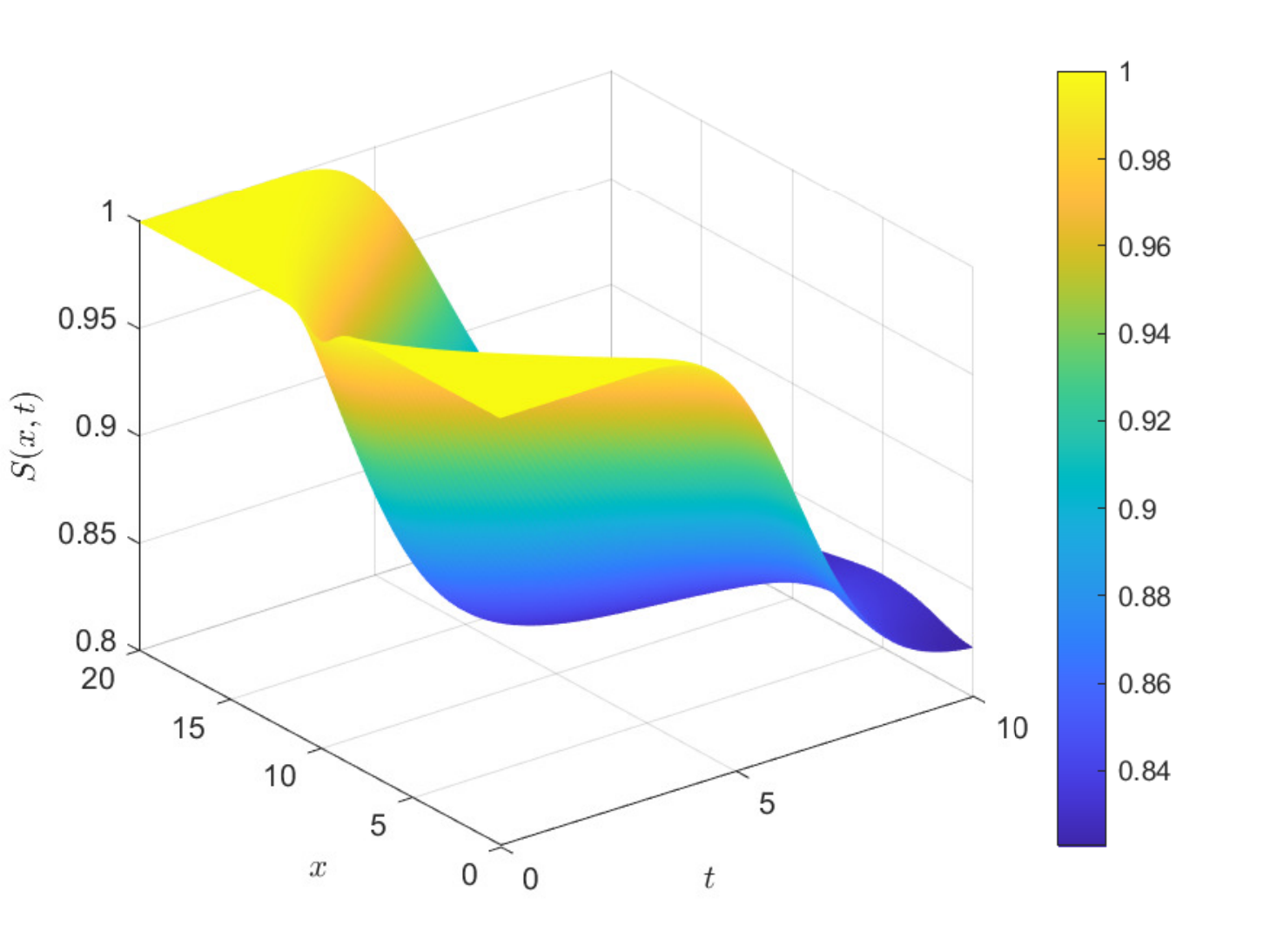}
\caption{(a)}
\label{fig.TC4.1_tau1e5_t10_xt_S}
\end{subfigure}
\begin{subfigure}{0.32\textwidth}
\includegraphics[width=1\linewidth]{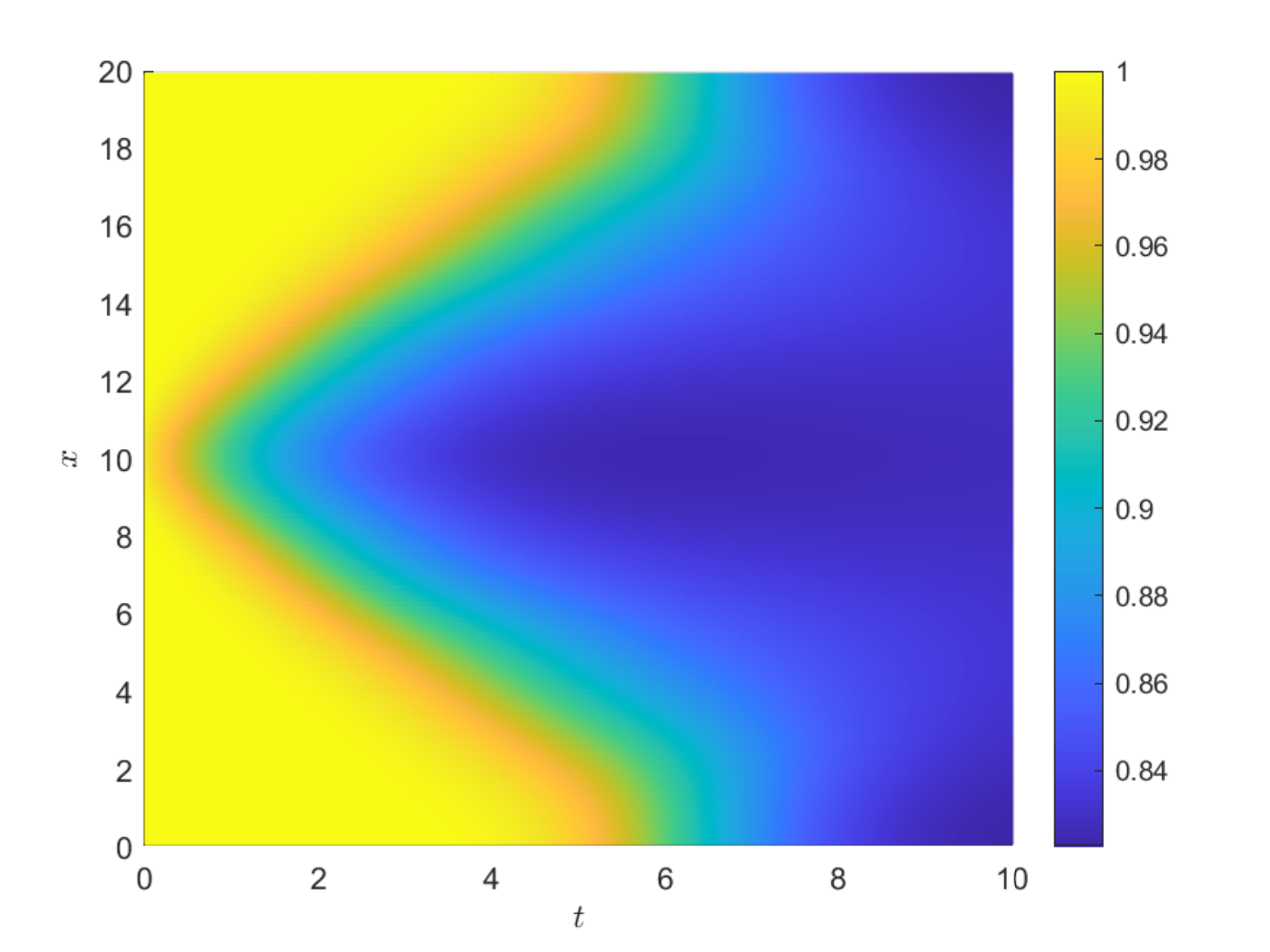}
\caption{(b)}
\label{fig.TC4.1_tau1e5_t10_xt_S_top}
\end{subfigure}
\begin{subfigure}{0.32\textwidth}
\includegraphics[width=1\linewidth]{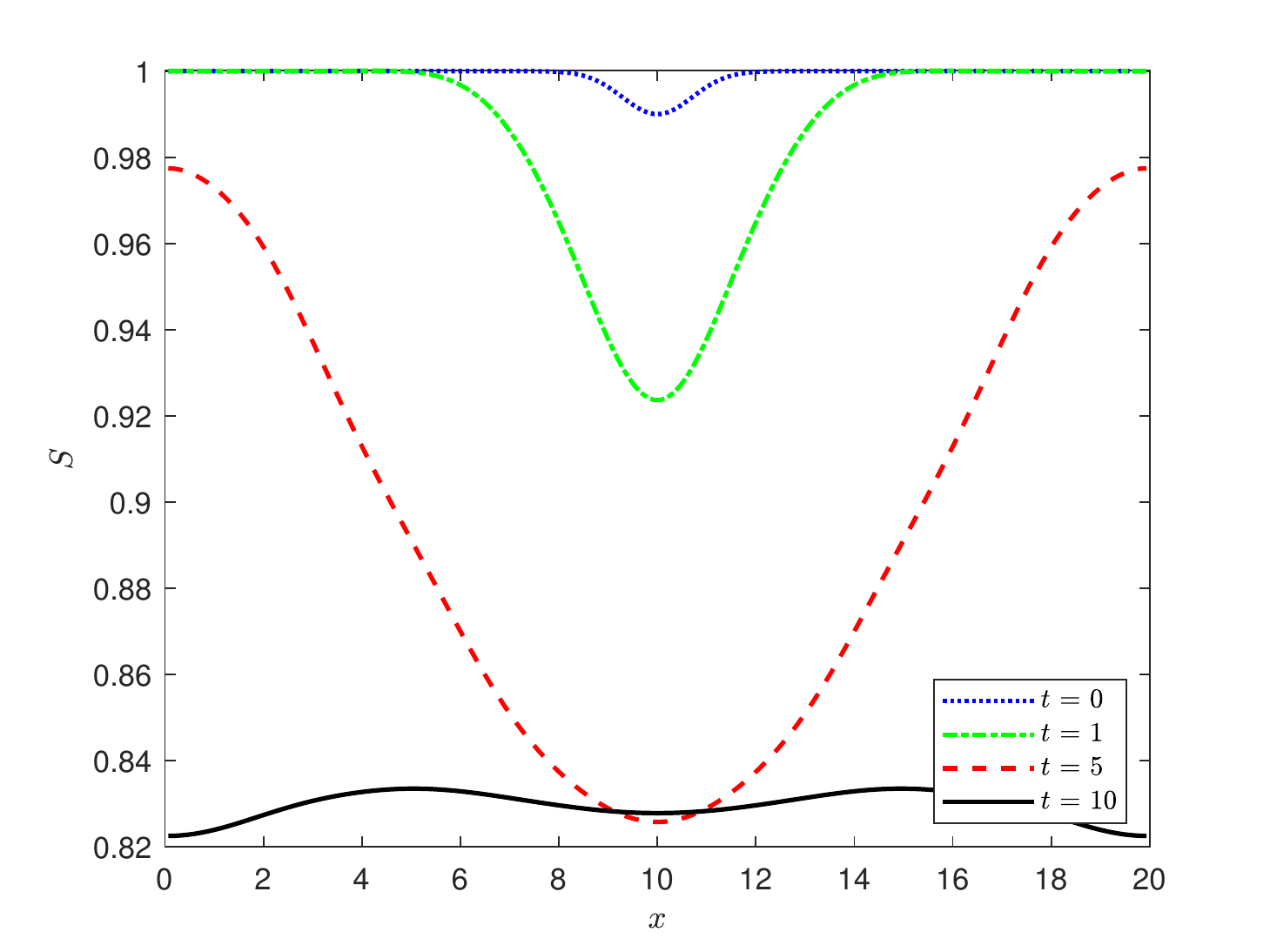}
\caption{(c)}
\label{fig.TC4.1_tau1e5_x_S}
\end{subfigure}
\begin{subfigure}{0.32\textwidth}
\includegraphics[width=1\linewidth]{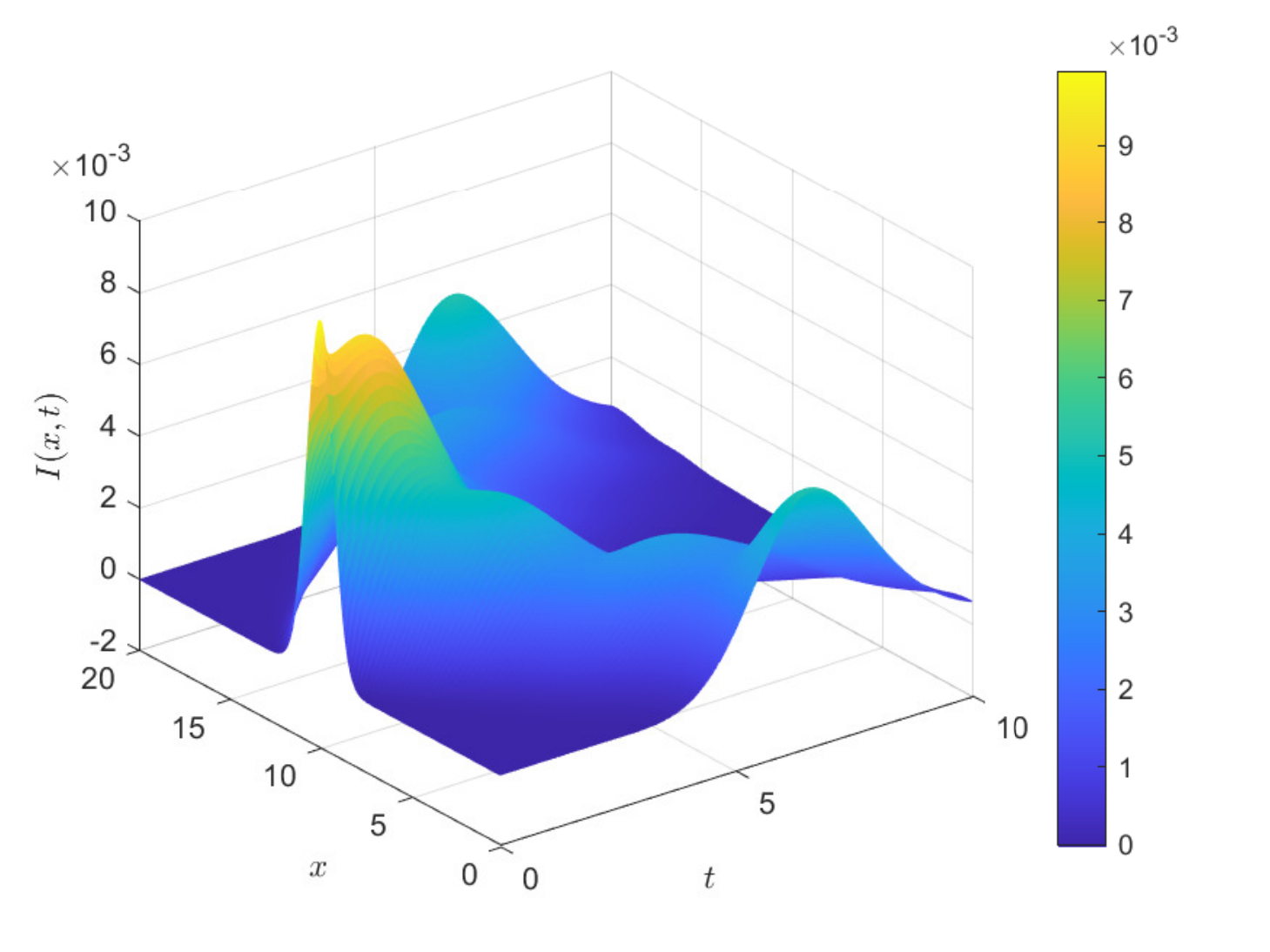}
\caption{(d)}
\label{fig.TC4.1_tau1e5_t10_xt_I}
\end{subfigure}
\begin{subfigure}{0.32\textwidth}
\includegraphics[width=1\linewidth]{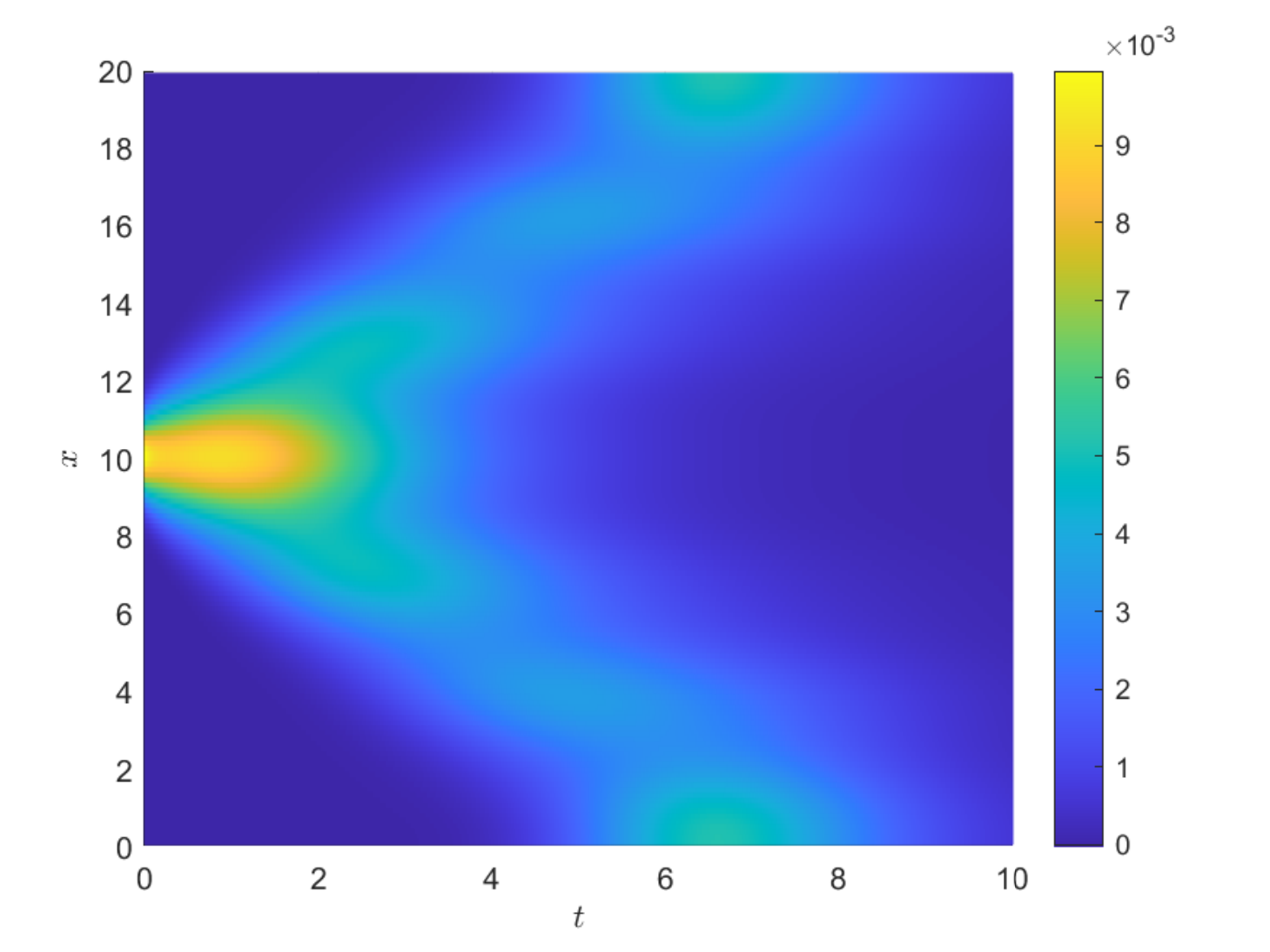}
\caption{(e)}
\label{fig.TC4.1_tau1e5_t10_xt_I_top}
\end{subfigure}
\begin{subfigure}{0.32\textwidth}
\includegraphics[width=1\linewidth]{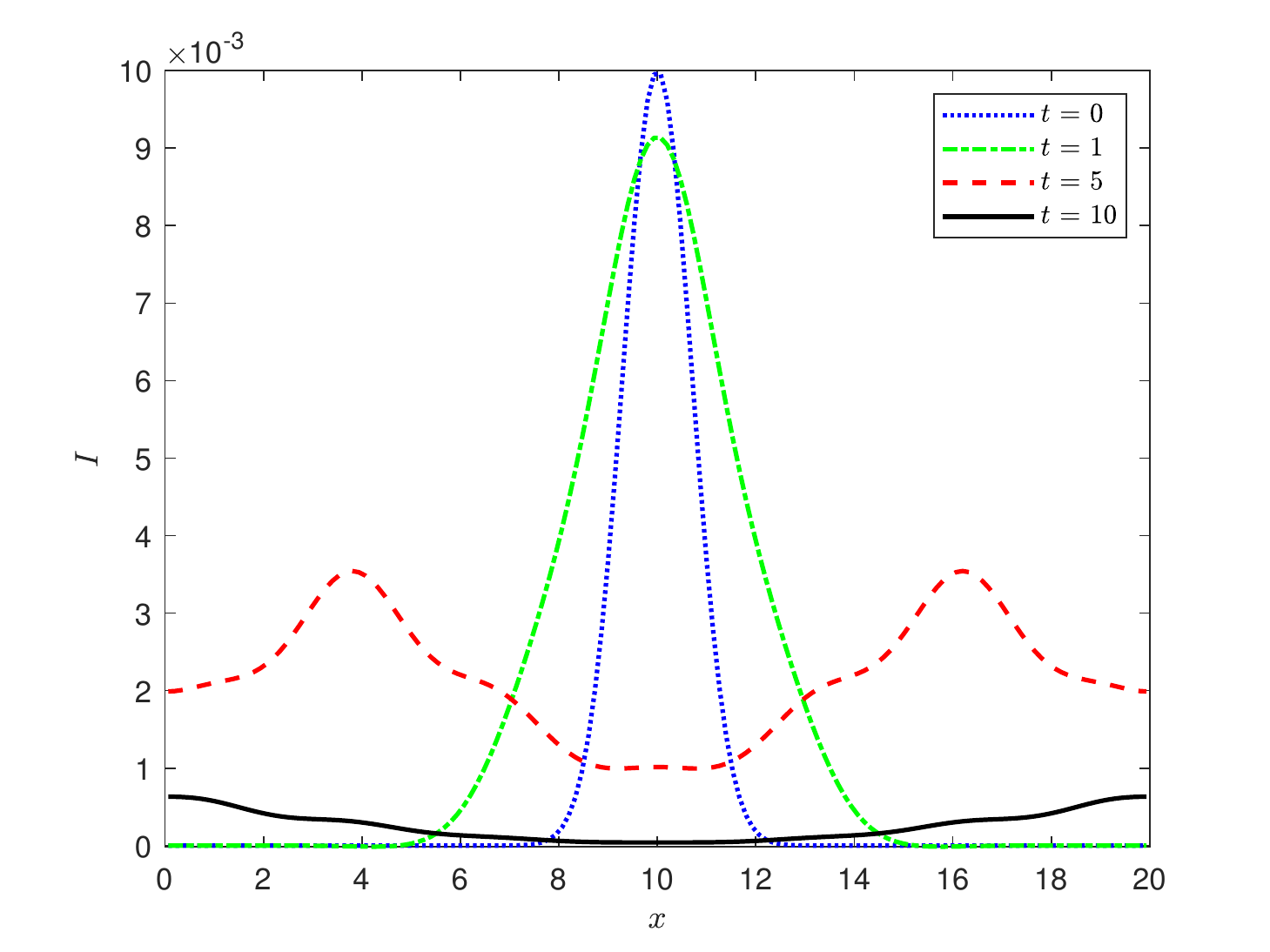}
\caption{(f)}
\label{fig.TC4.1_tau1e5_x_I}
\end{subfigure}
\caption{Numerical results of the spatially heterogeneous test case with parabolic configuration of relaxation times and characteristic velocities ($\tau = 10^{-5}, \lambda^2 = 10^5$) and reproduction number $R_0 >1$. Time and spatial evolution of $S$ presented in (a), (b) and (c); evolution of $I$ shown in (d), (e) and (f).}
\label{fig.TC4.1_tau1e5}
\end{figure}
\subsection{Accuracy analysis}
\label{section_accuracy}
To verify the order of accuracy of the scheme, even in stiff regimes, an accuracy analysis is conducted considering periodic boundary conditions and the following initial conditions
\begin{equation*}
S(x,0) = 0.5\left( 1 + \mathrm{sin}\frac{2\pi x}{L}\right), \qquad I(x,0) = 1 - S(x,0),  \qquad R(x,0) = 0.0 ,
\end{equation*}
with $x \in [-1;1]$, hence a domain length $L = 2$, and null initial fluxes. In these tests, the contact rate is $\beta = 10.0$ and the recovery rate is $\gamma = 4.0$, with $k=0$ as the classical bilinear case. In each simulation, relaxation times and characteristic velocities are fixed equal for all the compartments of the population $\lambda_S=\lambda_I=\lambda_R=\lambda$, taking into account three different cases, corresponding to a hyperbolic system ($\tau = 1.0$, $\lambda^2 = 1.0$), a mildly diffusive system ($\tau = 10^{-2}$, $\lambda^2 = 10^2$) and a purely diffusive system ($\tau = 10^{-6}$, $\lambda^2 = 10^6$). The stability condition is satisfied imposing $\Delta t = \Delta x \, \max\left\{\CFL/\lambda; \nu \Delta x \right\}$. The final time of the simulations is $t_{end} = 0.1$. The refinement of the computational grid is made with a factor 3, to work with embedded grids, and the time step $\Delta t$ decreases with $\Delta x$ accordingly to the stability condition. Values obtained with $N_x = 1215$ cells are taken as reference solutions. \par
The analysis is performed with both the IMEX-BPR(4,4,2) scheme in the AP-explicit form following eqs.~\eqref{eq:vh}-\eqref{eq:imexcs}, and the IMEX-BPR(4,4,2) scheme in the AP-implicit form presented in eqs.~\eqref{eq:vh2}-\eqref{eq:imexcs2}. When using the AP-explicit form, in the stability condition $\CFL = 0.9$ and $\nu = 0.5$; when using the AP-implicit form, $\CFL = 0.9$ and $\nu = 0.5/\Delta x$.\par
Results are presented for variables $S$, $I$, $J_S$ and $J_I$ in Table~\ref{tab:accuracy_1} for the AP-explicit scheme and in Table~\ref{tab:accuracy_2} for the AP-implicit scheme. Errors are reported in terms of $L^1$ norms and related order of accuracy, evaluated as
\begin{equation*}
\mathcal{O}\left( L^1\right) = \mathrm{log}_3\left(\frac{||E_{\Delta x}||}{||E_{\Delta x/3}||} \right),
\end{equation*}
with $||E_{\Delta x}||$ the relative error computed with grid size $\Delta x$. We observe that the second-order of accuracy of the method is satisfied by both versions of the scheme uniformly in all regimes. It is worth to underline that the expected accuracy is maintained in the limit of diffusive regime for all the variables, even for the fluxes, thanks to the GSA property of the scheme and because the following additional conditions are satisfied by the IMEX scheme \eqref{eq:tables} (see \cite{boscarino2017}):
\begin{equation}
b^TA^{-2}\tilde A e = 1, \qquad b^TA^{-2}\tilde A c^2 = 1, \qquad b^TA^{-2}\tilde A A c = 1/2, \qquad b^TA^{-2}\tilde A \tilde A c = 1/2 .
\end{equation}
\subsection{Spatially heterogeneous environments}
\label{section_spatially_heterogeneous_tests}
Following \cite{wang2020}, we analyze the behavior of the model concerning spatially heterogeneous environments, taking into account a spatially variable contact rate
\begin{equation*}
\beta(x) = \hat \beta \left(1 + 0.05 \, \mathrm{sin} \frac{13 \pi x}{20} \right) .
\end{equation*}
Initial conditions are imposed as follows
\begin{equation*}
S(x,0) = 1 - I(x,0), \qquad I(x,0) = 0.01\,e^{-(x-10)^2}, \qquad R(x,0) = 0.0 ,
\end{equation*}
with fluxes $J_S(x,0) = J_I(x,0) = J_R(x,0) = 0.0$ and zero-flux boundary conditions. No social distancing or control effects are considered in the incidence function, $k=0$. Simulations are performed in two different scenarios. In the first one an overall value in the domain of the initial reproduction number $R_0 = 0.808 < 1$ is considered, choosing $\hat \beta = 8.0$ and $\gamma = 10.0$, corresponding to an infection which is not able to start spreading. In the second scenario, the initial reproduction number is  $R_0 = 1.111 > 1$, given by the choice $\hat \beta = 11.0$ and $\gamma = 10.0$, which identifies an infection that can persist in a new host population. Moreover, for each scenario, two different sets of relaxation times are considered, to concern both the hyperbolic and the parabolic limit of the system of equations. In the hyperbolic configuration, the relaxation times of all the compartments of individuals are $\tau = 1.0$, with the square of the characteristic velocities $\lambda^2 = 1.0$; while in the parabolic configuration $\tau = 10^{-5}$ and $\lambda^2 = 10^5$. Simulations are run with the AP-explicit scheme, with $N_x = 150$ cells in a domain having length $L = 20$, with final time $t_{end} = 10$.\par
Numerical results of each one of the two tests performed for the scenario representing an infectious disease characterized by $R_0 <1$ are shown in Figs.~\ref{fig.TC4_tau1} and \ref{fig.TC4_tau1e5}. We can observe that when the reproduction number is less than 1, the effect of the spatial variability of the contact rate vanishes and the amount of infected individuals converges to zero very rapidly as time evolves. When comparing the trend of the solution obtained imposing $\tau = 1.0, \lambda^2 = 1.0$ with the one obtained considering $\tau = 10^{-5}, \lambda^2 = 10^5$, different dynamics of the infectious spread are noticed, which represents the tendency of the system towards more and more diffuse behavior as relaxation times reach values close to zero and characteristic speeds close to infinity.\par
In Figs.~\ref{fig.TC4.1_tau1} and \ref{fig.TC4.1_tau1e5}, numerical results for the scenario with $R_0$ slightly $>1$ are reported. Here, a temporary persistence of the infectious can be noticed, with oscillations that reflect the sinusoidal form of the spatially variable contact rate. In this case, differences of the dynamics of the epidemics in the two configurations of the relaxation times are more accentuated. In particular, observing the evolution of susceptible individuals, it can be seen that in the purely diffusive case the amount of susceptible tends to a much lower equilibrium value than in the hyperbolic case, with almost all the individuals of the system infected by the disease.
\begin{figure}[t!]
\centering
\begin{subfigure}{0.32\textwidth}
\includegraphics[width=1\linewidth]{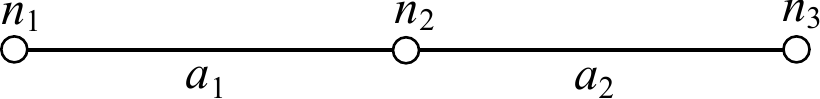}
\caption{}
\label{fig.network1_scheme}
\end{subfigure}
\begin{subfigure}{0.32\textwidth}
\includegraphics[width=1\linewidth]{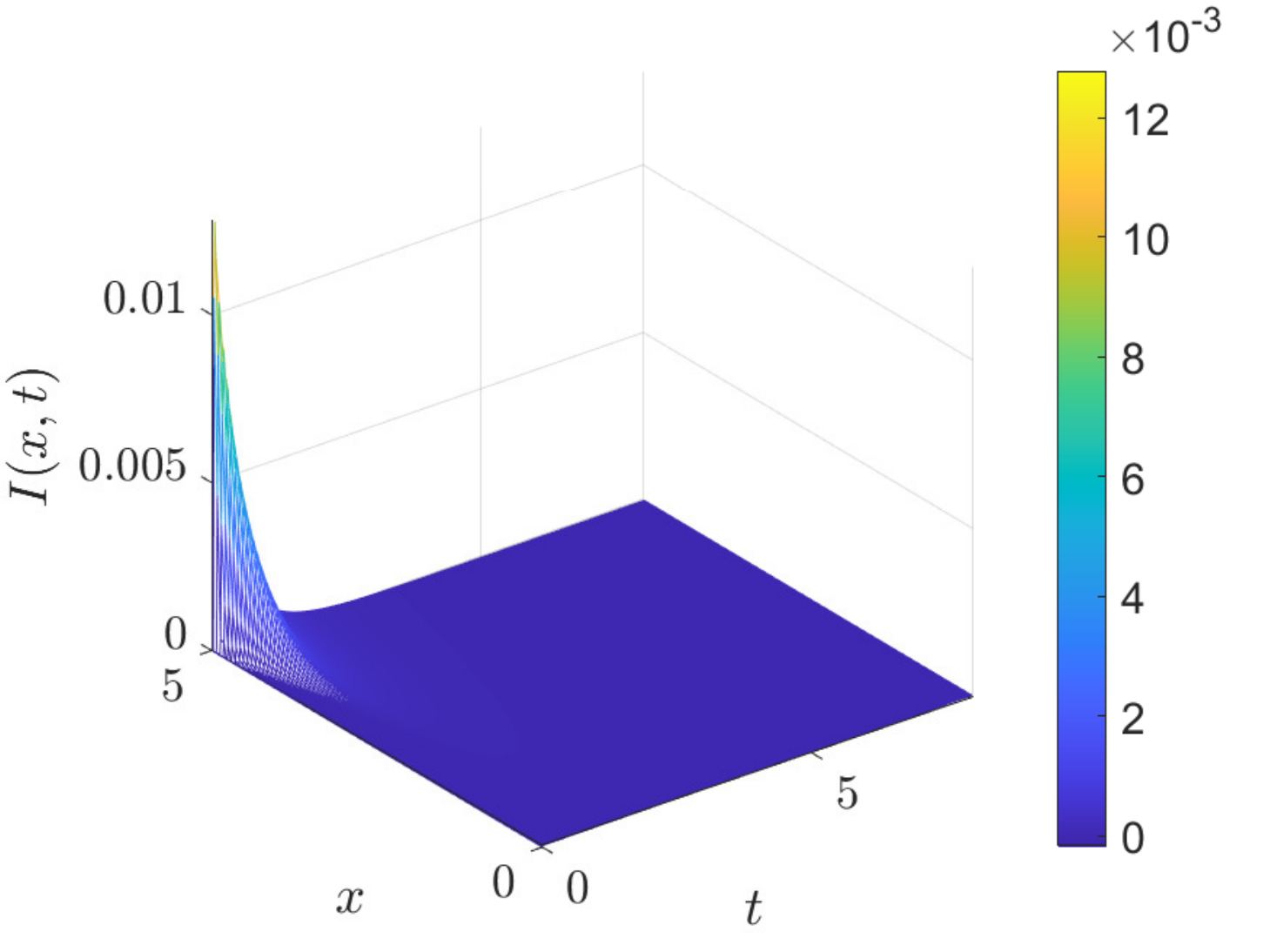}
\caption{($a_1$)}
\label{fig.Network1_arc1}
\end{subfigure}
\begin{subfigure}{0.32\textwidth}
\includegraphics[width=1\linewidth]{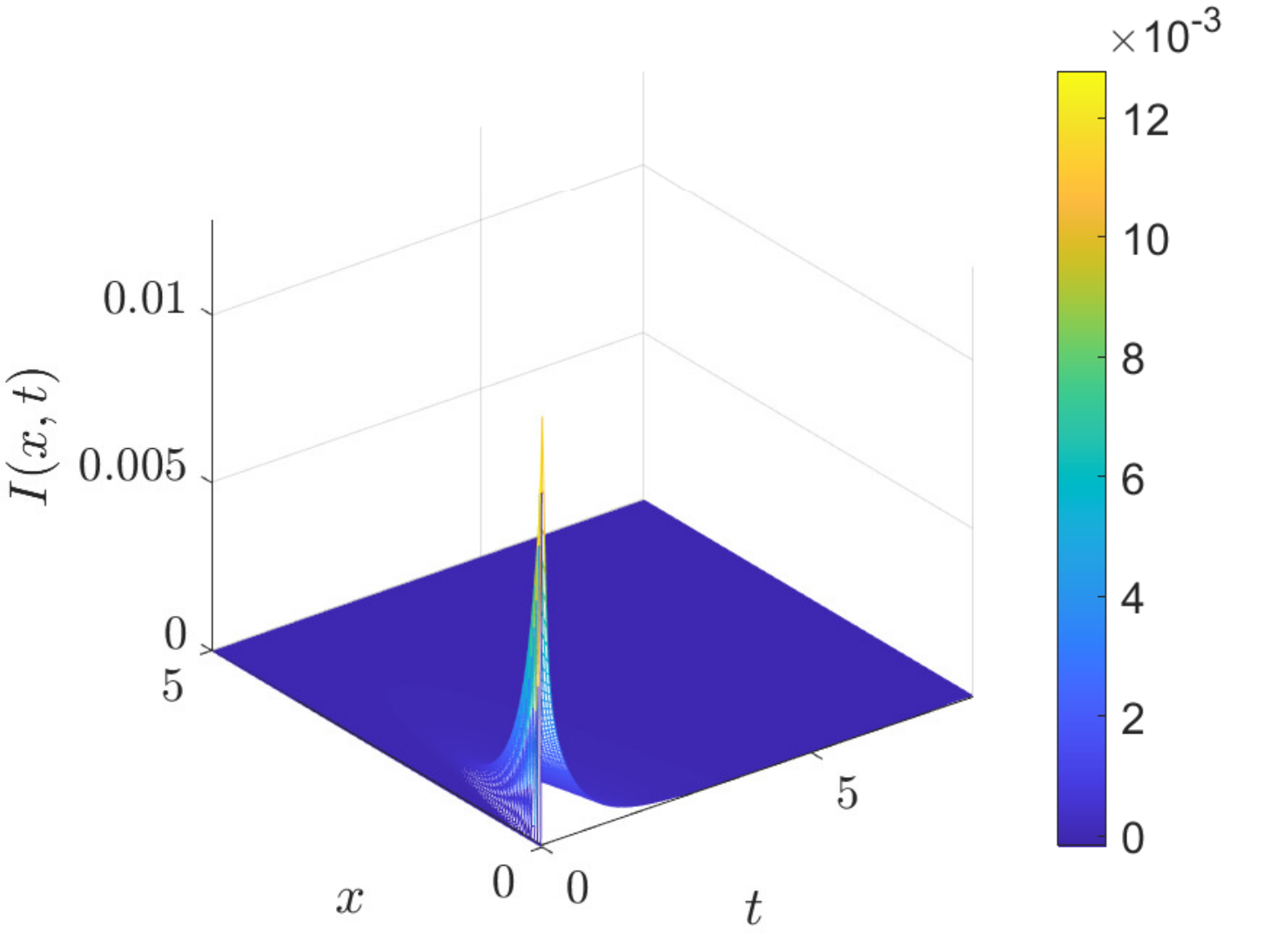}
\caption{($a_2$)}
\label{fig.Network1_arc2}
\end{subfigure}
\begin{subfigure}{0.32\textwidth}
\includegraphics[width=1\linewidth]{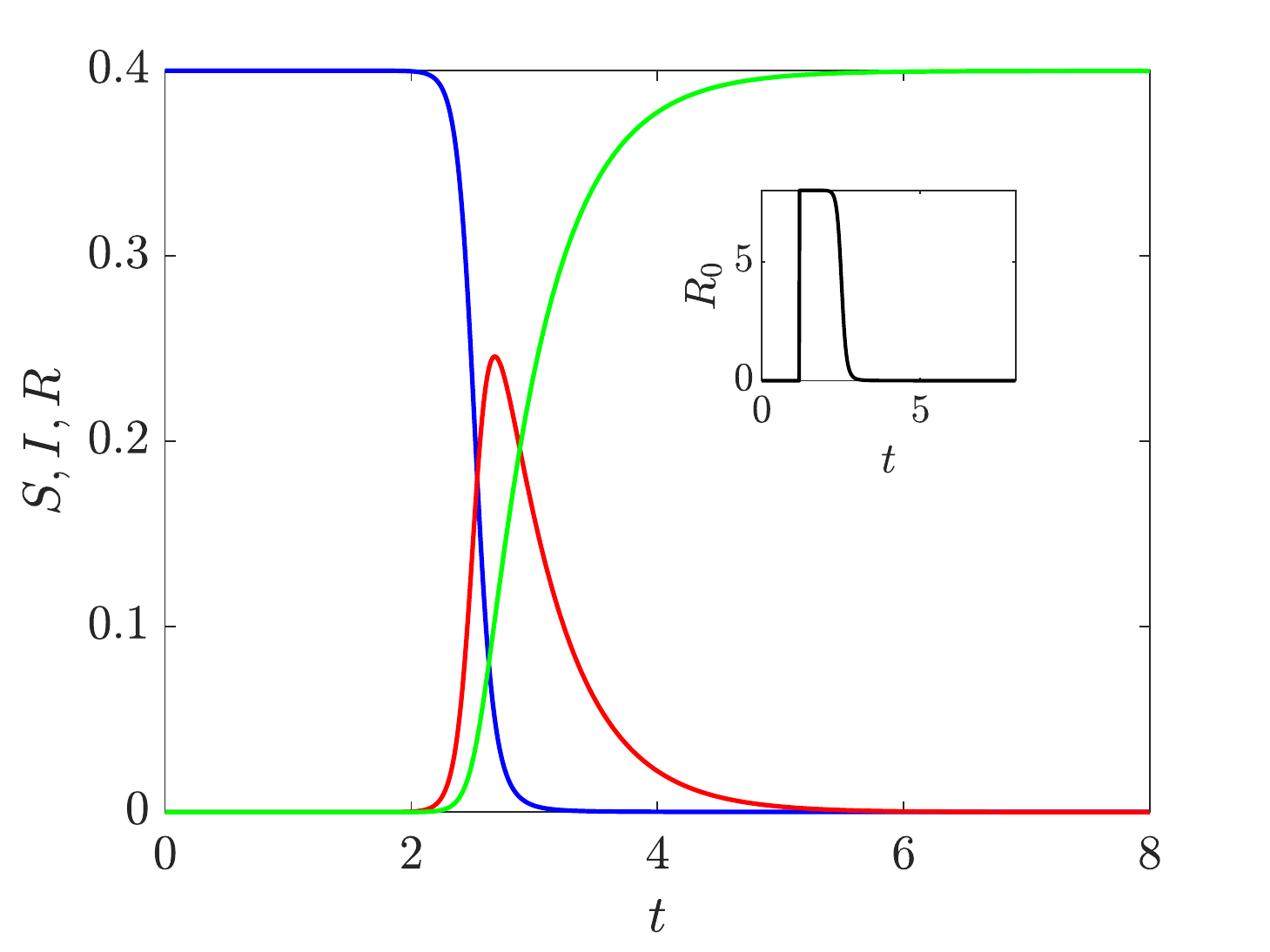}
\caption{($n_1$)}
\label{fig.Network1_node1}
\end{subfigure}
\begin{subfigure}{0.32\textwidth}
\includegraphics[width=1\linewidth]{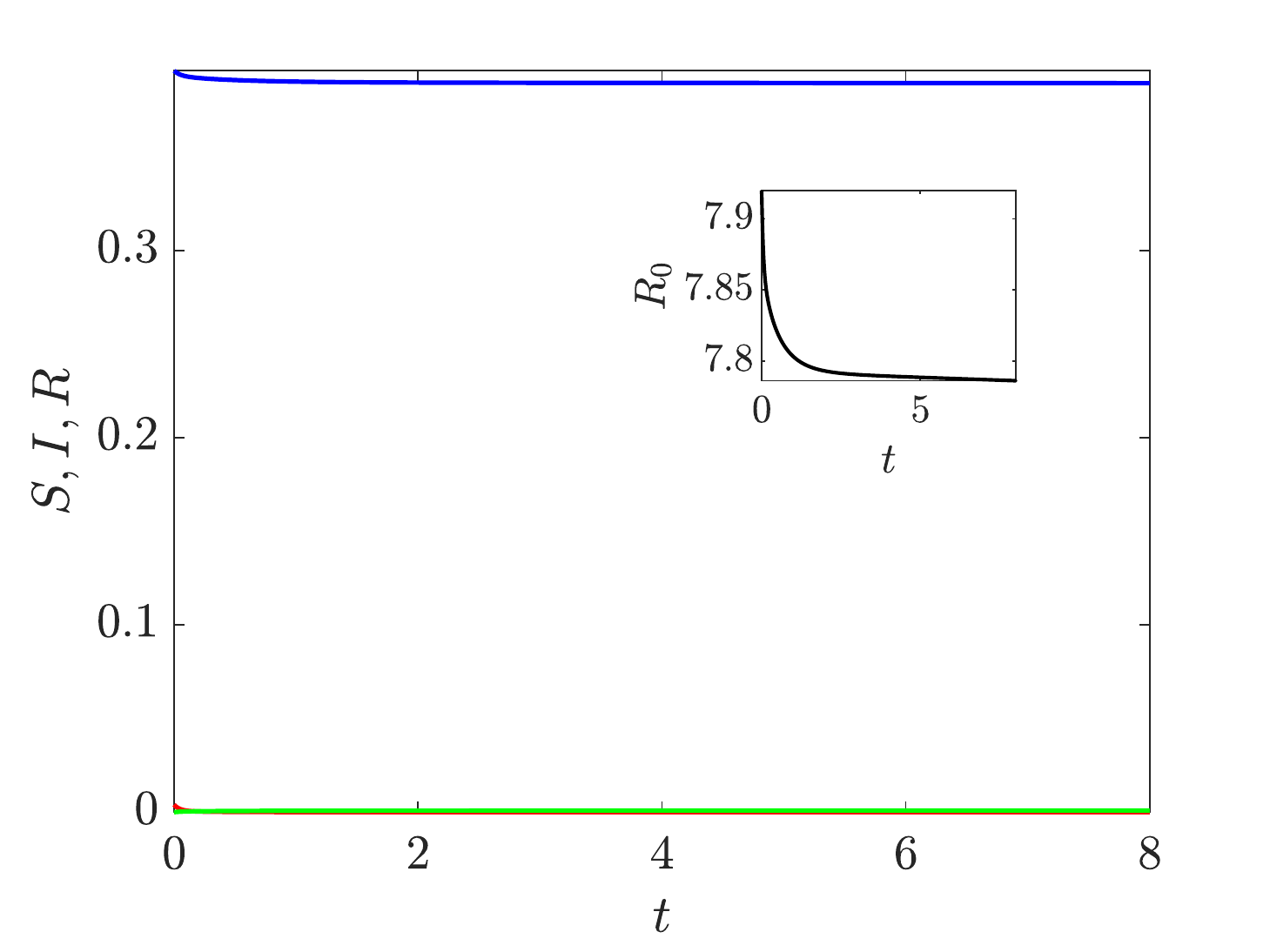}
\caption{($n_2$)}
\label{fig.Network1_node2}
\end{subfigure}
\begin{subfigure}{0.32\textwidth}
\includegraphics[width=1\linewidth]{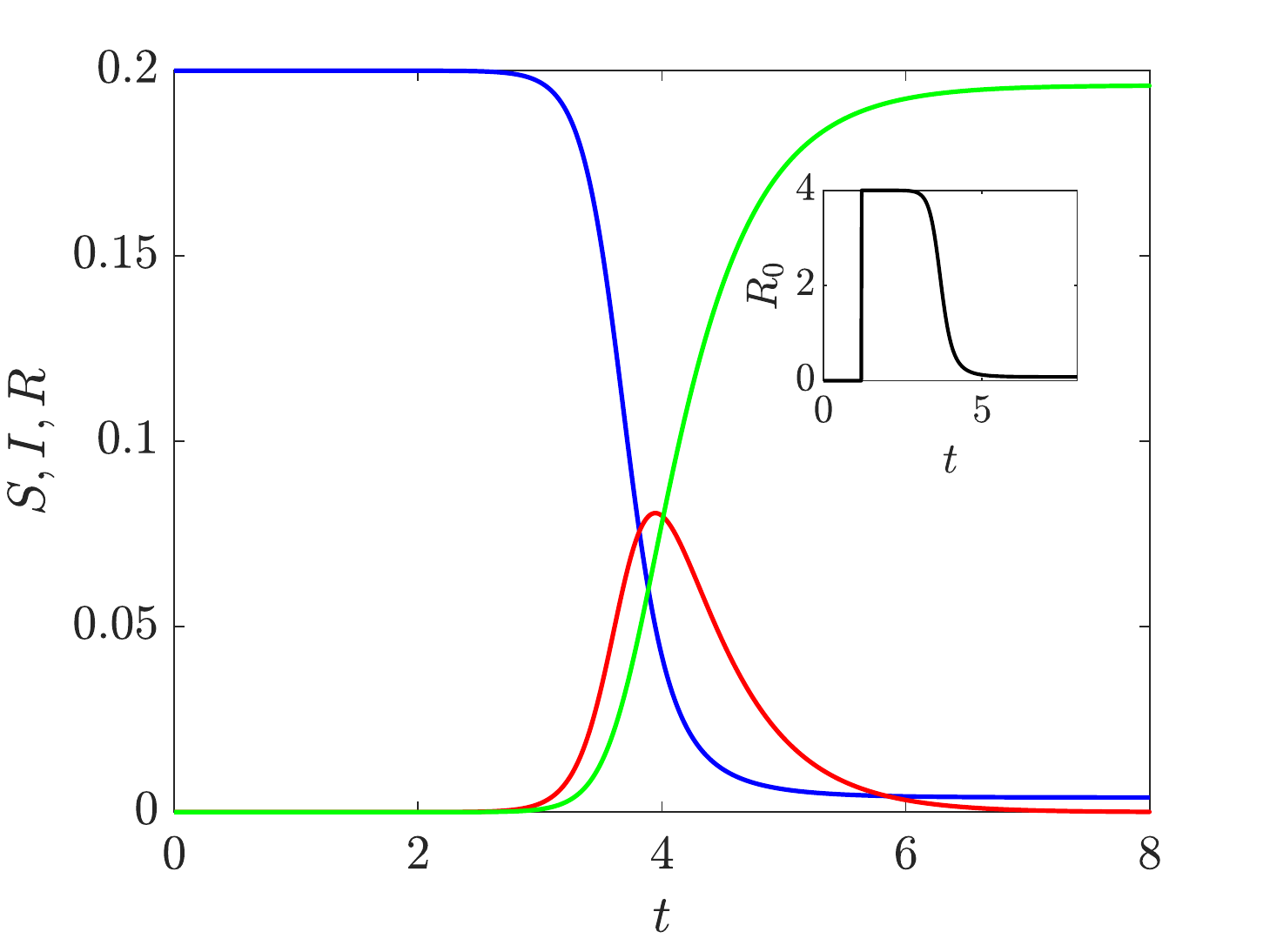}
\caption{($n_3$)}
\label{fig.Network1_node3}
\end{subfigure}
\caption{Numerical results of Test 1a, in which there is a symmetric temporal and spatial dynamics of infectious individuals $I$ along arcs $a_1$ and $a_2$ and stationary susceptible and recovered people. Time evolution of $S$ (blue), $I$ (red) and $R$ (green) at node $n_1$, node $n_2$ (from which the infectious disease starts spreading symmetrically along the two arcs) and at node $n_3$. In the same plots, the evolution of the coefficient $R_0$ is also shown.}
\label{fig.Network1}
\end{figure}
\begin{figure}[t!]
\centering
\begin{subfigure}{0.32\textwidth}
\includegraphics[width=1\linewidth]{network1_scheme}
\caption{}
\end{subfigure}
\begin{subfigure}{0.32\textwidth}
\includegraphics[width=1\linewidth]{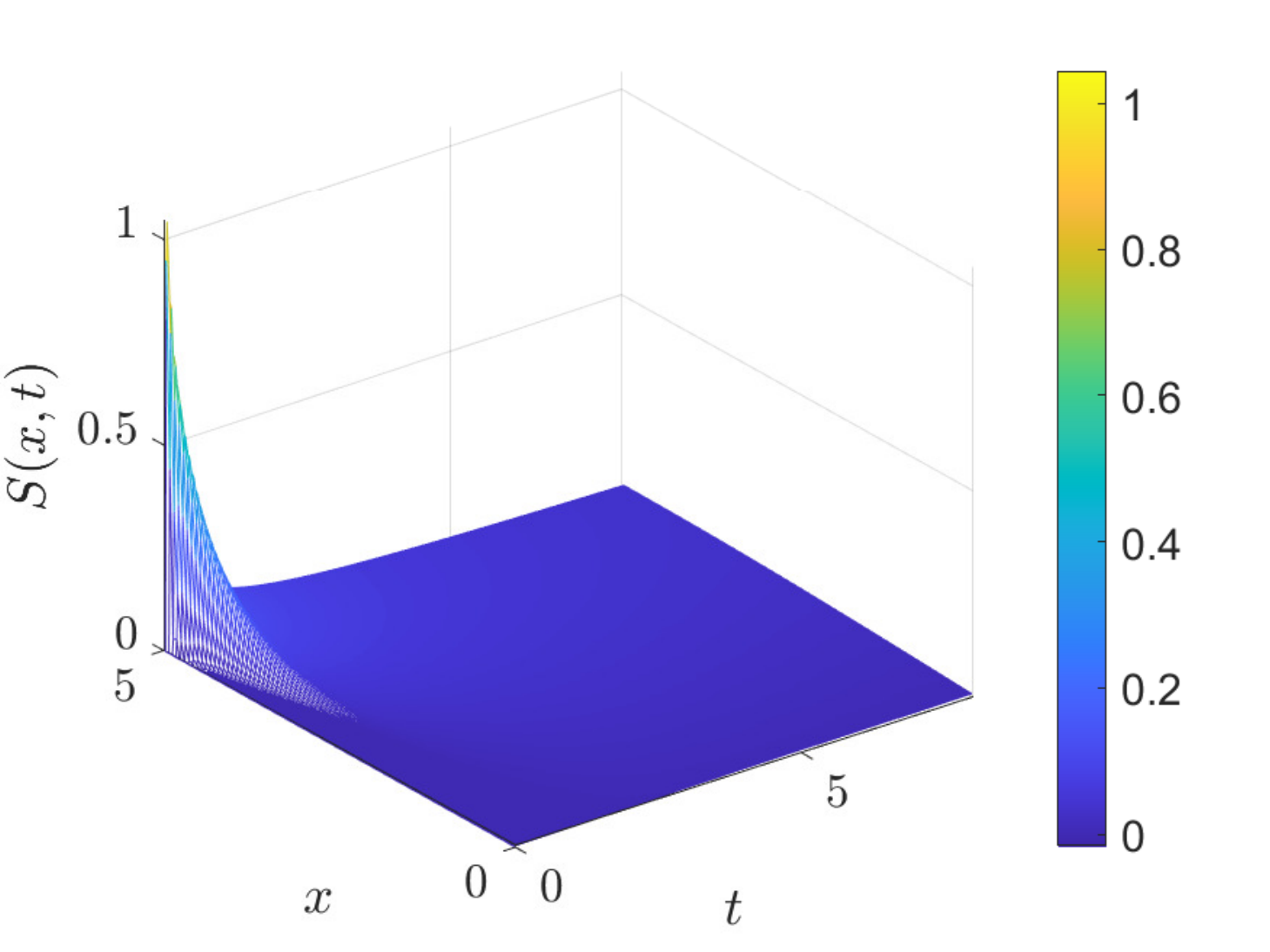}
\caption{($a_1$ - $S$)}
\label{fig.Network1_arc1_new_S}
\end{subfigure}
\begin{subfigure}{0.32\textwidth}
\includegraphics[width=1\linewidth]{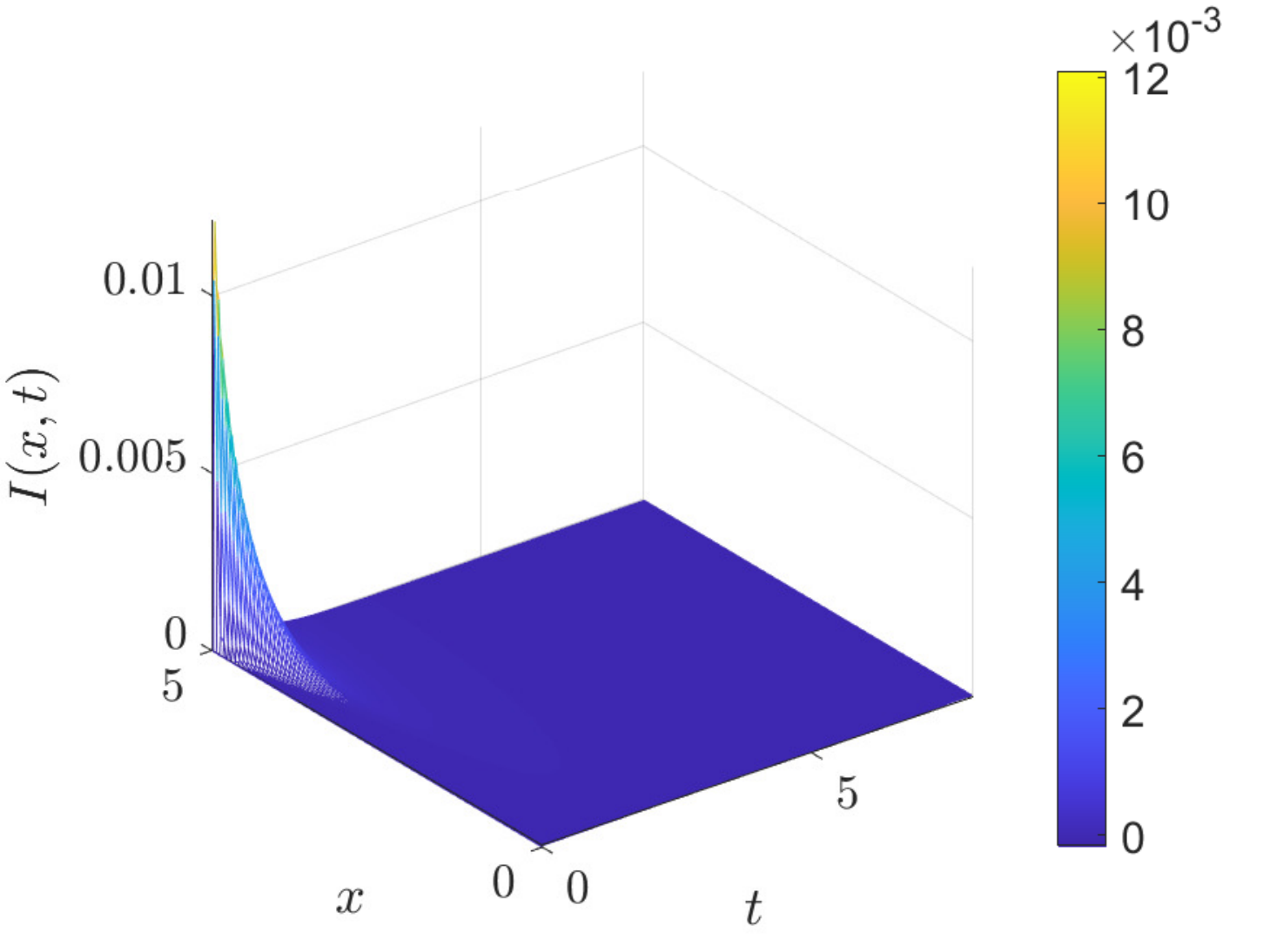}
\caption{($a_1$ - $I$)}
\label{fig.Network1_arc1_new_I}
\end{subfigure}
\begin{subfigure}{0.32\textwidth}
\includegraphics[width=1\linewidth]{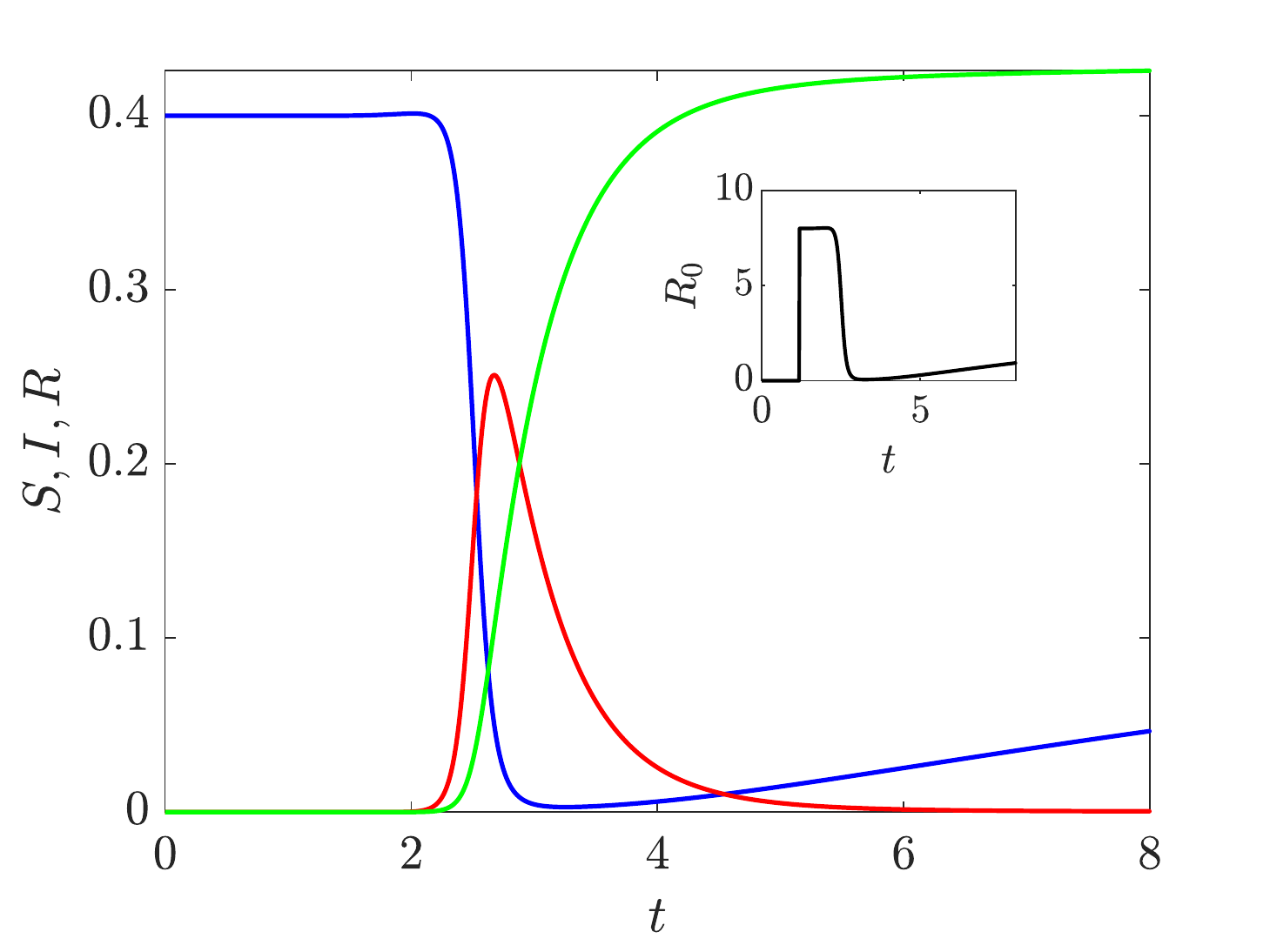}
\caption{($n_1$)}
\label{fig.Network1_node1_new}
\end{subfigure}
\begin{subfigure}{0.32\textwidth}
\includegraphics[width=1\linewidth]{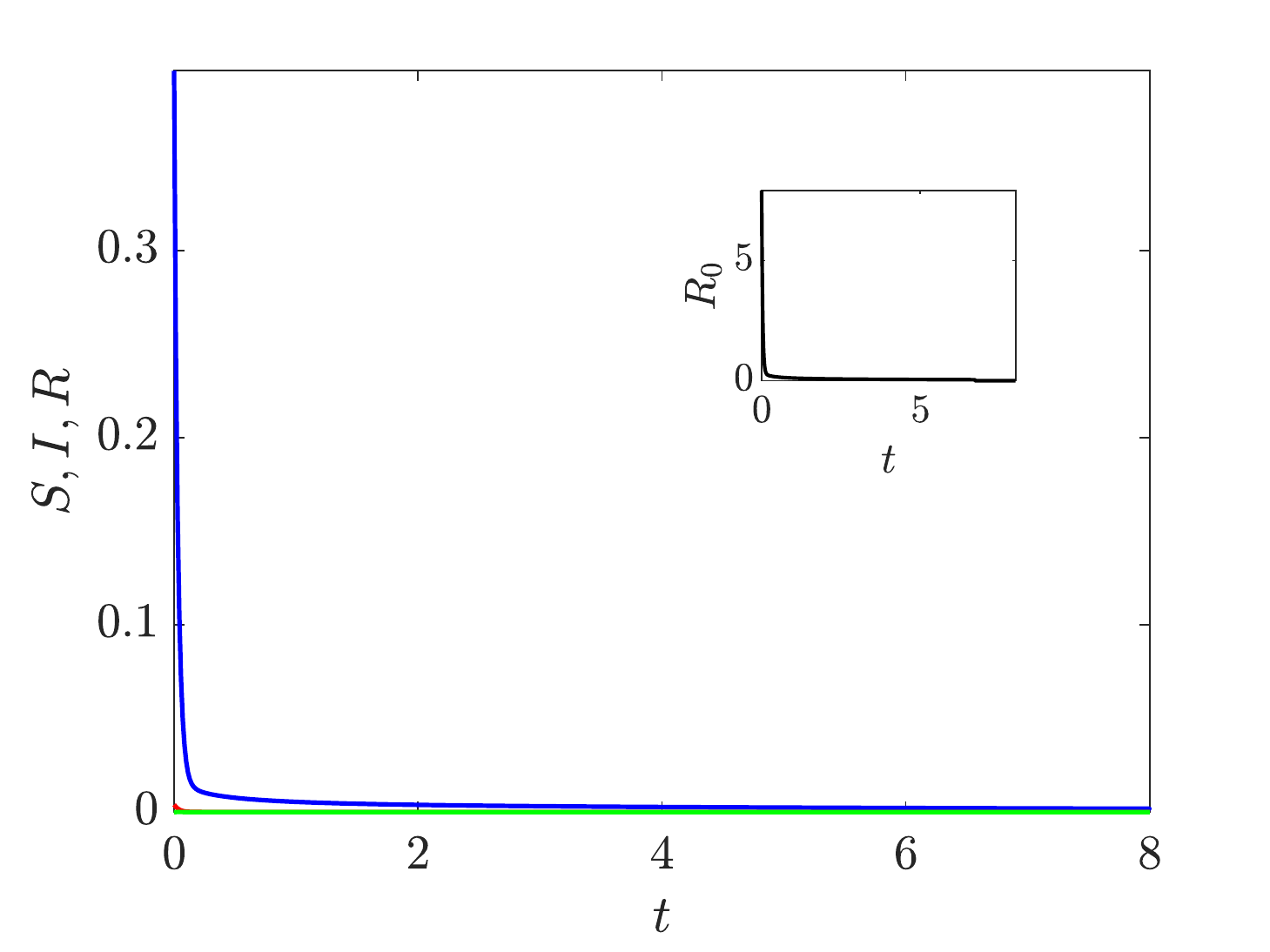}
\caption{($n_2$)}
\label{fig.Network1_node2_new}
\end{subfigure}
\begin{subfigure}{0.32\textwidth}
\includegraphics[width=1\linewidth]{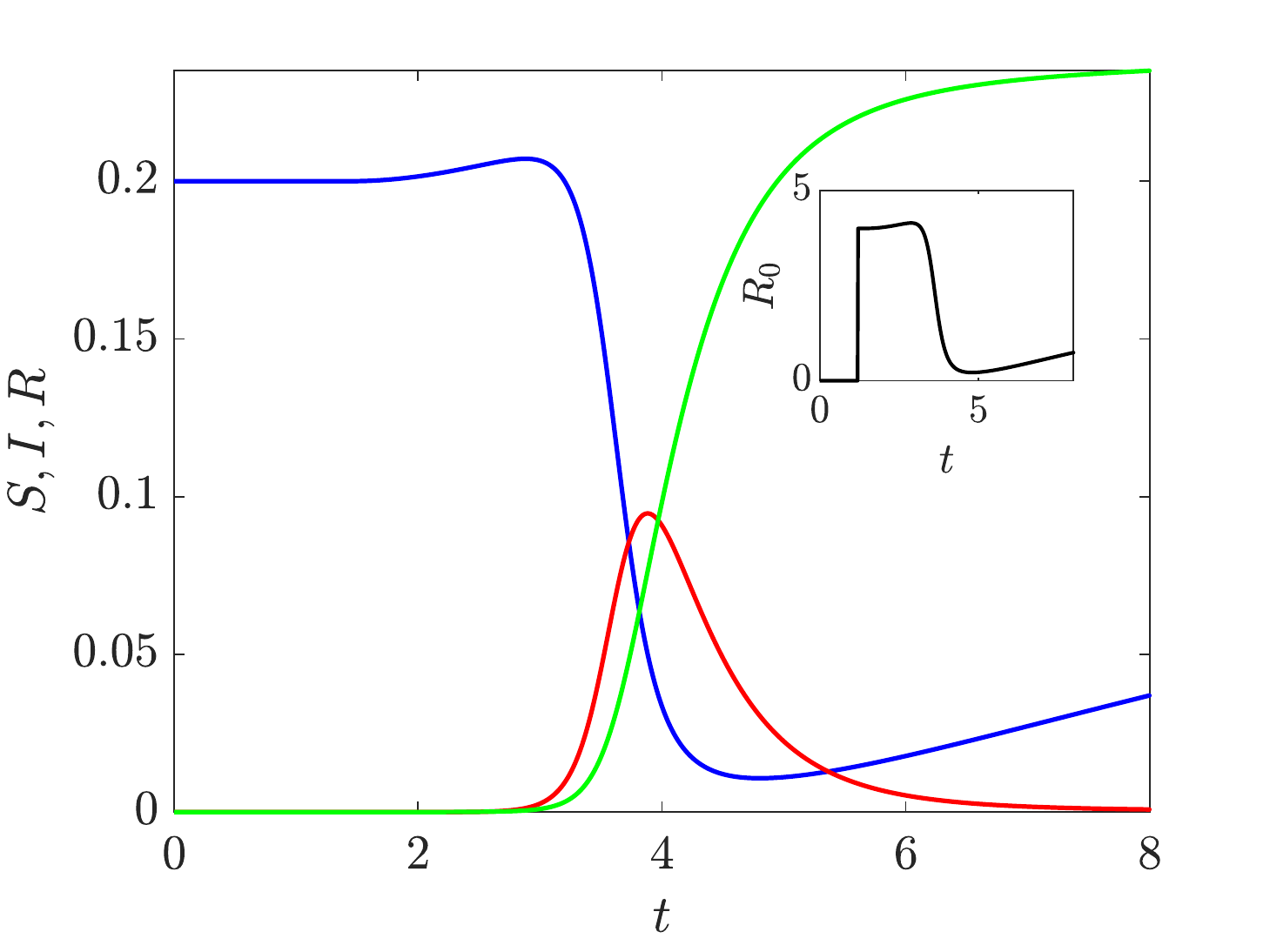}
\caption{($n_3$)}
\label{fig.Network1_node3_new}
\end{subfigure}
\caption{Numerical results of Test 1b, in which there is a symmetric temporal and spatial dynamics of susceptible $S$, infectious $I$ and recovered $R$ individuals along arcs $a_1$ and $a_2$. Time evolution of $S$ (blue), $I$ (red) and $R$ (green) at node $n_1$, node $n_2$ (from which the infectious disease starts spreading symmetrically along the two arcs) and at node $n_3$. In the same plots, the evolution of the coefficient $R_0$ is also shown.}
\label{fig.Network1_new}
\end{figure}
\begin{figure}[t!]
\centering
\begin{subfigure}{0.32\textwidth}
\includegraphics[width=1\linewidth]{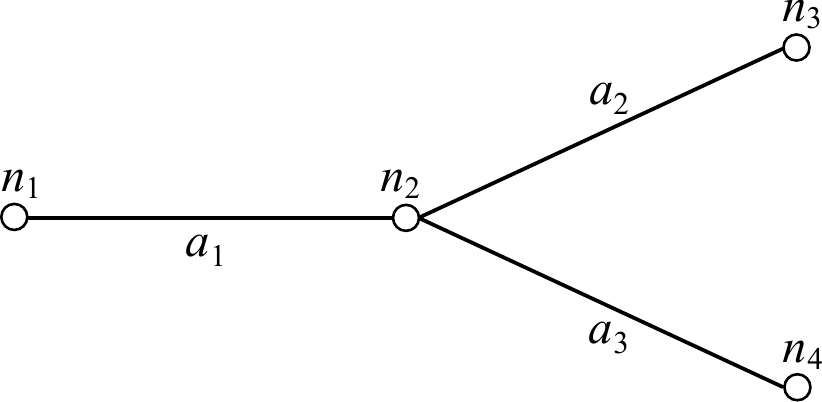}
\caption{}
\label{fig.networkY_scheme}
\end{subfigure}
\begin{subfigure}{0.32\textwidth}
\includegraphics[width=1\linewidth]{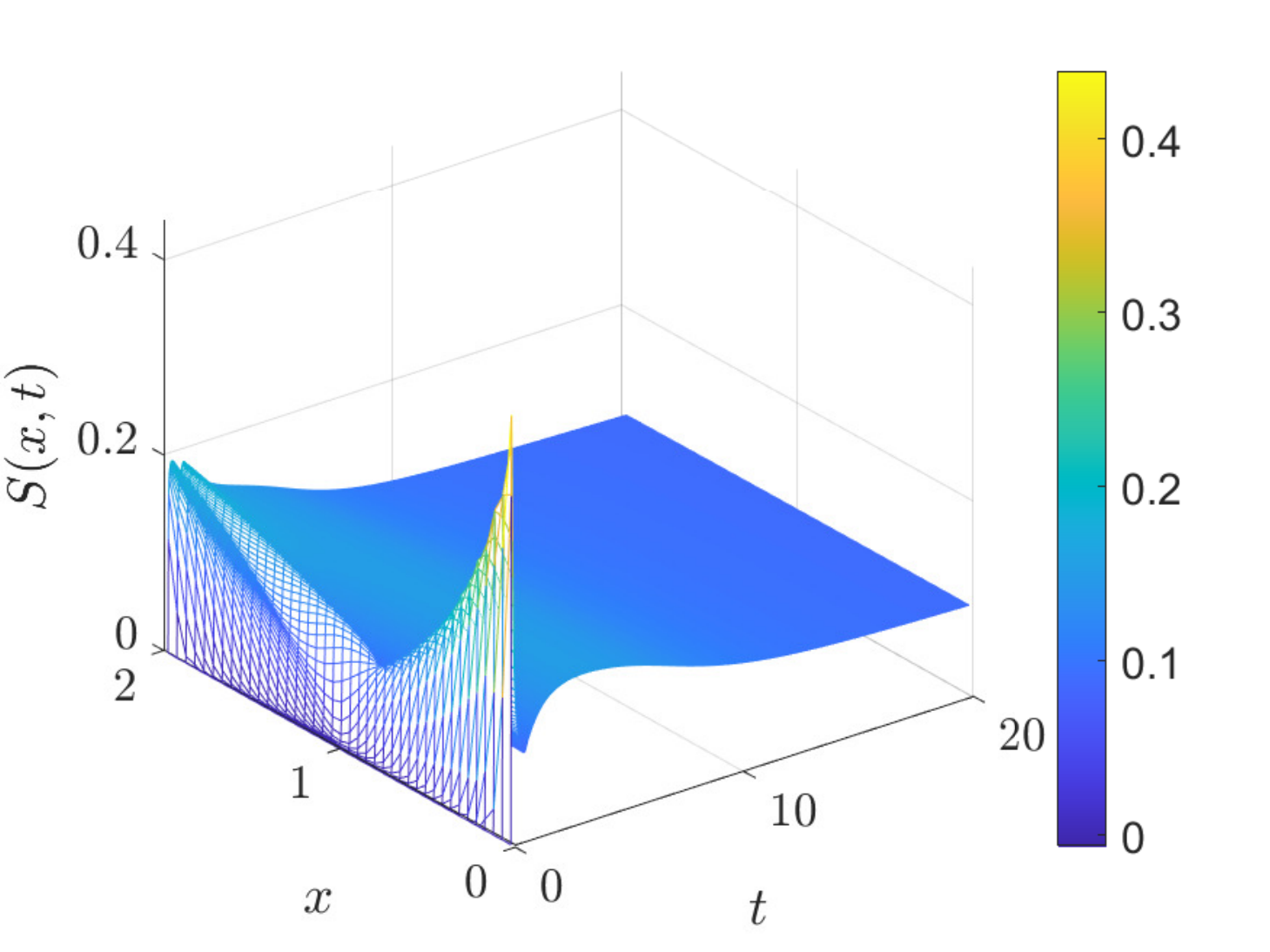}
\caption{($a_1$ - $S$)}
\label{fig.NetworkY_arc1_xt_S}
\end{subfigure}
\begin{subfigure}{0.32\textwidth}
\includegraphics[width=1\linewidth]{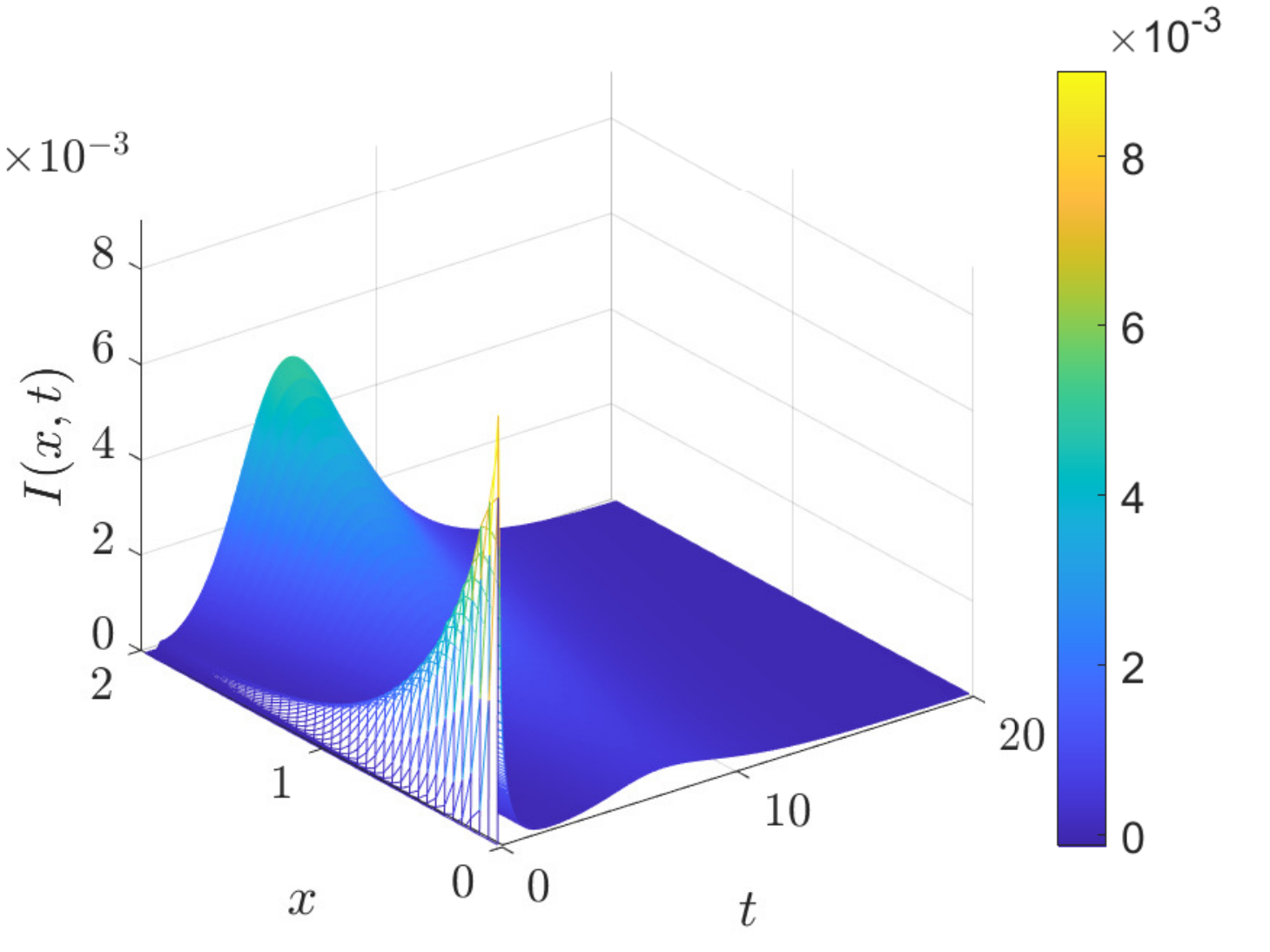}
\caption{($a_1$ - $I$)}
\label{fig.NetworkY_arc1_xt_I}
\end{subfigure}
\begin{subfigure}{0.32\textwidth}
\includegraphics[width=1\linewidth]{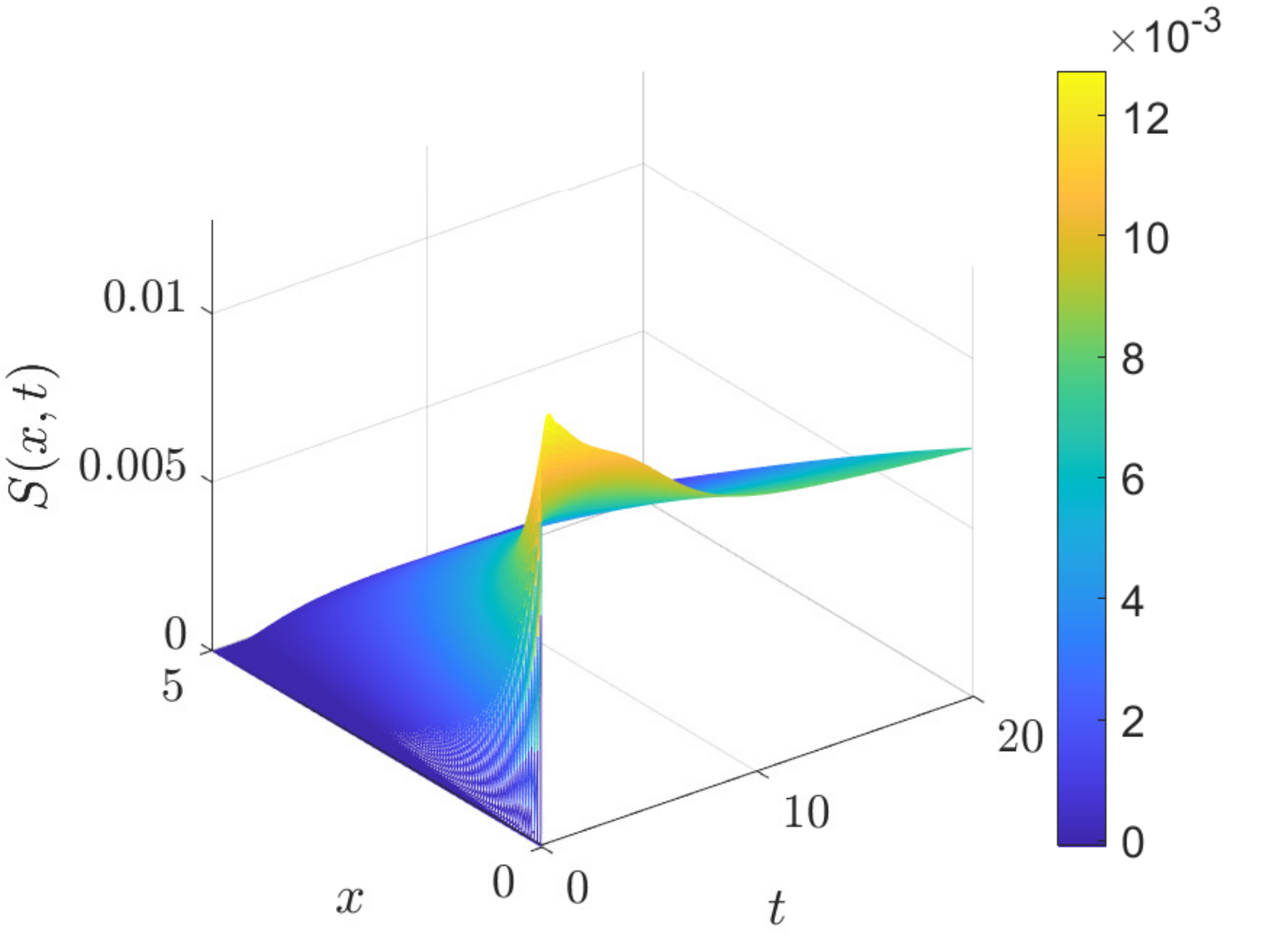}
\caption{($a_2$ - $S$)}
\label{fig.NetworkY_arc2_xt_S}
\end{subfigure}
\begin{subfigure}{0.32\textwidth}
\includegraphics[width=1\linewidth]{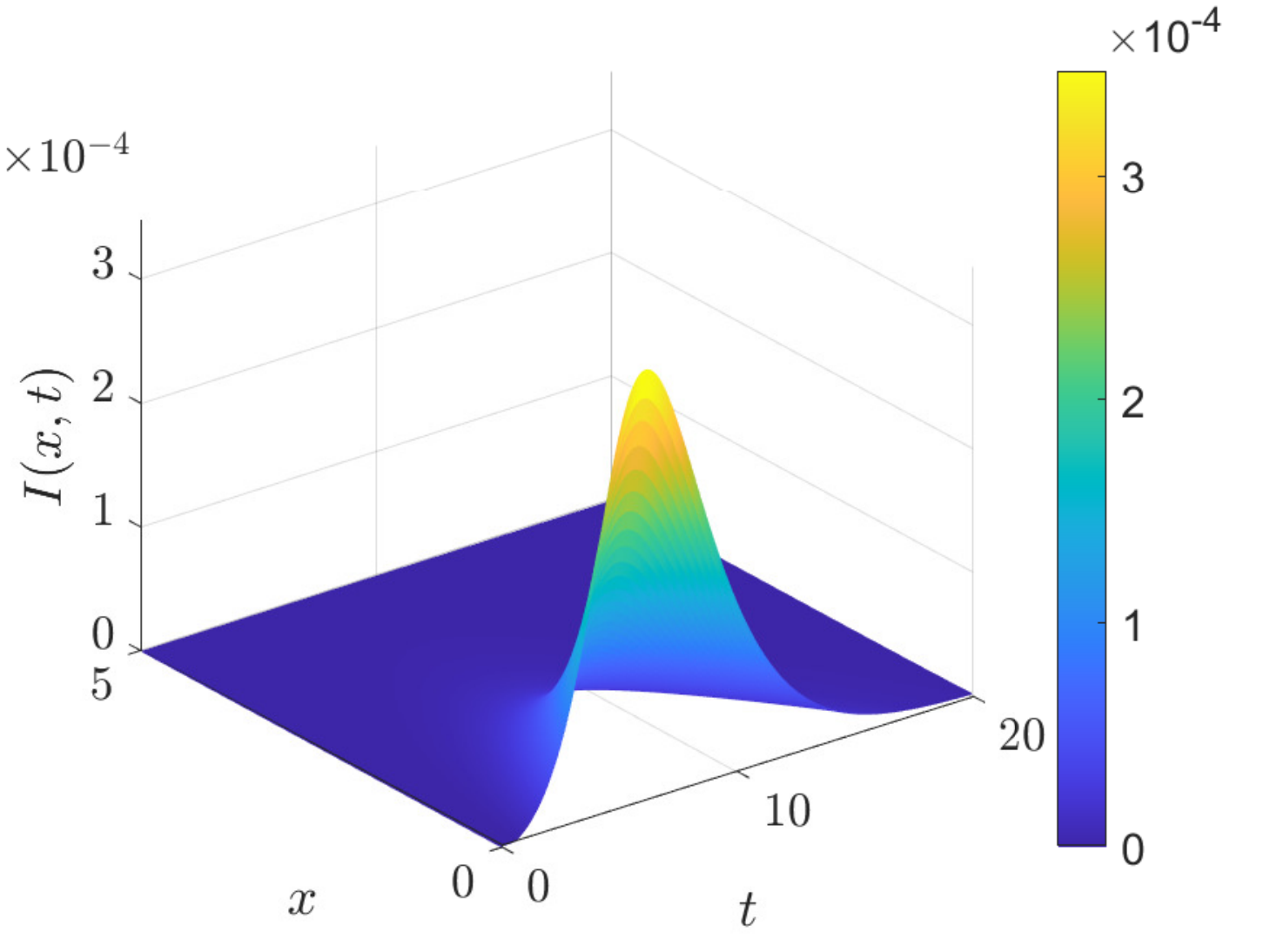}
\caption{($a_2$ - $I$)}
\label{fig.NetworkY_arc2_xt_I}
\end{subfigure}
\begin{subfigure}{0.32\textwidth}
\includegraphics[width=1\linewidth]{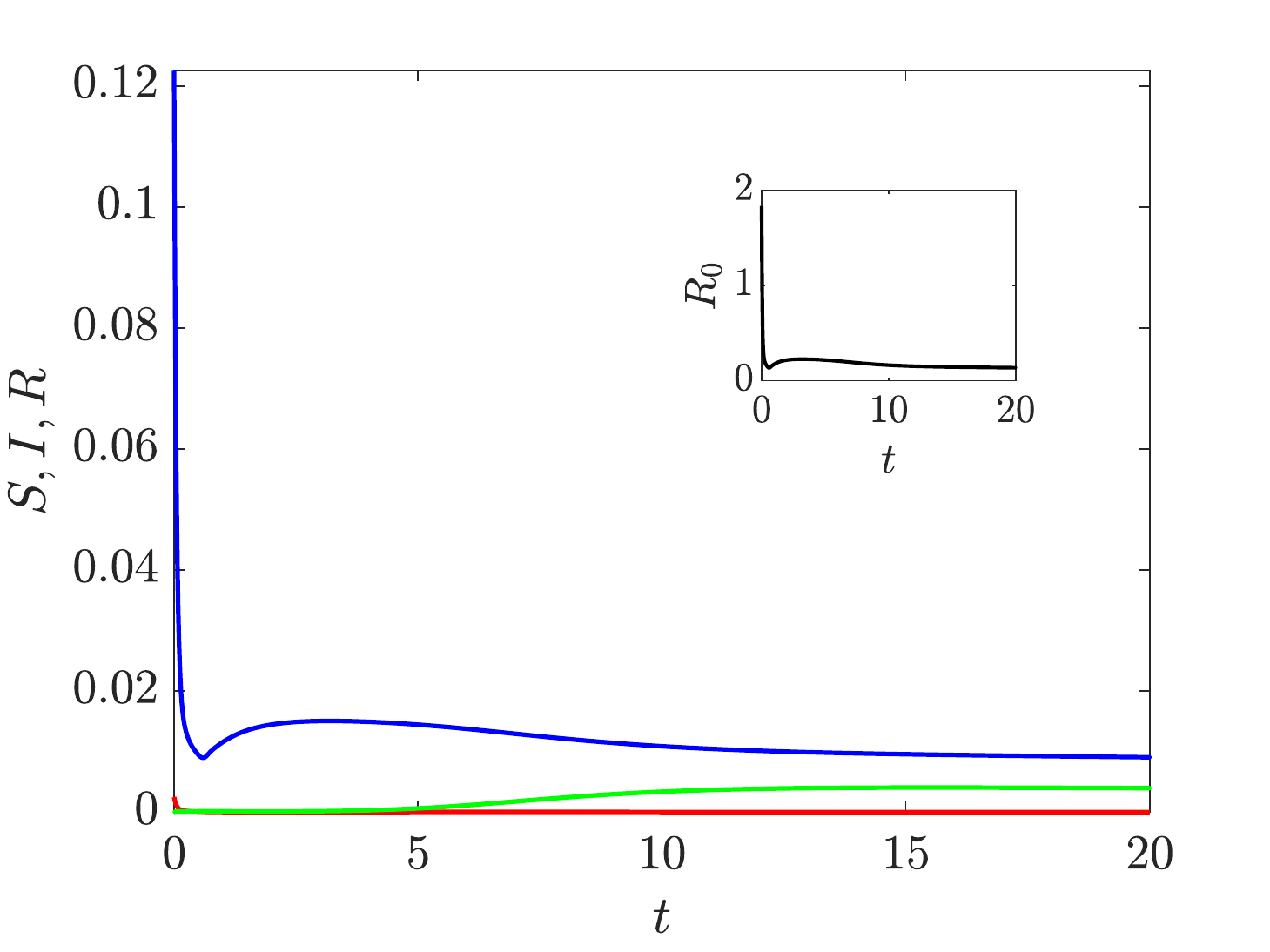}
\caption{($n_1$)}
\label{fig.NetworkY_node1}
\end{subfigure}
\begin{subfigure}{0.32\textwidth}
\includegraphics[width=1\linewidth]{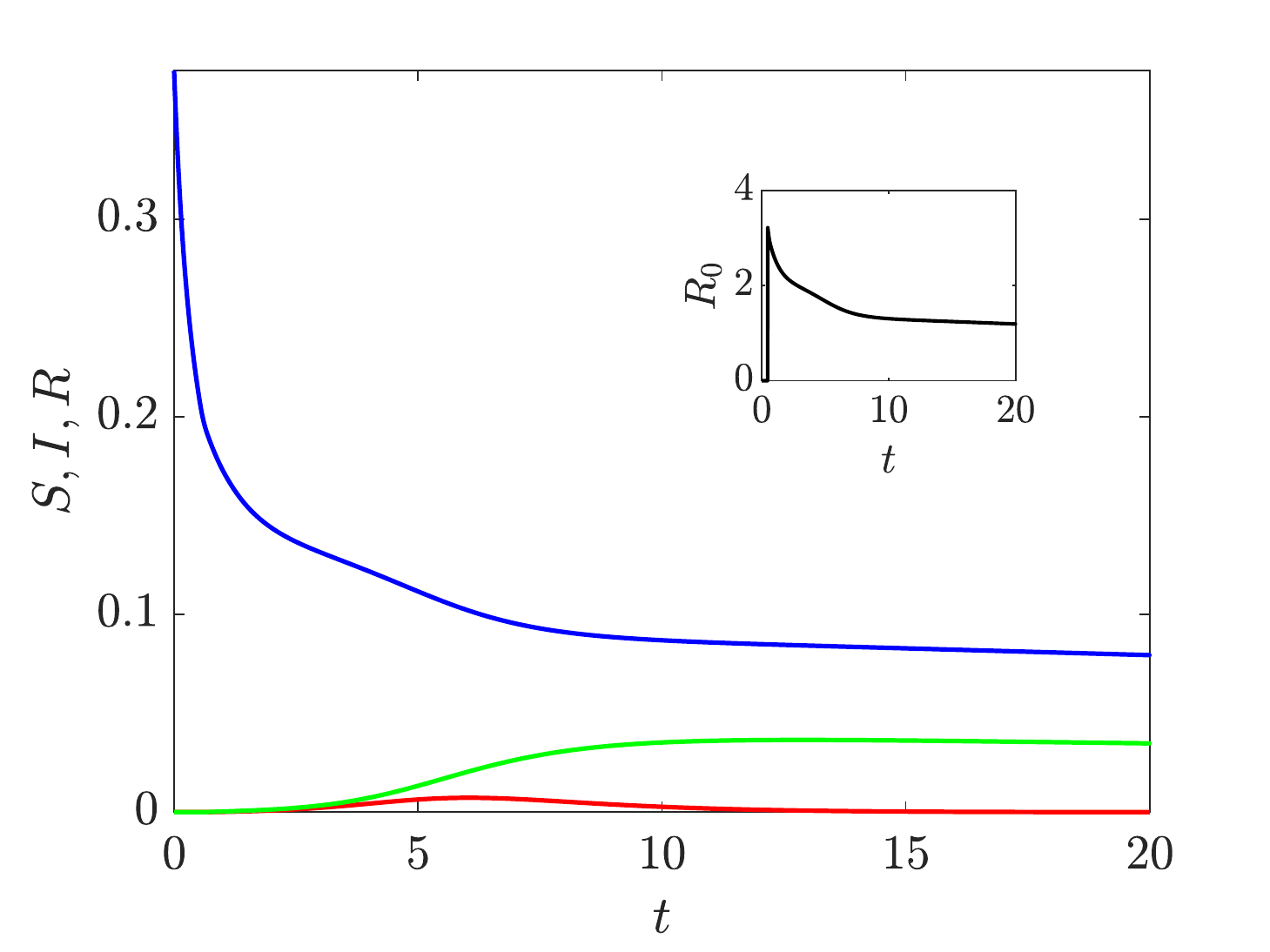}
\caption{($n_2$)}
\label{fig.NetworkY_node2}
\end{subfigure}
\begin{subfigure}{0.32\textwidth}
\includegraphics[width=1\linewidth]{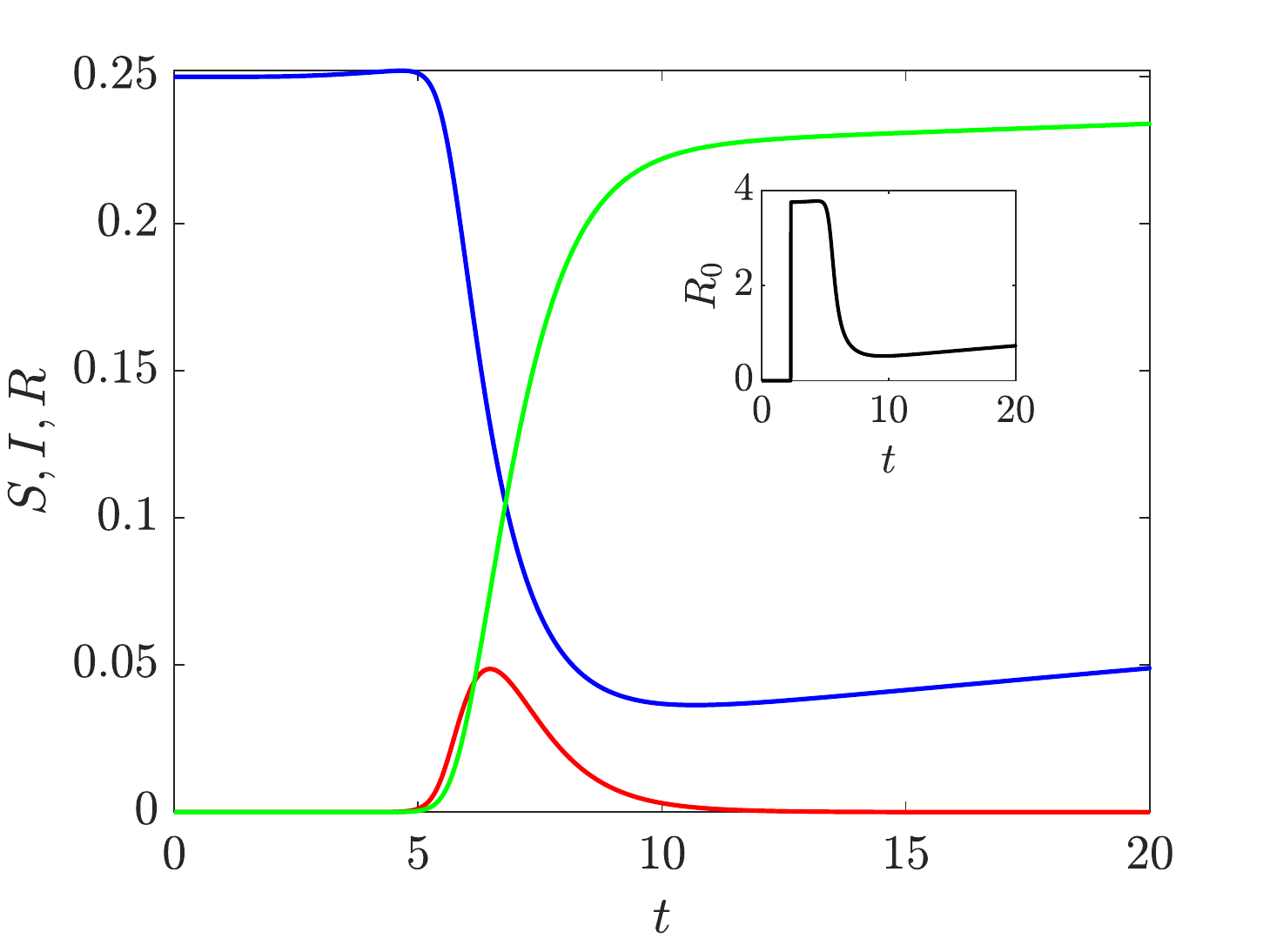}
\caption{($n_3$)}
\label{fig.NetworkY_node3}
\end{subfigure}
\begin{subfigure}{0.32\textwidth}
\includegraphics[width=1\linewidth]{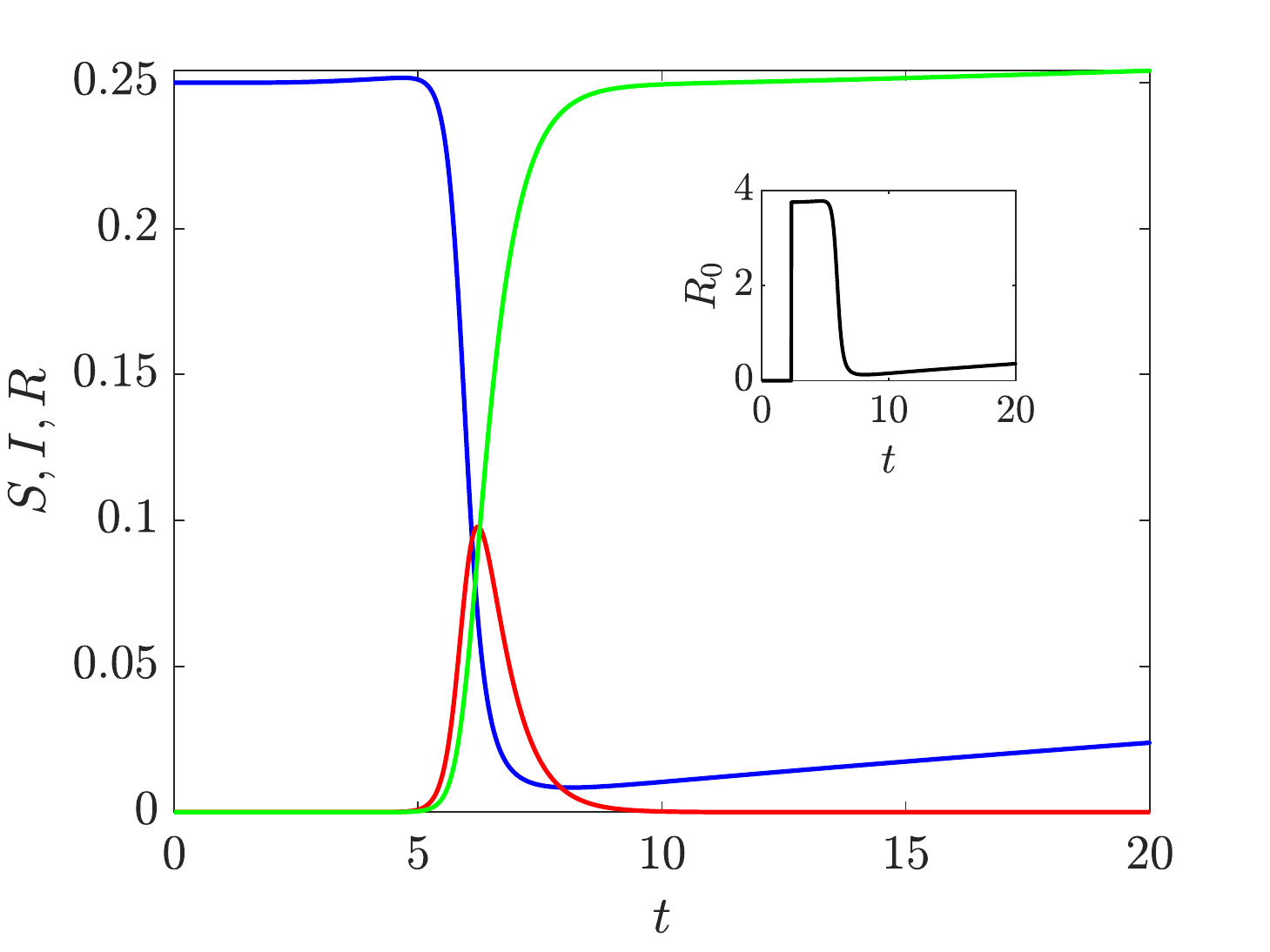}
\caption{($n_4$)}
\label{fig.NetworkY_node4}
\end{subfigure}
\caption{Numerical results of Test 2a, with time and spatial dynamics of susceptible $S$ and infectious $I$ individuals along arcs $a_1$ and $a_2$ that are initially without population. Results concerning arc $a_3$ are omitted because very similar to those of $a_2$. Time evolution of $S$ (blue), $I$ (red) and $R$ (green) dynamics at node $n_1$ (small city), from which the infectious disease starts spreading, node $n_2$ (big city), node $n_3$ and $n_4$ (medium-size cities). In the same plots, the evolution of the coefficient $R_0$ is also shown.}
\label{fig.NetworkY}
\end{figure}
\begin{figure}[t!]
\centering
\begin{subfigure}{0.48\textwidth}
\includegraphics[width=1\linewidth]{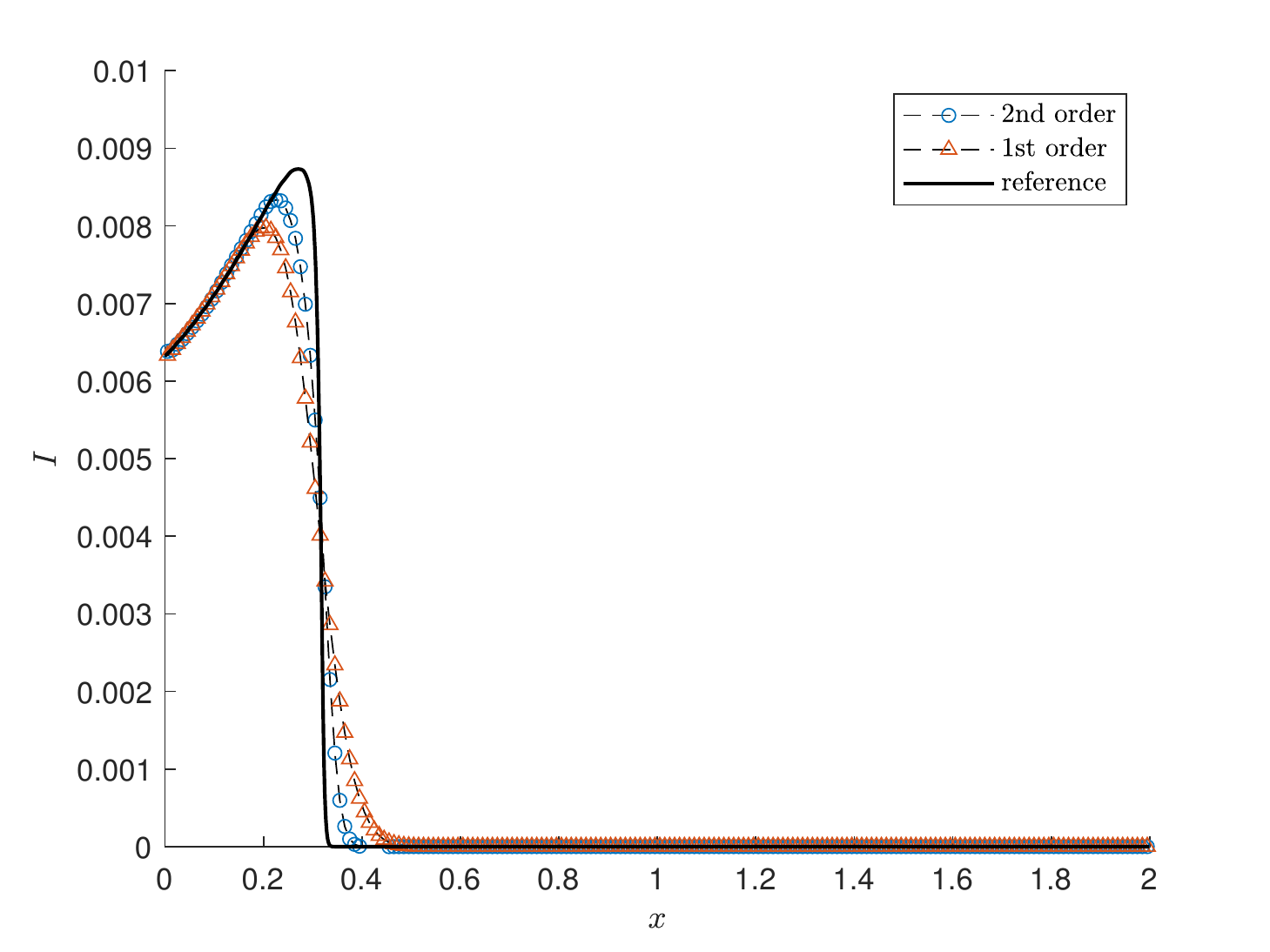}
\caption{($t = 0.1$)}
\label{fig.networkY_arc1_x_I_t1}
\end{subfigure}
\begin{subfigure}{0.48\textwidth}
\includegraphics[width=1\linewidth]{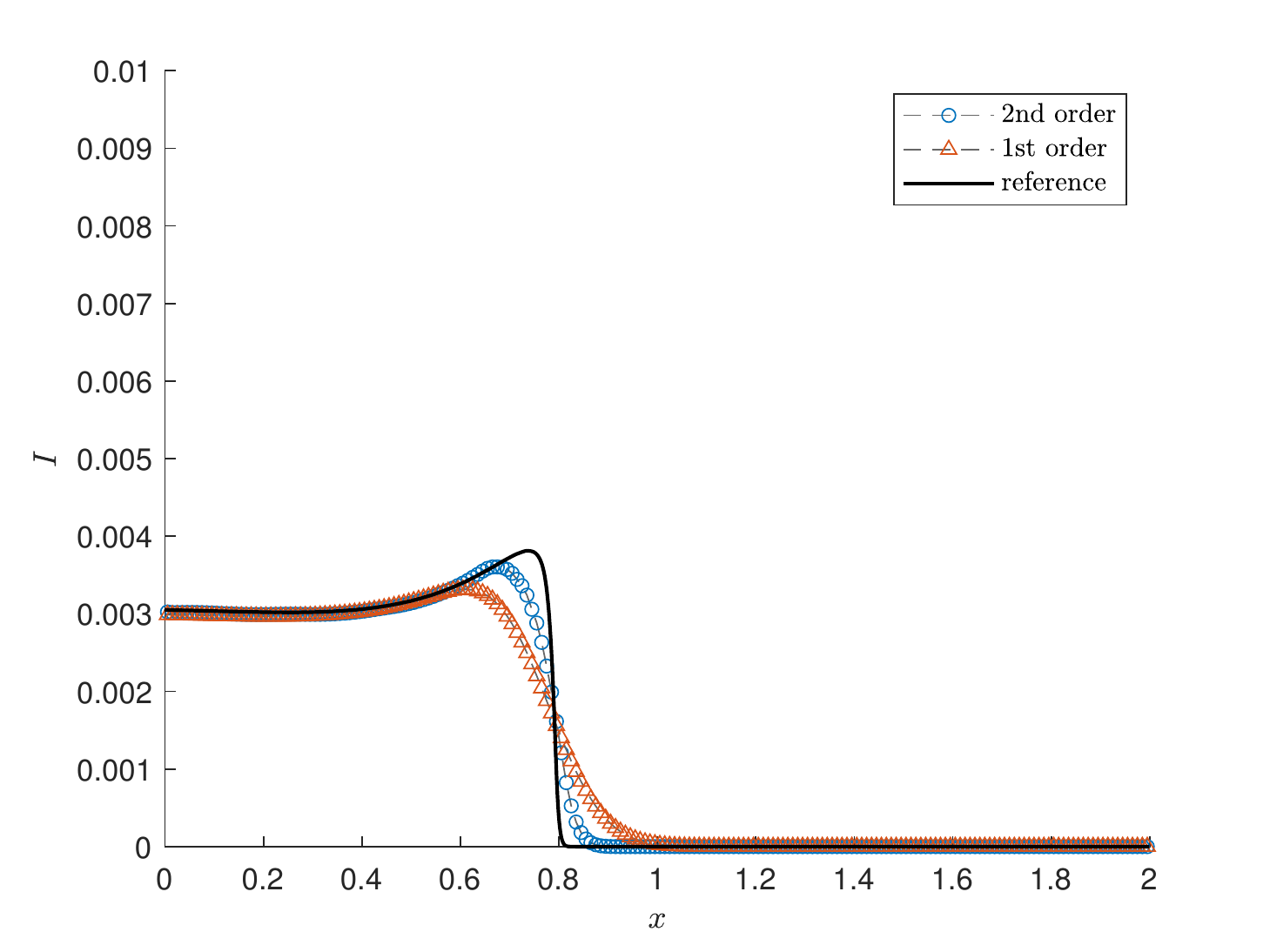}
\caption{($t = 0.25$)}
\label{fig.networkY_arc1_x_I_t2}
\end{subfigure}
\caption{Spatial dynamics, at two different time steps, of infectious individuals along arc $a_1$ in Test 2a (see Fig. \ref{fig.NetworkY}), with the spread of the disease starting from node $n_1$. Results in solid line are the reference solution ($\Delta x = 0.00125$); results in dashed line with circles are those obtained with the second-order IMEX scheme here proposed ($\Delta x = 0.01$); results in dashed line with triangles are those obtained with the corresponding first-order scheme ($\Delta x = 0.01$).}
\label{fig.NetworkY_arc1_x}
\end{figure}
\begin{figure}[t!]
\centering
\begin{subfigure}{0.32\textwidth}
\includegraphics[width=1\linewidth]{networkY_scheme}
\caption{}
\label{fig.networkY_scheme_v1}
\end{subfigure}
\begin{subfigure}{0.32\textwidth}
\includegraphics[width=1\linewidth]{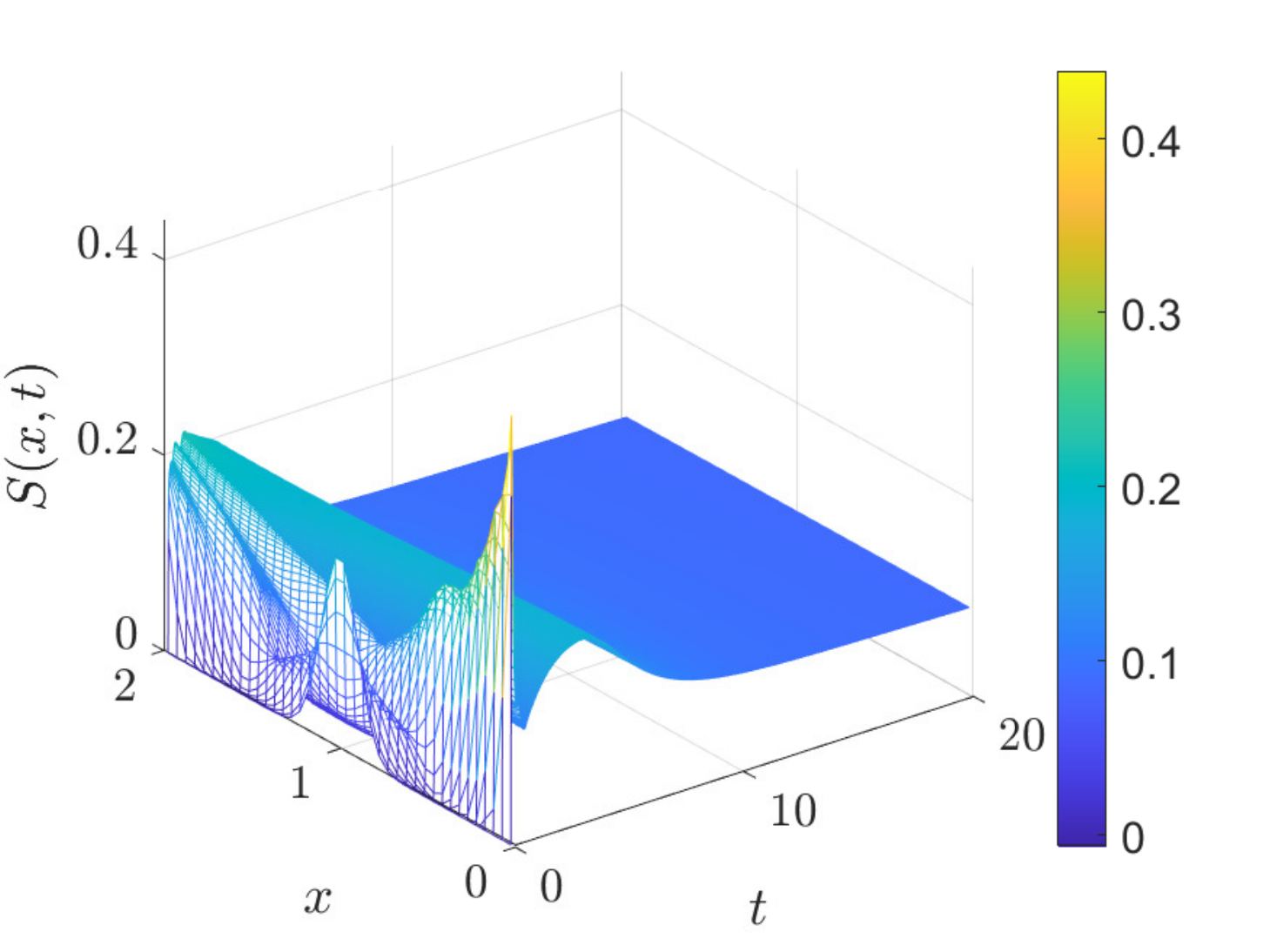}
\caption{($a_1$ - $S$)}
\label{fig.NetworkY_arc1_xt_S_v1}
\end{subfigure}
\begin{subfigure}{0.32\textwidth}
\includegraphics[width=1\linewidth]{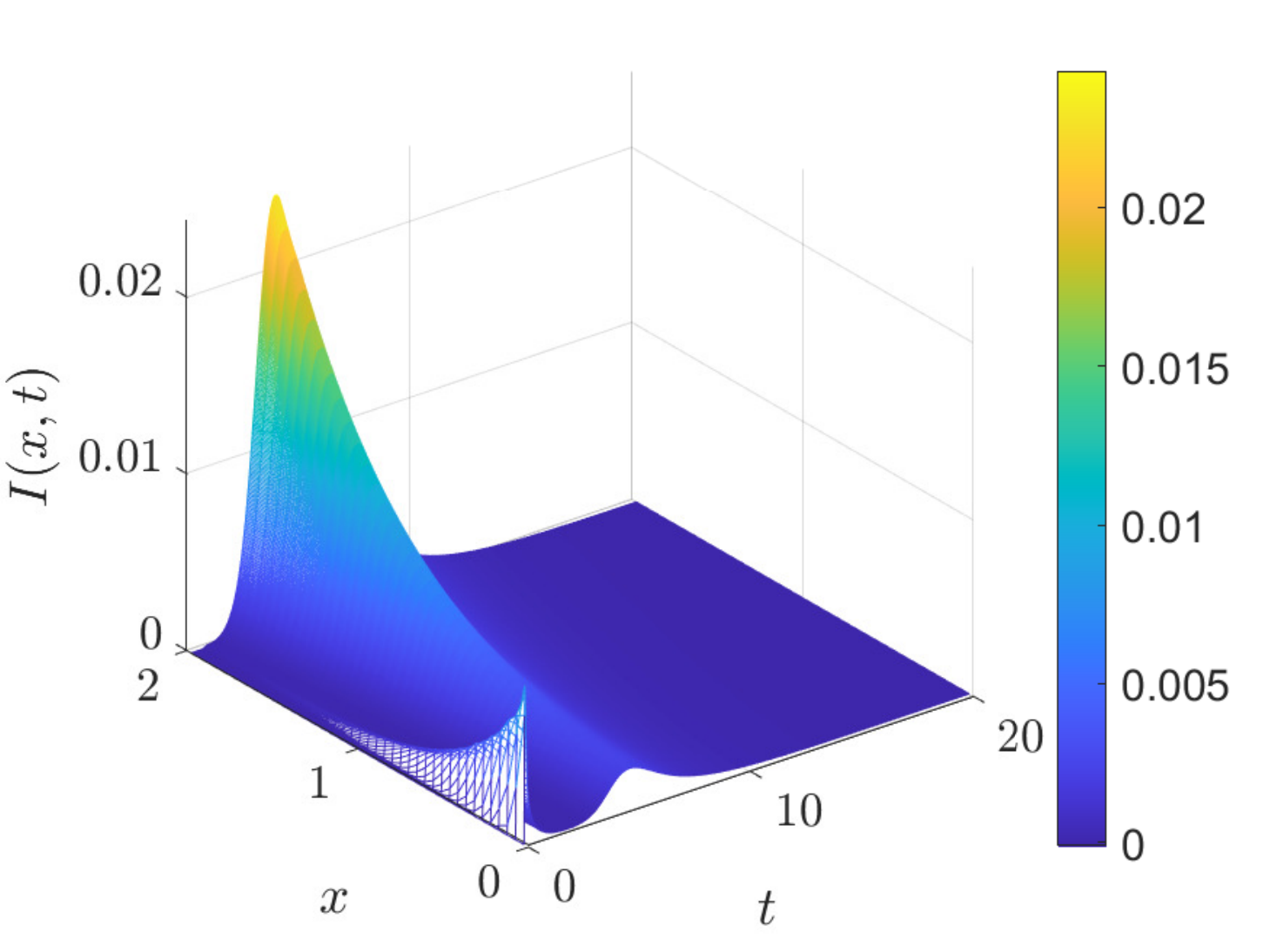}
\caption{($a_1$ - $I$)}
\label{fig.NetworkY_arc1_xt_I_v1}
\end{subfigure}
\begin{subfigure}{0.32\textwidth}
\includegraphics[width=1\linewidth]{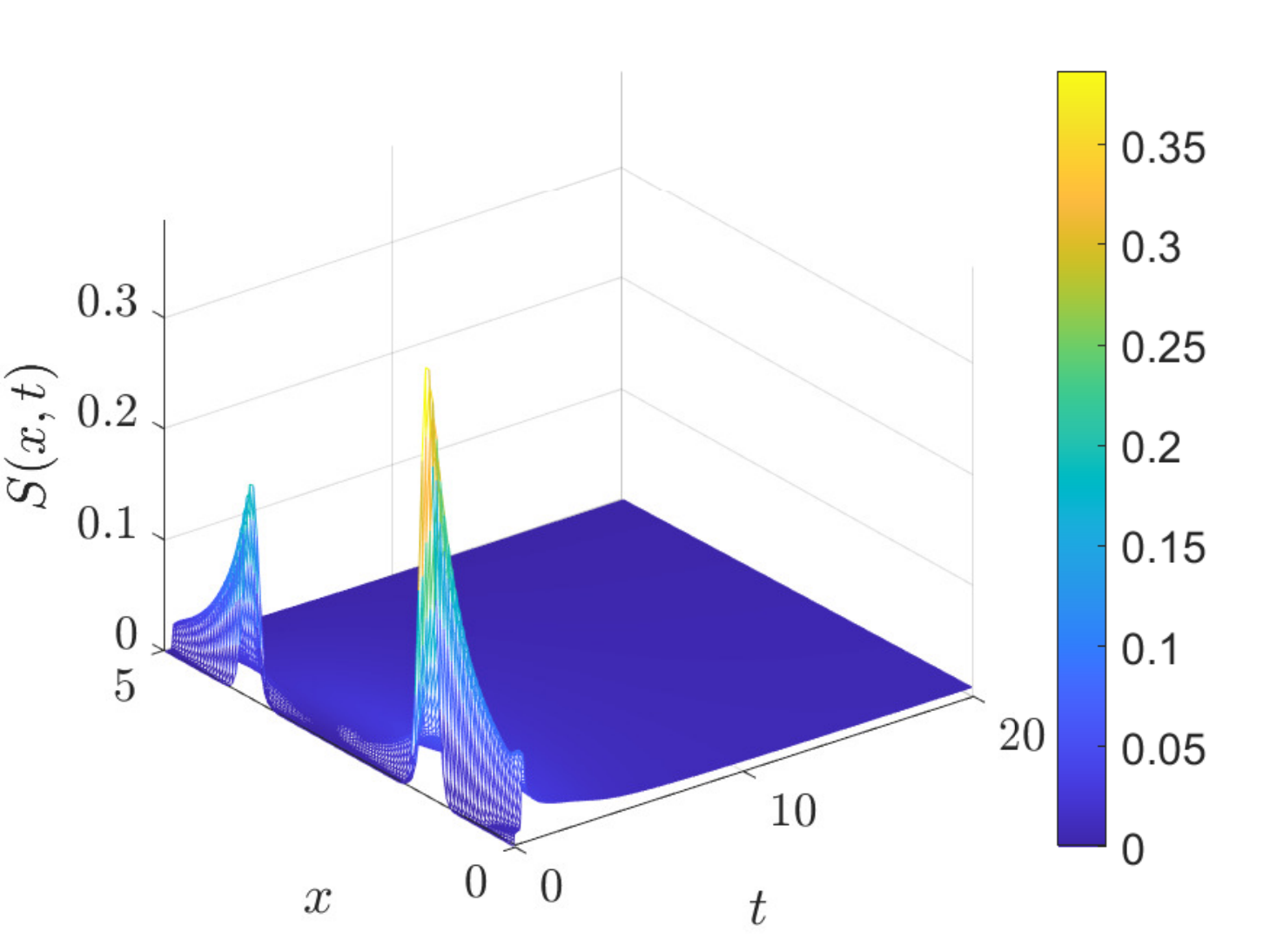}
\caption{($a_2$ - $S$)}
\label{fig.NetworkY_arc2_xt_S_v1}
\end{subfigure}
\begin{subfigure}{0.32\textwidth}
\includegraphics[width=1\linewidth]{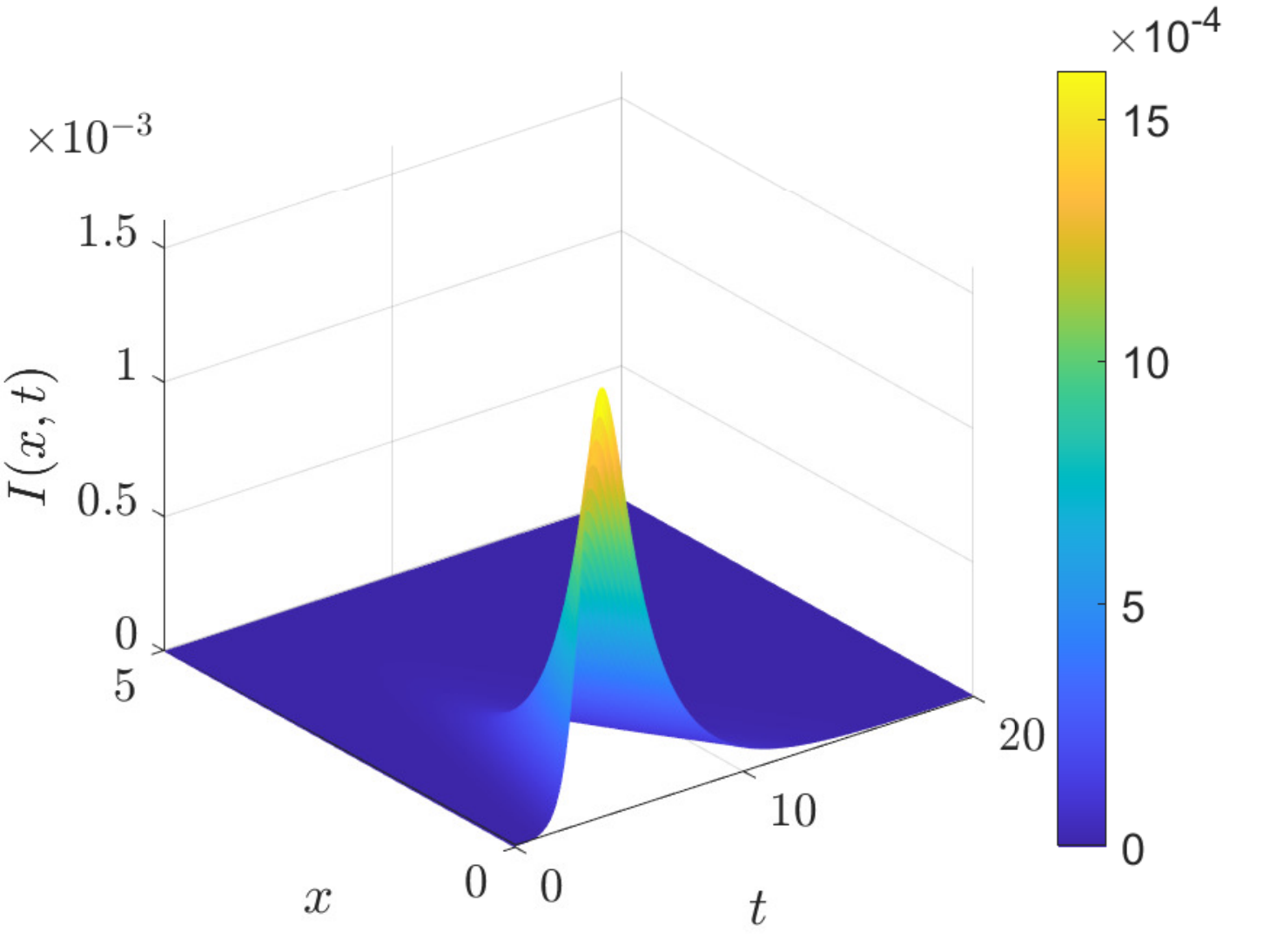}
\caption{($a_2$ - $I$)}
\label{fig.NetworkY_arc2_xt_I_v1}
\end{subfigure}
\begin{subfigure}{0.32\textwidth}
\includegraphics[width=1\linewidth]{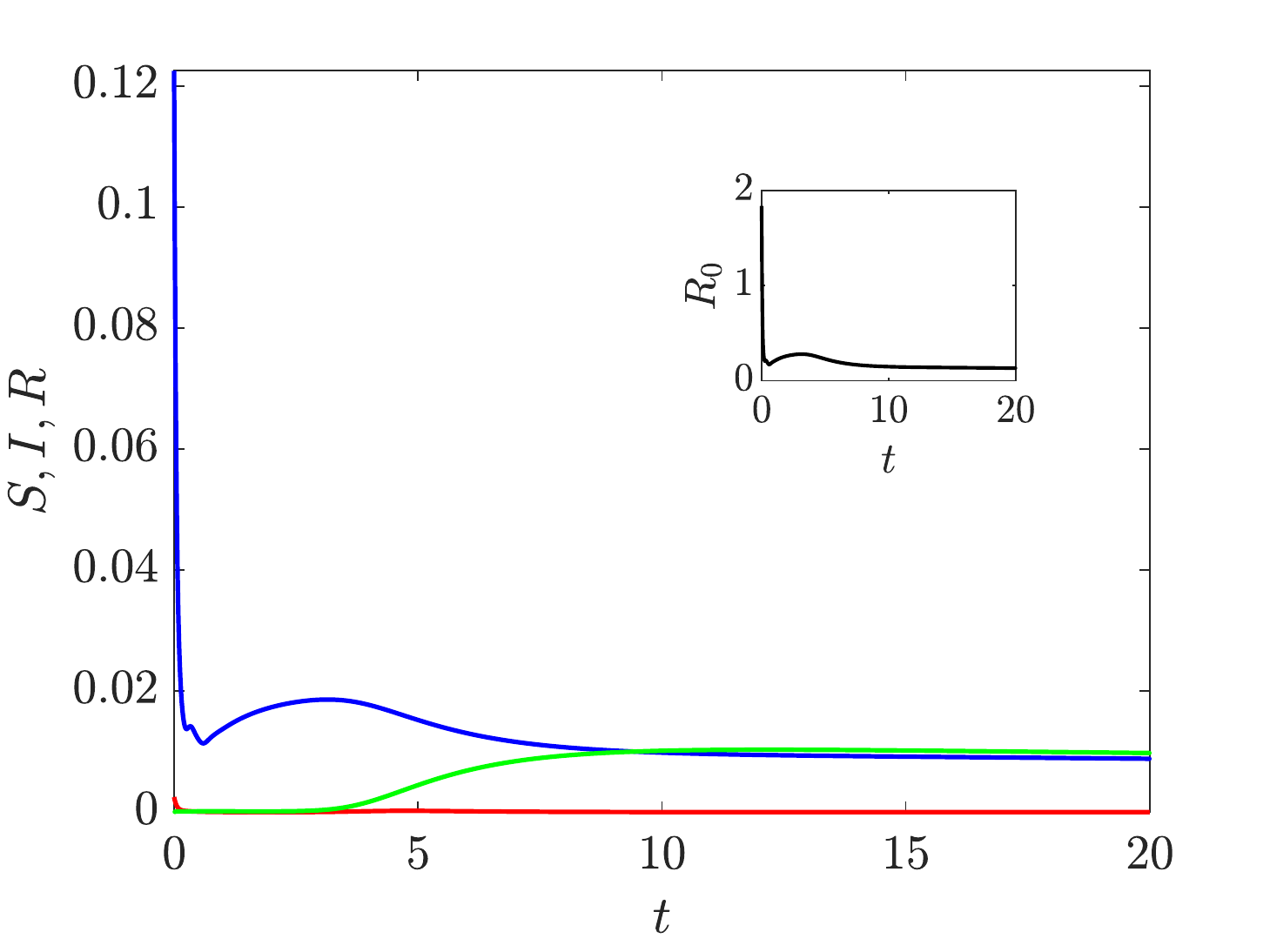}
\caption{($n_1$)}
\label{fig.NetworkY_node1_v1}
\end{subfigure}
\begin{subfigure}{0.32\textwidth}
\includegraphics[width=1\linewidth]{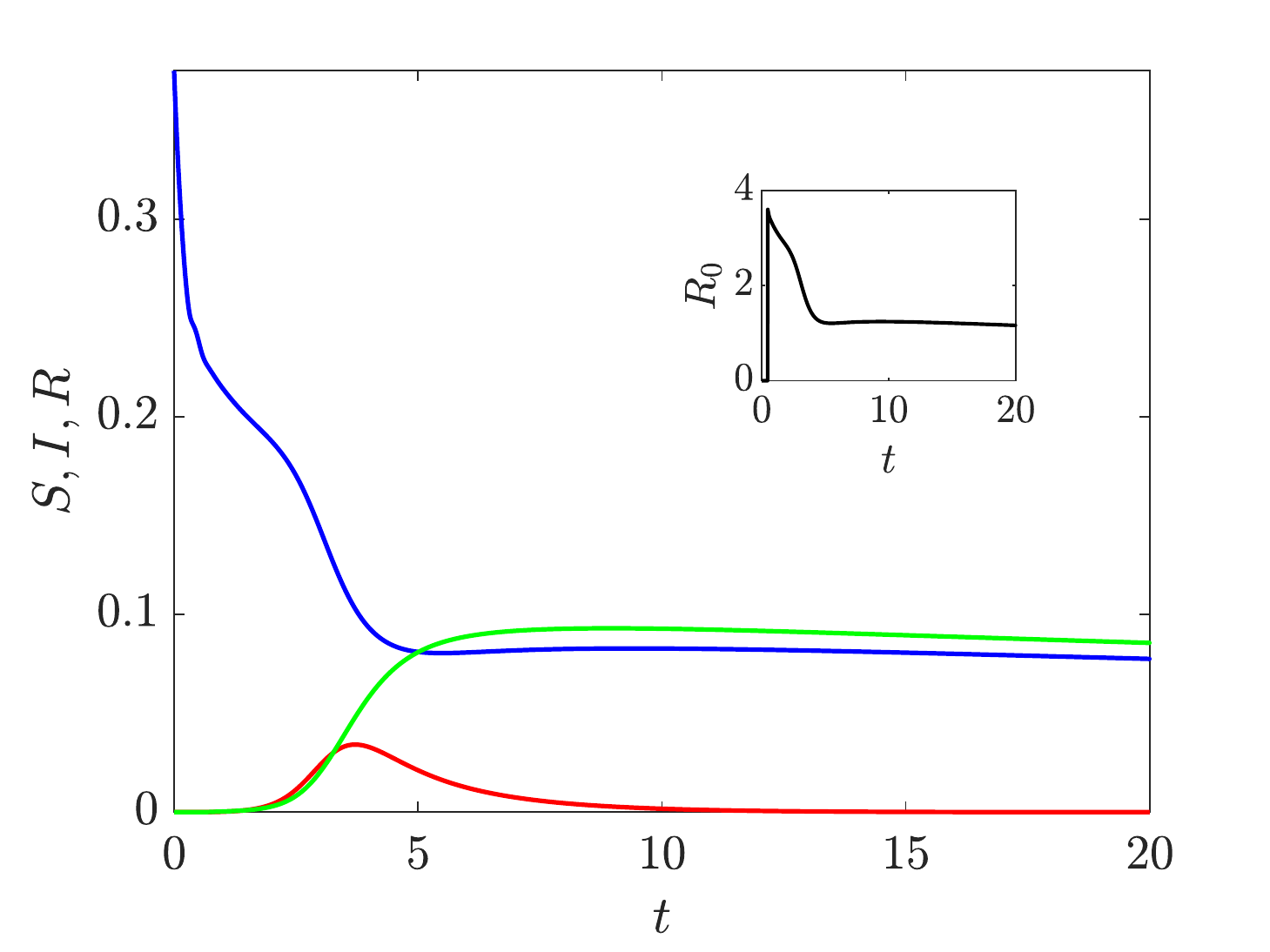}
\caption{($n_2$)}
\label{fig.NetworkY_node2_v1}
\end{subfigure}
\begin{subfigure}{0.32\textwidth}
\includegraphics[width=1\linewidth]{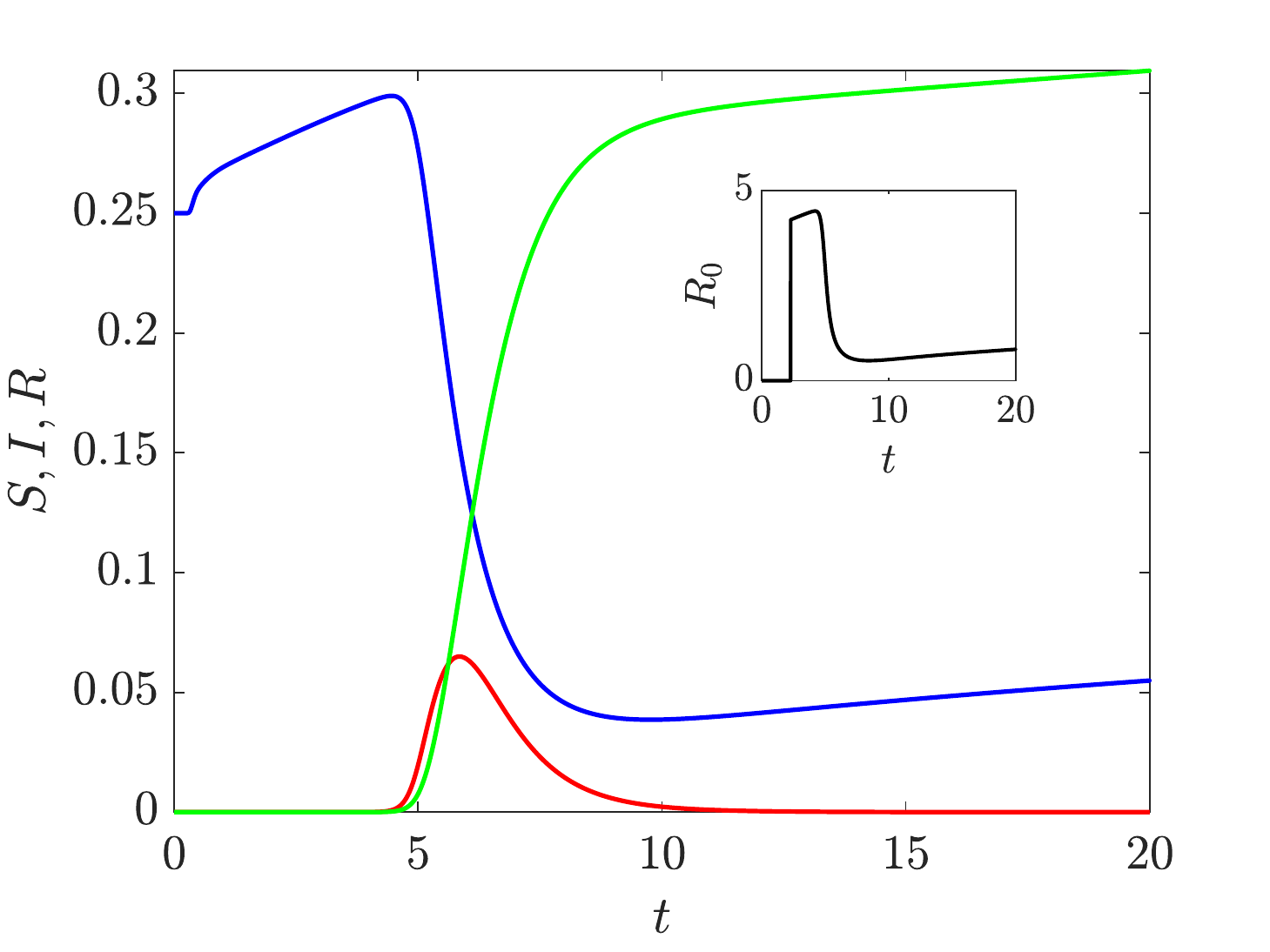}
\caption{($n_3$)}
\label{fig.NetworkY_node3_v1}
\end{subfigure}
\begin{subfigure}{0.32\textwidth}
\includegraphics[width=1\linewidth]{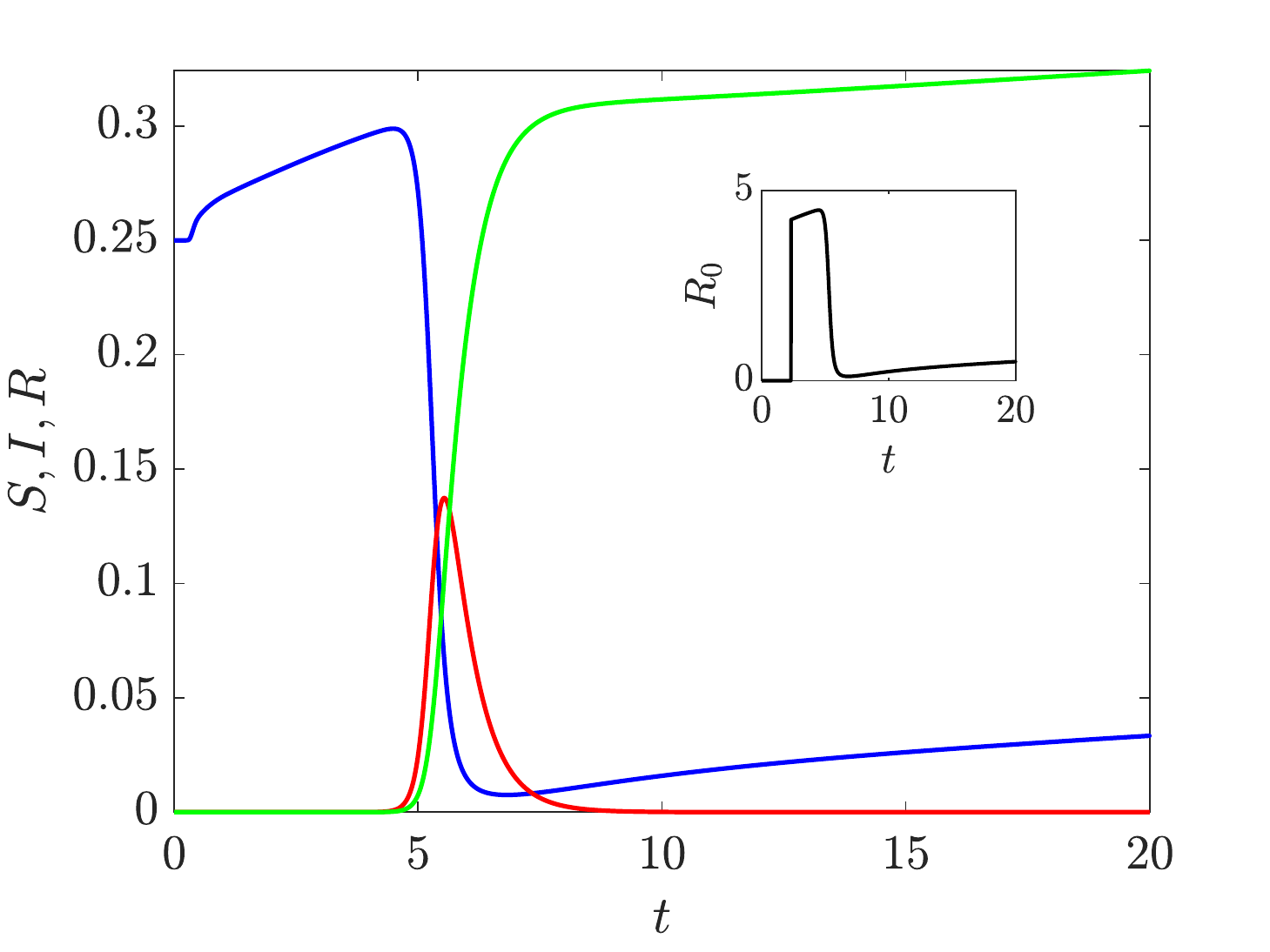}
\caption{($n_4$)}
\label{fig.NetworkY_node4_v1}
\end{subfigure}
\caption{Numerical results of Test 2b, with time and spatial dynamics of susceptible $S$ and infectious $I$ individuals along arcs $a_1$ and $a_2$ that have an initial amount of population (Gaussian distributed). Results concerning arc $a_3$ are omitted because very similar to those of $a_2$. Time evolution of $S$ (blue), $I$ (red) and $R$ (green) dynamics at node $n_1$ (small city), from which the infectious disease starts spreading, node $n_2$ (big city), node $n_3$ and $n_4$ (medium-size cities). In the same plots, the evolution of the coefficient $R_0$ is also shown.}
\label{fig.NetworkY_v1}
\end{figure}
\subsection{Network cases}
\label{section_network_tests}
To assess the effects of the mobility of individuals on networks with respect to the spread of an infectious disease, two tests with a total of five scenarios are performed concerning different simple networks. 
\subsubsection{A three-node network}
In Test 1, we consider the spread of an epidemic, characterized by $\beta = 2.0$ and $\gamma = 2.0$, in a network composed by 3 nodes connected by 2 bidirectional arcs in series, as shown in Fig.~\ref{fig.Network1}, having length $L_1 = L_2 = 5$ and discretized with a grid size $\Delta x = 0.05$. No social distancing or control effects are taken into account, fixing $k=0$. With this network we want to simulate the spread of an epidemic that starts from node $n_2$ (which represents a big city) and reaches nodes $n_1$ (a city as big as the one represented by $n_2$) and $n_3$ (which identifies a rather small city). For this reason, initial conditions are null for each variable in the two arcs, while at nodes we fix
\begin{align*}
&S(n_1,0) = 0.400, & I(n_1,0)&= 0.0, & &R(n_1,0) = 0.0,\\
&S(n_2,0) = 0.396, & I(n_2,0)&= 0.004, & &R(n_2,0) = 0.0,\\
&S(n_3,0) = 0.200, & I(n_3,0)&= 0.0, & &R(n_3,0) = 0.0 ,
\end{align*}
with zero initial fluxes for all the compartments at each location. Transmission conditions at each arc-node interface, satisfying conditions \eqref{eq.conservationFluxes_L} and \eqref{eq.conservationFluxes_0}, are given in Table \ref{tab:network1}. It can be noticed that a symmetric transmission of the infectious on arcs $a_1$ and $a_2$ from $n_2$ is imposed.\par
\begin{table}[b!] 
\caption{Transmission coefficients used in Test 1 (Fig.~\ref{fig.Network1}), given for each node-arc interface $\mathcal{I}$.} \label{tab:network1} 
\centering
\begin{tabular}{ l | c c }
\hline
\multirow{2}*{$\mathcal{I}(n_1, a_1)$} 	&$\alpha_{n_1,a_1} =1.00$  &$\alpha_{n_1,n_1} =1.00$\\
											&$\alpha_{a_1,a_1} =0.00$  &$\alpha_{a_1,n_1} =0.00$ \\
\hline
\multirow{2}*{$\mathcal{I}(a_1, n_2)$}	&$\alpha_{a_1,n_2} =0.50$  &$\alpha_{a_1,a_1} =0.50$ \\
											&$\alpha_{n_2,n_2} =0.50$  &$\alpha_{n_2,a_1} =0.50$ \\
\hline 
\multirow{2}*{$\mathcal{I}(n_2, a_2)$}	&$\alpha_{n_2,a_2} =0.50$  &$\alpha_{n_2,n_2} =0.50$\\
											&$\alpha_{a_2,a_2} =0.50$  &$\alpha_{a_2,n_2} =0.50$\\
\hline 
\multirow{2}*{$\mathcal{I}(a_2, n_3)$}	&$\alpha_{a_2,n_3} =0.00$  &$\alpha_{a_2,a_2} =0.00$\\
											&$\alpha_{n_3,n_3} =1.00$  &$\alpha_{n_3,a_2} =1.00$\\
\hline 
\end{tabular} 
\end{table} 
\smallskip
\paragraph{\it Test 1a}
The first scenario is run assuming that only the infectious individuals are moving along the network, with $\lambda_I^2 = 10$ and $\tau_I = 0.1$, while susceptible and recovered subjects do not leave the origin node (city). This choice, albeit unrealistic, allows us to observe in a cleaner way the effects of transport and spread of the infection related to the mobility of people. In Fig.~\ref{fig.Network1} we can notice a slight decay of susceptible individuals at node $n_2$ (from which the epidemic outbreak starts developing), because of the initial percentage of infectious people that is there at the beginning of the simulation and that, then, leaves the node very fast. In this way, a very small percentage of susceptible population of the node is infected and the majority of it remains unharmed at the equilibrium state. A different evolution of the epidemic is observed at nodes $n_1$ and $n_3$. Here it is clearly highlighted how the model associates a higher incidence function to the node with a larger population. In fact, node $n_1$ has twice as many susceptible subjects as node $n_3$, information that is reproduced by the model as an epidemic that at $n_1$ has an incidence double that of $n_3$ (even though infected subjects travel along arcs $a_1$ and $a_2$ in a total symmetry, as shown in Fig.~\ref{fig.Network1}). Indeed, the entire population at node $n_1$ is infected, while a very little percentage of susceptible at node $n_3$ remains uninfected, even if here individuals are the half of $n_1$. In addition, we can see that the epidemic peak itself is reached much earlier at $n_1$.\par
\begin{table}[b!] 
\caption{Transmission coefficients used in Test 2a and 2b (Fig.~\ref{fig.NetworkY} and Fig.~\ref{fig.NetworkY_v1}), given for each node-arc interface $\mathcal{I}$.} \label{tab:networkY} 
\centering
\begin{tabular}{ l | c c c }
\hline
\multirow{2}*{$\mathcal{I}(n_1, a_1)$} 	&$\alpha_{n_1,a_1} =1.00$  &$\alpha_{n_1,n_1} =0.50$\\
											&$\alpha_{a_1,a_1} =0.00$  &$\alpha_{a_1,n_1} =0.50$ \\
\hline
\multirow{2}*{$\mathcal{I}(a_1, n_2)$} 	&$\alpha_{a_1,n_2} =0.05$  &$\alpha_{a_1,a_1} =0.10$\\
											&$\alpha_{n_2,n_2} =0.95$  &$\alpha_{n_2,a_1} =0.90$ \\
\hline
\multirow{2}*{$\mathcal{I}(n_2, a_2, a_3)$} 	
							&$\alpha_{n_2,a_3} =0.50$  &$\alpha_{n_2,a_2} =0.50$   &$\alpha_{n_2,n_2} =0.995$\\
							&$\alpha_{a_2,a_3} =0.50$  &$\alpha_{a_2,a_2} =0.00$  &$\alpha_{a_2,n_2} =0.003$\\
							&$\alpha_{a_3,a_3} =0.00$  &$\alpha_{a_3,a_2} =0.50$  &$\alpha_{a_3,n_2} =0.002$\\
\hline
\multirow{2}*{$\mathcal{I}(a_2, n_3)$} 	&$\alpha_{a_2,n_3} =0.00$  &$\alpha_{a_2,a_2} =0.00$\\
											&$\alpha_{n_3,n_3} =1.00$  &$\alpha_{n_3,a_2} =1.00$ \\
\hline
\multirow{2}*{$\mathcal{I}(a_3, n_4)$} 	&$\alpha_{a_3,n_4} =0.00$  &$\alpha_{a_3,a_3} =0.00$\\
											&$\alpha_{n_4,n_4} =1.00$  &$\alpha_{n_4,a_3} =1.00$ \\
\hline
\end{tabular} 
\end{table} 
\smallskip
\paragraph{\it Test 1b}
In the second scenario, we allow mobility also of susceptible and recovered subjects, to whom the same velocity and relaxation time of infectious individuals is assigned ($\lambda^2 = 10$, $\tau = 0.1$). The additional mobility of susceptible individuals appears evident when comparing the evolution at $n_2$ in Fig.~\ref{fig.Network1} to the one in Fig.~\ref{fig.Network1_new}. In this second simulation, indeed, hardly anyone remains at node $n_2$ and the whole population moves symmetrically towards nodes $n_1$ and $n_3$. In these last nodes, the epidemic spreads in a similar way to that seen in the previous scenario, with the difference that a percentage of the healthy population continues to arrive even when the disease is in regression. Furthermore, it is worth to be mentioned that in this case study, considering the network as a whole, more individuals are infected: almost 67\% against the 60\% of infected population obtained in the previous simulation. This result is explained by the fact that the mobility added to the susceptible in this case does not avoid contagion to those who had previously remained stationary at $n_2$. 
\subsubsection{A four-node network with social distancing}
In Test 2, we consider the spread of an infectious disease in a more complex network, composed by 4 nodes and 3 bidirectional arcs, as shown in Fig.~\ref{fig.NetworkY}. Arcs have length $L_1 = 2$ and $L_2 = L_3 = 5$ and the whole network is discretized with a grid size $\Delta x = 0.05$. In these tests, the epidemic outbreak starts from $n_1$ (small city) and propagates along the whole network, reaching first $n_2$ (big city) and then $n_3$ and $n_4$ (medium-size cities). 

\smallskip
\paragraph{\it Test 2a}
In a first scenario, initial conditions are again null in all the arcs, while at nodes are imposed as follows:
\begin{align*}
&S(n_1,0) = 0.1225, & I(n_1,0)&= 0.0025, & &R(n_1,0) = 0.0,\\
&S(n_2,0) = 0.3750, & I(n_2,0)&= 0.0, & &R(n_2,0) = 0.0,\\
&S(n_3,0) = 0.2500, & I(n_3,0)&= 0.0, & &R(n_3,0) = 0.0 ,\\
&S(n_4,0) = 0.2500, & I(n_4,0)&= 0.0, & &R(n_4,0) = 0.0 ,
\end{align*}
with zero initial fluxes for all the classes at each location. Transmission conditions at each interface are given in Table \ref{tab:networkY} and it can be seen that these coefficients define a more complex dynamics with respect to the one of the previous tests. We set the epidemic parameters $\beta = 1.5$ and $\gamma = 2.0$ in the whole network. Furthermore, to verify the effects of a control measure of social distancing on the population, we impose $k = 1$ at $n_3$, leaving $k = 0$ in the rest of the network. The same transport parameters are given to all the three compartments, fixing again $\lambda^2 = 10$ and $\tau = 0.1$. In Fig.~\ref{fig.NetworkY} numerical results of the evolution of the infectious disease at each node are presented. From the outbreak location, $n_1$, the population starts traveling through arc $a_1$, to reach node $n_2$. The former receives only later in time a return of part of the susceptible individuals arriving from $n_2$. The resulting mobility of susceptible and infectious people along arc $a_1$ is shown in the two initial plot of Fig.~\ref{fig.NetworkY}. At node $n_2$ it can be seen an epidemiological dynamics representative of a node of transit, in which individuals remain partially untouched by the disease. Waves of motion along arc $a_2$ are represented in the first two plots in the second line of Fig.~\ref{fig.NetworkY}. These waves are very similar to those observed in arc $a_3$, as a consequence of the transmission coefficients chosen (reason why graphical results regarding $a_3$ are omitted). There are few more people moving in the direction of $n_3$ than there are traveling towards $n_4$, with both the nodes accommodating an initial susceptible population of equal size. Nevertheless, the spread of the infectious disease is not comparable in these two cities, due to the control measure imposed at $n_3$ ($k=1$) and not at $n_4$ ($k=0$). In fact, it is here confirmed the impact of the enforcement of social distancing measures, which permit to significantly lower the epidemic peak at $n_3$ and prevent a larger part of the population from being infected. This is also presented in terms of reproduction number $R_0$, whose evolution is shown in the same plots. 
Moreover, in Fig.~\ref{fig.NetworkY_arc1_x}, it can be observed a comparison of the results of the infectious wave, spreading similarly to a shock wavefront along arc $a_1$, at two different initial time steps, obtained with a spacial discretization characterized by $\Delta x = 0.01$ when adopting the here proposed second-order AP IMEX Runge-Kutta scheme or when choosing the corresponding first-order scheme. These numerical results are presented with respect to a reference solution obtained with the second-order scheme and a very refined grid ($\Delta x = 0.00125$). The higher spatial resolution of the results obtained with the second-order scheme is therefore confirmed, together with the relevance of the choice of high-order methods, especially for the description of spatial dynamics as those presented in this work.\par
\begin{table}[b!] 
\caption{Transmission coefficients used in Test 2c (Fig.~\ref{fig.NetworkYclosed}), given for each node-arc interface $\mathcal{I}$.} \label{tab:networkYclosed} 
\centering
\begin{tabular}{ l | c c c }
\hline
\multirow{2}*{$\mathcal{I}(n_1, a_1)$} 	&$\alpha_{n_1,a_1} =1.00$  &$\alpha_{n_1,n_1} =0.50$\\
											&$\alpha_{a_1,a_1} =0.00$  &$\alpha_{a_1,n_1} =0.50$ \\
\hline
\multirow{2}*{$\mathcal{I}(a_1, n_2)$} 	&$\alpha_{a_1,n_2} =0.05$  &$\alpha_{a_1,a_1} =0.10$\\
											&$\alpha_{n_2,n_2} =0.95$  &$\alpha_{n_2,a_1} =0.90$ \\
\hline
\multirow{2}*{$\mathcal{I}(n_2, a_2, a_3)$} 	
							&$\alpha_{n_2,a_3} =0.50$  &$\alpha_{n_2,a_2} =0.10$   &$\alpha_{n_2,n_2} =0.698$\\
							&$\alpha_{a_2,a_3} =2.50$  &$\alpha_{a_2,a_2} =0.00$  &$\alpha_{a_2,n_2} =1.50$\\
							&$\alpha_{a_3,a_3} =0.00$  &$\alpha_{a_3,a_2} =0.10$  &$\alpha_{a_3,n_2} =0.002$\\
\hline
\multirow{2}*{$\mathcal{I}(a_2, n_3)$} 	&$\alpha_{a_2,n_3} =0.00$  &$\alpha_{a_2,a_2} =0.00$\\
											&$\alpha_{n_3,n_3} =1.00$  &$\alpha_{n_3,a_2} =0.20$ \\
\hline
\multirow{2}*{$\mathcal{I}(a_3, a_4, n_4)$} 	
							&$\alpha_{a_3,n_4} =0.00$ 	&$\alpha_{a_3,a_4} =0.00$  &$\alpha_{a_3,a_3} =0.00$\\
							&$\alpha_{a_4,n_4} =0.005$ 	&$\alpha_{a_4,a_4} =0.00$  &$\alpha_{a_4,a_3} =0.00$\\
							&$\alpha_{n_4,n_4} =0.995$ 	&$\alpha_{n_4,a_4} =1.00$  &$\alpha_{n_4,a_3} =1.00$\\
\hline
\multirow{2}*{$\mathcal{I}(n_3, a_4)$} 	&$\alpha_{n_3,a_4} =1.00$  &$\alpha_{n_3,n_3} =1.00$\\
											&$\alpha_{a_4,n_3} =0.00$  &$\alpha_{a_4,a_4} =0.00$ \\
\hline
\end{tabular} 
\end{table} 

\smallskip
\paragraph{\it Test 2b}
Next, we conduct an experiment in which the population abundances outside the nodes are not initially null, which might represent a more realistic application, as the road connecting two cities would arguably be dotted with smaller towns and villages. Different Gaussian distributions of susceptible people, characterized by a variance $\sigma = 0.1$, are initially located along the arcs, while initial conditions here remain null for infectious and recovered people and their corresponding fluxes. In particular, we consider a Gaussian distribution centered in the middle of arc $a_1$ globally accounting for $S = 0.05$, which moves symmetrically towards $n_1$ and $n_2$; at one-quarter of $a_2$ and $a_3$ we collocate a Gaussian distribution enclosing $S = 0.10$, which moves towards $n_2$, and at three-quarter of the same two arcs we set a Gaussian distribution enclosing $S = 0.05$, which moves towards $n_3$ and $n_4$, respectively. The rest of initial conditions and parameters are left unvaried in the network with respect to the previous scenario.
The different results obtained in this second scenario, presented in Fig. \ref{fig.NetworkY_v1}, can be compared to those previously obtained in Test 2a (Fig. \ref{fig.NetworkY}), especially noticing in this case the presence of the Gaussian distributions of susceptible people along $a_1$ and $a_2$ in the first plots. It can also be observed that the presence of an additional amount of susceptible moving along the arcs affects the evolution of the disease especially at node $n_2$, which accommodate the incoming of a major quantity of population, allowing the epidemic to spread more consistently.

\begin{figure}[t!]
\centering
\begin{subfigure}{0.32\textwidth}
\includegraphics[width=1\linewidth]{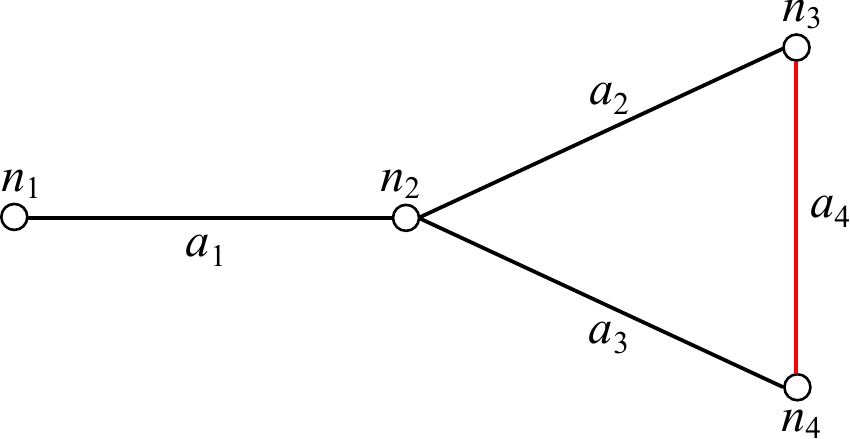}
\caption{}
\label{fig.networkYclosed_scheme}
\end{subfigure}
\begin{subfigure}{0.32\textwidth}
\includegraphics[width=1\linewidth]{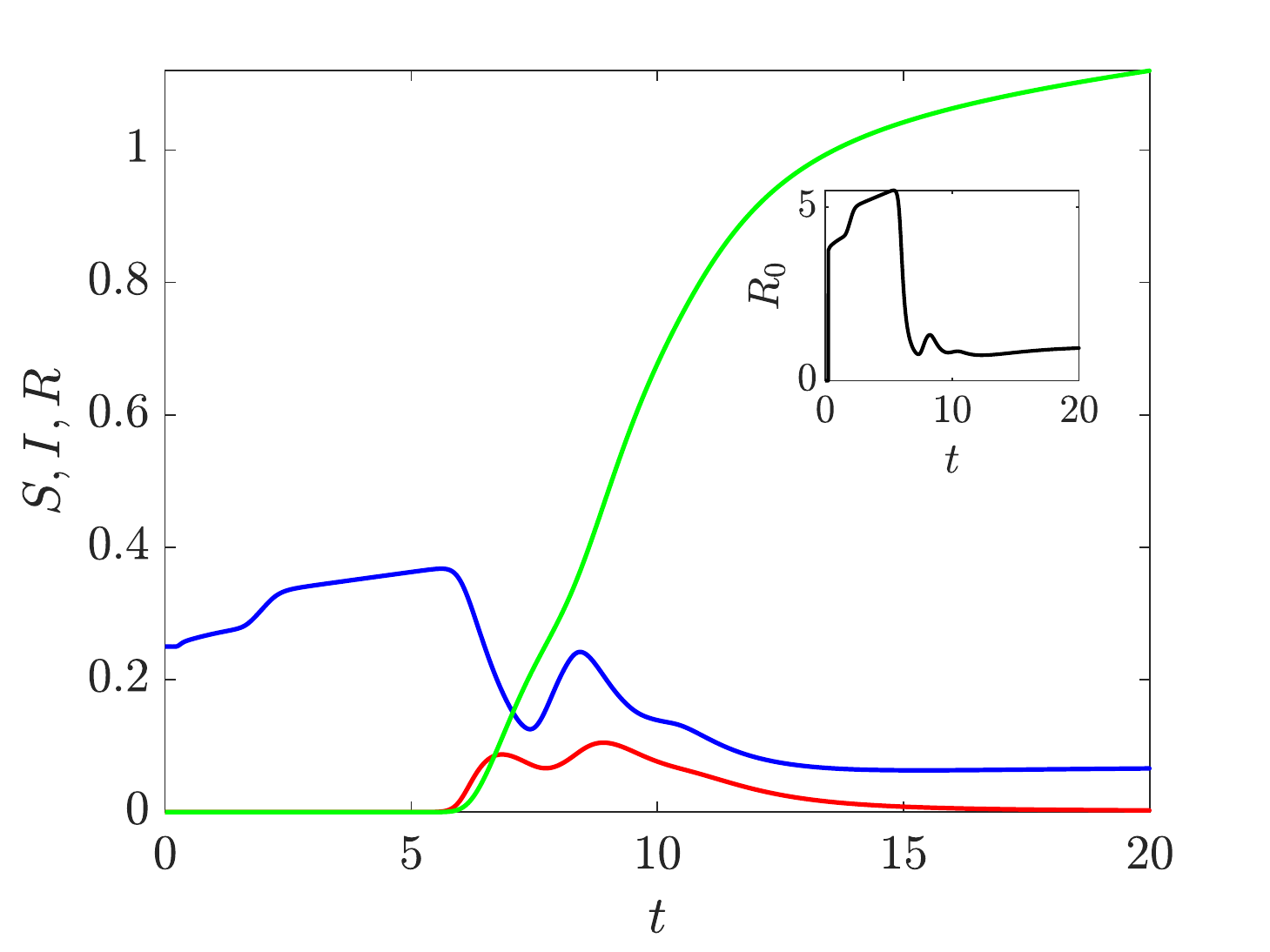}
\caption{($n_3$ - case 1)}
\label{fig.NetworkYclosed_node3}
\end{subfigure}
\begin{subfigure}{0.32\textwidth}
\includegraphics[width=1\linewidth]{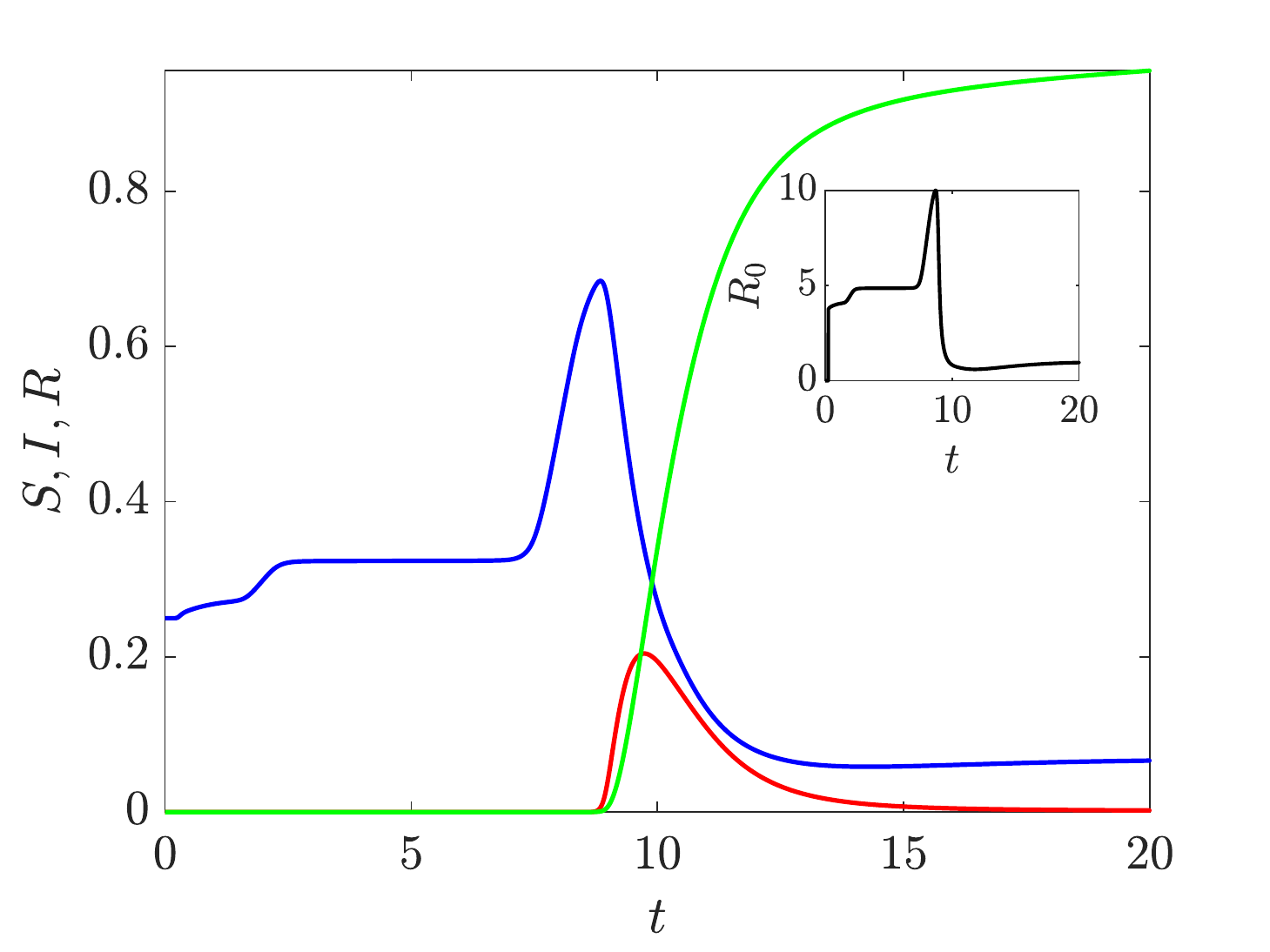}
\caption{($n_3$ - case 2)}
\label{fig.NetworkY_node3_v2}
\end{subfigure}
\caption{Numerical results of Test 2c at node $n_3$ when considering the presence of arc $a_4$ (case 1) and when leaving $n_3$ and $n_4$ disconnected (case 2). Time evolution of $S$ (blue), $I$ (red) and $R$ (green) dynamics and of the coefficient $R_0$.}
\label{fig.NetworkYclosed}
\end{figure}

\smallskip
\paragraph{\it Test 2c}
In the last test case, we initially change the properties of arc $a_2$, reducing the characteristic speed of all the SIR compartments to $\lambda^2 = 0.4$ and increase the relaxation time to $\tau = 40$, to simulate a slowdown along this connection. Furthermore, the amount of people leaving $n_2$ to reach $n_3$ is augmented by 100 times (in Test 2a and 2b there is 0.3\% of population, while in Test 2c there is 30\%).
Then, to assess the impact of the connectivity in transport networks, in this new scenario nodes $n_3$ and $n_4$ are connected by an additional arc $a_4$ (see Fig. \ref{fig.NetworkYclosed}), having length $L_4 = 2$, and an initial Gaussian distribution of susceptible, again characterized by $\sigma = 0.1$, is inserted in the middle of $a_4$ (presenting in total $S = 0.05$) and is set to spread symmetrically towards the two nodes. In this arc we fix $\lambda^2 = 10$ and $\tau = 0.1$. Transmission conditions of Test 2c are given in Table \ref{tab:networkYclosed} for each interface. Numerical results concerning the evolution of the epidemic at node $n_3$ are presented in Fig. \ref{fig.NetworkYclosed} when considering the presence of arc $a_4$ (case 1) and when leaving $n_3$ and $n_4$ disconnected (case 2). It can be observed that the simple presence of the connectivity between $n_3$ and $n_4$ highly affects the course of the infectious disease at the node of interest. Indeed, when $a_4$ is present, the development of the epidemic at $n_3$ is consistently anticipated in time due to the arrival of part of the population from $n_4$. Moreover, the later inlet of population arriving from $n_2$ causes a second peak of infection, even slightly higher than the first one. In the second case, instead, the infectious wave has a unique peak that manifests itself after the arrival of the sole population leaving $n_2$ to reach $n_3$. Clearly, it can be seen that also the evolution of the coefficient $R_0$ is affected by the presence of $a_4$.
Results in the rest of the network are not shown because similar to those obtained in Test 2b and not relevant in order to identify the effects of a different connectivity pattern in the network. 
\section{Conclusions}
\label{section_conclusions}
In this work, a novel SIR-type kinetic transport model for the spread of infectious diseases on networks is presented. The hyperbolic system describes at a macroscopic level the propagation of epidemics at finite speeds, recovering the classical one-dimensional reaction-diffusion model as relaxation times and characteristic speeds of each compartment of the population (susceptible, infectious and recovered individuals) tend to zero and infinity, respectively. The extension of the model to the treatment of networks that identify at each node particular limited locations (like a specific city or region), connected by paths (network arcs) along which individuals can move to reach different destinations, is also presented. In this context, at each node the compartmental SIR model with speed alignment describes the evolution of the epidemic in the areas of major interest, which are affected by the mobility of the population in the network, computed in the arcs through the here proposed SIR-type kinetic transport model. To ensure a correct coupling between nodes and arcs, proper transmission conditions are defined at each arc-node interface, which guarantee the conservation of the global mass of the system.

To solve the system of equations on each arc, a second-order IMEX finite volume method 
that is robust enough to correctly capture the asymptotic behavior of hyperbolic systems under different kinds of scaling is proposed. In particular, the numerical method satisfies the AP property in the stiff regime, hence in the parabolic diffusive limit. The expected accuracy of the numerical scheme is confirmed for all the variables of the problem by means of accuracy analysis, even when dealing with stiff regimes, characterized by relaxation times close to zero. Furthermore, the behavior of the model when considering spatially heterogeneous environments is investigated concerning different configurations of transport parameters (characteristic velocities and relaxation times). Results confirm that the effects of a spatial variability of the contact rate vanish when the reproduction number of the infectious disease, $R_0$, is less than 1, while a temporary persistence of the infectious is noticed when $R_0 >1$. Finally, in order to demonstrate the suitability of the proposed model to simulate the spread of epidemic diseases on networks, five tests are carried out concerning different simple networks. The impact of human mobility in these networks is assessed evaluating different mobility patterns, transmission coefficients at interfaces and social distancing interventions. Numerical results underline that these characteristics highly influence the course of an epidemic, therefore confirming that restrictions on the mobility of people and social distancing measures are very effective in reducing the spread of an infectious disease. In addition, it has been confirmed the paramount importance in transport networks of graph connectivity. Finally, the higher spatial resolution of results obtained with the here proposed second-order IMEX scheme with respect to those obtained with the corresponding first-order scheme is confirmed, together with the relevance of the choice of high-order methods, especially for the description of spatial dynamics as those here presented.

It is worth to mention that, even if out of the scope of the present work, in the model here proposed, all the parameters could potentially be set or estimated according to real data, namely: real amount of people and measured initial infectious in cities (many of which daily updated and available in various GitHub repository); mobility data tracked by Google systems and recently being released \cite{aktay2020}, which have been collected by geographical location in categories of retail and recreation, groceries and pharmacies, parks, transit stations, workplaces and residential, following the work done in \cite{vollmer2020}; or recurring to official national assessment of mobility flows, as used in the case of Italy in \cite{gatto2020}. This aspect is currently under study and will be part of the future developments.
Moreover, since data of the spread of epidemics are generally highly heterogeneous and affected by a great deal of uncertainty, future perspectives include the application of uncertainty quantification methods to assess the impact of stochastic inputs in the proposed SIR-type kinetic transport model. Finally, an extension of the model for the inclusion of the age structure of the population is foreseen, being an essential characteristic to correctly describe the impact of specific kinds of infectious diseases, like the COVID-19 pneumonia.
\section*{Acknowledgements}
This work was partially supported by MIUR (Ministero dell'Istruzione, dell'Universit\`a e della Ricerca) PRIN 2017, project \textit{``Innovative numerical methods for evolutionary partial differential equations and applications''}, code 2017KKJP4X. 
\bibliographystyle{abbrv}
\bibliography{SIRnetwork_Bertaglia2020_rev1}

\end{document}